\documentclass[a4paper,11pt]{article}
\pdfoutput=1
\usepackage{amsmath,amssymb,url}
\usepackage{graphicx,mathpazo,subfigure}
\usepackage{braket,euscript,hyperref,cite}
\usepackage{float}
\usepackage{graphicx}
\usepackage{braket,caption,a4wide}
\numberwithin{equation}{section}

\begin{document}
\title{{\bf No inverse magnetic catalysis in the QCD hard and soft wall models}}
\author{David Dudal$^{a,b}$\footnote{david.dudal@kuleuven.be}, Diego R.~Granado$^{c,b}$\footnote{diegorochagrana@uerj.br}, Thomas G.~Mertens$^{d,b}$\footnote{tmertens@princeton.edu}}
\date{}
\maketitle
\vspace{-1cm}
\begin{center}
{\footnotesize $^a$ KU Leuven Campus Kortrijk - KULAK, Department of Physics, Etienne Sabbelaan 53, 8500 Kortrijk, Belgium}\\
{\footnotesize $^b$ Ghent University, Department of Physics and Astronomy, Krijgslaan 281-S9, 9000 Gent, Belgium}\\
{\footnotesize $^c$ Departamento de F\'{i}sica Te\'{o}rica, Instituto de F\'{i}sica, UERJ - Universidade do Estado do Rio de Janeiro,
Rua S\~ao Francisco Xavier 524, 20550-013, Maracan\~a, Rio de Janeiro, Brasil }\\
{\footnotesize $^d$ Joseph Henry Laboratories, Princeton University, Princeton, NJ 08544, USA }
\end{center}

\begin{abstract}
In this paper, we study the influence of an external magnetic field in holographic QCD models where the backreaction is modeled in via an appropriate choice of the background metric. We add a phenomenological soft wall dilaton to incorporate better IR behavior (confinement). Elaborating on previous studies conducted by \cite{Mamo:2015dea}, we first discuss the Hawking-Page transition, the dual of the deconfinement transition, as a function of the magnetic field. We confirm that the critical deconfinement temperature can drop with the magnetic field.
Secondly, we study the quark condensate holographically as a function of the applied magnetic field and demonstrate that this model does not exhibit inverse magnetic catalysis at the level of the chiral transition. The quest for a holographic QCD model that qualitatively describes the inverse magnetic catalysis at finite temperature is thus still open.\\Throughout this work, we pay special attention to the different holographic parameters and we attempt to fix them by making the link to genuine QCD as close as possible. This leads to several unanticipated and so far overlooked complications (such as the relevance of an additional length scale $\ell_c$ in the confined geometry) that we discuss in detail.
\end{abstract}

\section{Introduction}
\label{sec:intro}
(De)confinement and chiral symmetry breaking/restoration are important features of quantum chromodynamics (QCD). A good way to describe them consistently is very hard due to the nonperturbative character of these phenomena. Along the years, several tools have been developed in order to access this regime. A recent paradigm rests on the AdS/CFT correspondence, well suited to nonperturbative regimes of strongly coupled gauge theories as QCD. \\

\noindent The AdS/CFT correspondence maps a strongly coupled conformal theory living in flat space-time into a weakly coupled theory living in a higher dimensional anti-de Sitter space. In \cite{maldacena,Aharony:1999ti} the conformal field theory is $\mathcal{N}=4$ supersymmetric Yang-Mills. However, most of the interesting strongly coupled systems found in nature (such as QCD) do not have conformal symmetry. QCD, for example, is neither supersymmetric nor conformal: its nonzero running coupling constant shows that the conformal symmetry in QCD is broken. QCD definitely needs a (dynamical) mass scale to explain its spectrum. Several holographic models that exhibit this breaking have been constructed (so-called AdS/QCD models). One can basically distinguish two approaches: the first (top-down) approach utilizes stringy constructions to access field theory \cite{Karch:2002sh,polchinski,Sakai:2004cn,Sakai:2005yt,
Kruczenski:2003be, Kruczenski:2003uq,Erdmenger:2007cm}; whereas the other (bottom-up) approach involves phenomenological models \cite{Erlich:2005qh,Karch:2006pv,Karch:2010eg,dePaula:2008fp,
Gursoy:2007cb,Gursoy:2007er,Gursoy:2008bu,Gursoy:2009jd,Jarvinen:2011qe,Alho:2012mh}, where we constrain the bulk theory as to reproduce the desirable features of QCD. \\

\noindent Our goal in this work is to analyze more closely the implementation of a background magnetic field in QCD in a holographic set-up. Multiple studies have been performed in the past where this magnetic field is modeled as a bulk diagonal flavor gauge field whose matrix elements are proportional to the electric charge of each of the quark flavors. The holographic dictionary then guarantees a correct coupling to a magnetic field in the boundary theory. One typically takes a holographic geometry, then one solves Maxwell's equations on this background to obtain a magnetic field solution, and finally diverse quantities (correlation functions, spectral functions, thermodynamic quantities etc.) are holographically computed in this geometric and gauge background \cite{Johnson:2008vna,Callebaut:2013ria,Ballon-Bayona:2013cta,Callebaut:2011ab,Callebaut:2013wba,Dudal:2014jfa,Dudal:2015kza,Sadofyev:2015hxa,Ali-Akbari:2013txa,Ali-Akbari:2015bha,Rougemont:2014efa,Rougemont:2015oea}. However, the backreaction of the magnetic field on the geometry itself is usually neglected. Several years ago, D'Hoker and Kraus solved the Einstein-Maxwell system with asymptotic AdS boundary conditions and a constant magnetic field in the bulk \cite{D'Hoker:2009mm,D'Hoker:2009bc}. This model hence cures these earlier deficiencies. \\

\noindent We are interested in modifying this model in the infrared to account for the correct phenomenological predictions of QCD. In this paper we will follow the path of phenomenological AdS/QCD models. Among several options available on the holographic market, there are two well-known models namely, the hard wall  \cite{Erlich:2005qh} and soft wall model \cite{Karch:2006pv,Karch:2010eg}. Both of these models can generate essential features of confinement and chiral symmetry breaking, but they utilize different strategies. In the hard wall model, one introduces a cut-off in the gravitational geometry that confines the space. Recently, Mamo examined the Hawking-Page phase transition in the D'Hoker-Kraus background with hard wall cut-off, related to the confinement/decon\-finement transition in the boundary theory \cite{Mamo:2015dea}. In the soft wall model, one introduces an extra field that explicitly breaks the conformal symmetry in the IR regime of the theory. Regarding confinement, both models experience some issues. The hard wall model is, for example not capable of reproducing the linear Regge trajectory and the soft wall model fails in the sense that the Wilson loop vacuum expectation value does not present an area law \cite{Karch:2010eg}. \\

\noindent Regarding the chiral phase transition, each model also has its own drawback. The soft wall model directly relates the bare quark mass $m$ with the chiral condensate $\braket{\bar{\psi}\psi}$ in the sense that $\braket{\bar{\psi}\psi}\sim m$.\footnote{Explicit symmetry breaking and spontaneous symmetry breaking are not allowed to be described separately. Such can be overcome by playing with adding potentials for both dilaton and the scalar field representing the chiral condensate, at the cost of more complicated equations \cite{Gherghetta:2009ac,Bartz:2014oba,Fang:2015ytf,Chelabi:2015gpc}. Throughout this work, we will mean by $\braket{\bar{\psi}\psi}$ only a single flavor. In practice, we will look at the degenerate up and down sector, and the total condensate (which we will denote by $\braket{\bar{Q}Q}$) should then be twice what we determine.} Such a relation does not exist in QCD.  In the hard wall model the parameters $m$ and $\braket{\bar{\psi}\psi}$ can not be determined dynamically (at least in the confined phase): they act as independent constraints that one has to impose on the theory. We will come back to this model in the end and demonstrate that it is quite pathological when considering the chiral dynamics. Despite the drawback mentioned before, the soft wall model is the one capable of providing the desired results. \\

\noindent So the model we will work with is the geometry and gauge field background obtained by D'Hoker and Kraus, supplemented ``by hand'' with the soft wall dilaton field to model in confinement, this completely analogous to the rationale behind the original soft wall model construction \cite{Karch:2006pv}. We remark at the outset already that the resulting model does not solve Einstein's equations. However, the soft wall model can be viewed as a phenomenological model and a first step towards obtaining intuition and insight into the effects that might occur in real QCD. It seems that by including the soft wall dilaton field, we are taking a step back again. D'Hoker and Kraus finally obtained a fully backreacted solution, while we again ignore parts of the backreaction (of the dilaton). Note though that this actually can be viewed as a piecewise process towards the final answer: we include the magnetic field in a more satisfying way and we improve this model in the infrared by including a soft wall. We hence expect that this model is a step forward towards real QCD with magnetic fields. We will come back to the issue on how to relate the bulk and boundary magnetic field later in this paper. Notice that the D'Hoker-Kraus solution describes the holographic dual for magnetized $\mathcal{N}=4$ supersymmetric Yang-Mills with hence adjoint flavors in the boundary theory. However, soft-wall models are utilized to understand real QCD (with fundamental flavors). Within the same philosophy, we employ our soft-wall modified D'Hoker-Kraus solution with the hope of understanding magnetized QCD with fundamental flavors. \\

\noindent The QCD deconfinement and chiral transition phase diagram under the influence of the magnetic field has been studied before using a myriad of approaches, next to the already quoted papers let us also refer to e.g.~\cite{Gusynin:1994re,Miransky:2002rp,Mizher:2010zb,Fraga:2008um,Gatto:2010pt,Gatto:2010qs,Osipov:2007je,Kashiwa:2011js,Klimenko:1992ch,Fraga:2013ova,Alexandre:2000yf,Filev:2007gb,Albash:2007bk,Bergman:2008sg,
Evans:2010xs,Alam:2012fw,Preis:2010cq,Filev:2010pm,Callebaut:2011zz,Bolognesi:2011un,Bali:2011qj,Bali:2012zg,Bali:2014kia,Bonati:2014ksa,Bonati:2013lca,D'Elia:2011zu,D'Elia:2010nq,Ilgenfritz:2013ara,Ilgenfritz:2012fw,Frasca:2011zn,Fukushima:2012xw,Ferreira:2013oda,McInnes:2015kec} or \cite{Kharzeev:2012ph,Miransky:2015ava} for recent reviews. The interest in this was revived since it became clear that strong magnetic fields are most likely generated during the early stages of noncentral heavy ion collisions and with a lifetime that persists into the quark-gluon plasma phase \cite{Kharzeev:2007jp,Skokov:2009qp,Bzdak:2011yy,Deng:2012pc,Tuchin:2013apa,Tuchin:2013ie,McLerran:2013hla}.

\noindent Despite the fact that the recent lattice results \cite{Bali:2012zg,Bali:2011qj,Ilgenfritz:2013ara} indicate an \textit{inverse magnetic catalysis} (the critical temperature decreases under the influence of the magnetic field, at least in the explored regime of magnetic fields and temperature), most of the (holographic) QCD phase diagram models predict \textit{magnetic catalysis} \cite{Miransky:2002rp,Johnson:2008vna,Filev:2010pm,Callebaut:2011zz,Bolognesi:2011un,Rougemont:2015oea}. Non-holographic approaches towards inverse magnetic catalysis can be found in \cite{Fraga:2012fs,Fraga:2012ev,Fukushima:2012kc,Ayala:2014iba,Ayala:2014gwa,Ayala:2015bgv,Farias:2014eca,Ferreira:2014kpa,Costa:2015bza,Mueller:2015fka}.\\

\noindent In this paper we study the influence of a magnetic field in both chiral and confinement/decon\-finement phase transitions using phenomenological AdS/QCD models. In Section \ref{sect2} we review the elements of the magnetized background geometry, both confining and deconfining, in Einstein-Maxwell theory in 5D used in \cite{Mamo:2015dea}, which is based on work of D'Hoker and Kraus \cite{D'Hoker:2009mm,D'Hoker:2009bc}. We describe how to embed the latter into a soft wall model intended to describe magnetized QCD. Throughout this process, we will see that the confined geometry actually contains a free dimensionful parameter $\ell_c$ that affects physical quantities. We will fix it later on in this work. Another feature that will be explored is the region of validity of the solution itself and how the hard and soft walls actually save the models in the end. This section is supplemented by material collected in the Appendix in which the structure of the black hole solution is explored. We want to remark that (to the best of our knowledge) we are the first to explore these backgrounds to this level of accuracy.  \\
Armed with this knowledge, we study in Section \ref{sect4} in detail the Hawking-Page transition in both the hard wall and soft wall model setting, thereby obtaining the magnetic field dependent confinement/deconfinement transition. The hard wall analysis is a revisiting of \cite{Mamo:2015dea} in which case we add some clarifications, the soft wall results are new. In both cases, we recover that the critical deconfinement temperature drops with increasing magnetic field, at least for reasonable values of the length scale $\ell_c$ that we will introduce. In Section \ref{sect5}, we include for completeness the thermodynamical stability analysis. We continue in Section \ref{sect6} to scrutinize the chiral condensate, symmetry breaking and related restoration at finite temperature and magnetic field, thereby extending the earlier (zero magnetic field) results of \cite{Colangelo:2011sr}. It becomes clear there is no sign of inverse magnetic catalysis. We end with our conclusion in Section \ref{sect7}. We have relegated several computational details to a series of Appendices, in which we also analyze the horizon structure of the D'Hoker-Kraus black hole solution, the relative normalization of the magnetic field in bulk vs.~boundary, next to how a meaningful finite (renormalized) chiral condensate can be derived.

\section{Holographic set-up}\label{sect2}
\subsection{Einstein-Maxwell action and its magnetized AdS black hole solution}
In this section we set the stage by describing the action and classical solution found in \cite{D'Hoker:2009mm,D'Hoker:2009bc}. The Einstein-Maxwell action is given by:\footnote{$M$ stands for Minkowski signature.}
\begin{equation}
S^M=S^M_{\text{bulk}}+S^M_{\text{bndy}},
\label{action}
\end{equation}
where the bulk piece $S^M_{\text{bulk}}$ is:
\begin{equation}
S^M_{\text{bulk}}=\frac{1}{16\pi G_5}\int d^5x\sqrt{-g}\left(R-F^{MN}F_{MN}+\frac{12}{L^2}\right),
\label{sbulk}
\end{equation}
with $\sqrt{-g}=\sqrt{-\det g_{\mu\nu}}$, $F_{MN}$ is the electromagnetic field strength, $R$ is the Ricci scalar and $\Lambda=-\frac{12}{L^2}$ is the negative cosmological constant. The second piece is the boundary action $S^M_{\text{bndy}}$ consisting of the Gibbons-Hawking surface term and holographic counterterms to cancel the UV divergence (close to the AdS boundary). These are introduced as boundary terms. This action $S^M_{\text{bndy}}$ is of the following form:
\begin{equation}
S^M_{\text{bndy}}=\frac{1}{8\pi G_5}\int d^4x\sqrt{-\gamma}\left(K-\frac{3}{L}-\frac{L}{2}F^{\mu\nu}F_{\mu\nu}\left(\ln \frac{r}{L}\right)\right)\Bigg|_{r_\lambda}.
\label{sbndy}
\end{equation}
The 5D solution will be written in coordinates ($t$, $x$, $y$, $z$, $r$) where the radial holographic coordinate $r$ will be introduced shortly; the boundary is found at $r=0$. $r_\lambda$ is introduced as a regulating UV cut-off for the divergence at $r=0$. $\gamma$ denotes the determinant  of the induced metric $\gamma_{\mu\nu}$ at $r\to\infty$:
\begin{equation}
\gamma_{\mu\nu}\equiv\text{diag}\left(g_{tt},g_{xx},g_{yy},g_{zz}\right)
\end{equation}
and $K$ is the trace of the extrinsic curvature: $K:=\gamma^{\mu\nu}K_{\mu\nu}=-\sqrt{g^{rr}}\frac{\partial_r\sqrt{\gamma}}{\sqrt{\gamma}}$. \\

\noindent The equations of motion obtained from \eqref{sbulk} are:
\begin{eqnarray}
\label{ricciequation}
R_{MN}&=&-\frac{4}{L^2}g_{MN}-\frac{1}{3}F^{PQ}F_{PQ}g_{MN}+2F_{MP}F_N{}^P,\\
\nabla^MF_{MN}&=&0.
\label{maxwelleq}
\end{eqnarray}

\noindent Next we describe the D'Hoker-Kraus solution. The black hole metric (perturbative in $B$) that was found in \cite{D'Hoker:2009mm,D'Hoker:2009bc} is:\footnote{It is found by setting the charge density $\rho=0$ for the solution in section 6 of \cite{D'Hoker:2009bc}.}
\begin{equation}
\label{bhmetric}
ds_{bh}^2=\frac{L^2}{r^2}\left(-f(r)dt^2+q(r)dz^2+h(r)\left(dx^2+dy^2\right)+\frac{dr^2}{f(r)}\right)+\mathcal{O}(B^4),
\end{equation}
where $L$ is the AdS radius.

The coefficient functions appearing in this metric are:
\begin{eqnarray}
f(r)&=&1-\frac{r^4}{r_h^4}+\frac{2}{3}\frac{B^2r^4}{L^2}\ln \left(\frac{r}{\ell_d}\right)+\mathcal{O}(B^4), \\
q(r)&=&1+\frac{8}{3}\frac{B^2}{L^2}\int_{+\infty}^{1/r}dx \frac{\ln(r_h x)}{x^3\left(x^2-\frac{1}{r_h^4x^2}\right)}+\mathcal{O}(B^4), \\
h(r)&=&1-\frac{4}{3}\frac{B^2}{L^2}\int_{+\infty}^{1/r}dx \frac{\ln(r_h x)}{x^3\left(x^2-\frac{1}{r_h^4x^2}\right)}+\mathcal{O}(B^4),
\end{eqnarray}
and a constant magnetic field $B$ in the $z$-direction $F_{xy}$ indeed solves the Maxwell equations \eqref{maxwelleq}.
Some comments are in order at this point. In $f(r)$ we have introduced an extra length parameter $\ell_d$ that is a priori a completely independent scale in the problem: for any choice of $\ell_d$, this metric solves Einstein's equations with a constant magnetic field up to order $B^2$.  \\
The factor of $r_h$ in $\ln(r_h x)$ is chosen such that no singularity is encountered at $r=r_h$. \\

\noindent It should be noted that this solution differs from the one utilized in \cite{Mamo:2015dea} in that the functions $q(r)$ and $h(r)$ are different; even more so: the metric given in \cite{Mamo:2015dea} is not even a solution to Einstein's equations to the relevant order in $B$. However, it turns out that (luckily) this on its own does not influence the results obtained there. \\

\noindent From the Einstein equation one can then find the Ricci scalar as:
\begin{equation}
R=-\frac{20}{L^2}+\frac{2}{3}B^2g^{xx}g^{yy}.
\label{ricciscalar}
\end{equation}

\noindent A closely related background can be found by letting $r_h\to\infty$. This corresponds to a magnetized AdS solution. This is actually more subtle than one might imagine at first sight. Up to order $B^2$, a solution is
\begin{equation}
\label{thmetric}
ds_{th}^2=\frac{L^2}{r^2}\left(-f(r)dt^2+q(r)dz^2+h(r)\left(dx^2+dy^2\right)+\frac{dr^2}{f(r)}\right)+\mathcal{O}(B^4),
\end{equation}
where in this case
\begin{eqnarray}
f(r)&=&1+\frac{2}{3}\frac{B^2r^4}{L^2}\ln \left(\frac{r}{\ell_c}\right)+\mathcal{O}(B^4), \\
q(r)&=&1+\frac{8}{3}\frac{B^2}{L^2}\int_{+\infty}^{1/r}dx \frac{\ln(\ell_Y x)}{x^5}+\mathcal{O}(B^4), \\
h(r)&=&1-\frac{4}{3}\frac{B^2}{L^2}\int_{+\infty}^{1/r}dx \frac{\ln(\ell_Y x)}{x^5}+\mathcal{O}(B^4).
\end{eqnarray}
For small enough $B$, this metric indeed has no horizons.  In this case however, the length scale $\ell_c$ is of direct physical relevance. We will later on fix this parameter to find the best match with actual magnetized QCD by matching to the confined chiral condensate. \\
Since this represents the confined phase, we can expect the Hawking-Page temperature to be also sensitive to $\ell_c$ (as it uses input from both confined and deconfined phases).\\
The length scale $\ell_Y$ on the other hand is completely irrelevant for anything we might compute using this metric up to order $B^2$ in this work. \\
As is well known, the thermal AdS and the AdS black hole represent both phases of the confinement/deconfinement phase transition. The above solutions hence represent the analogues of these when a background magnetic field is turned on. Notice that we kept the AdS length $L$ explicit to keep track of dimensions.  \\

\noindent When considering both of these backgrounds as two phases in the same thermal ensemble, one requires the asymptotic geometry to match. This however is not sufficient to conclude that $\ell_d = \ell_c$ as the dominant asymptotic behavior of $f(r)$ is the same regardless of the independent choice of $\ell_c$ and $\ell_d$. \\

\noindent In Appendix \ref{sect3}, we have collected a technical analysis of the black hole described by the metric \eqref{bhmetric}, including its horizon structure in terms of the magnetic field $B$, the Hawking temperature, its extremal limit with temperature $T=0$ and the difference of the latter with the (needed) magnetized thermal AdS metric.

\noindent To make the transition to the physical boundary magnetic field requires some more thought. The above background simply describes a magnetic field embedded in AdS. In holography, it is known that one should model a magnetic field in the boundary theory by including a flavor-diagonal gauge field in the bulk. The above solution describes this for one flavor only. To proceed, we first embed this system into a larger one, more suitable to study chiral and confinement properties of the dual gauge theory. \\

\subsection{Embedding in the soft wall model}
The action that we envision of using is a generalization of the one written above, which includes multiple flavors and a soft wall dilaton:
\begin{equation}
S=S_{EH} + S_{\text{bndy}} + \frac{N_c}{16\pi^2}\int d^4x\int^{R_H}_0 dr e^{-\phi}\sqrt{-g}\text{Tr}\left[\left|DX\right|^2-m_5^2\left|X\right|^2-\frac{1}{4g_5^2}\left(F_L^2+F_R^2\right)\right].
\end{equation}
For the dilaton $\phi$, we make the standard choice \cite{Karch:2006pv}
\begin{equation}
\phi = cr^2.
\end{equation}
The scale $c$ is directly related to the QCD spectrum. \\

\noindent The background solution can be found by setting $X=0$ and $F_L = F_R \sim B$ and diagonal.\footnote{The magnetic field $B$ is included in the flavor \emph{vector} subgroup simply because an electromagnetic field couples to the Noether (vector) current $\bar{\psi}\gamma^{\mu}\psi$ of the fundamental quark fields.} Given this solution, the above action describes how gauge fluctuations (holographically dual to vector and axial currents) propagate. The $X$-field describes the quark condensate in soft wall models and the dilaton field $\phi$ ensures the IR effective cut-off of the model. Adding all of these additional fields enriches the structure that we are analyzing. The prefactors that we wrote down above have been fixed by comparing 2-point correlators in bulk and boundary \cite{Jugeau:2013zza,Colangelo:2008us,Krikun:2008tf}. Throughout this work, we will work with only two degenerate flavor indices (up and down) and study the chiral transition using the associated condensate.  \\ Our working hypothesis is thus a prolongation of the ``standard'' soft wall model. The
usual AdS space is dual to a theory of adjoint flavors. When a magnetic field is coupled
to this adjoint matter, the D'Hoker-Kraus magnetic AdS solution becomes the relevant
metric. Adding a soft wall in that space serves to model in confinement and to describe
QCD with fundamental (confined) flavors, with or without magnetic field depending on
the metric (normal AdS vs.~D'Hoker-Kraus). \\

\noindent The non-abelian (but diagonal) gauge field might worry the reader. Firstly we remark that this is still a solution to the coupled Einstein-Maxwell system, where the energy density of the magnetic field sourcing the Einstein equations gets contributions from the different flavors. The Maxwell equations are again trivially satisfied for each gauge component. What remains to be done then is to make the link between this effective magnetic field sourcing the Einstein equations, and the real physical 4D magnetic field as measured in the boundary QCD-like theory. \\
A related issue is that the magnetic field $B$ has mass dimension 1 in 5D. However, the physical 4D magnetic field  $\mathcal{B}$ should have mass dimension 2 (GeV$^2$). In order to obtain the physical magnetic field from the one in \eqref{euclideanbulkaction}, it turns out we need to rescale it such that: $\mathcal{B}=1.6\frac{B}{L}$. This is explained in detail in Appendix \ref{normalization} where we are particularly careful in making this transition. The main idea to write down such a formula, is to use the fact that the flavor gauge field has a fixed holographic coupling constant, and we insist on embedding the magnetic field in the flavor gauge field in the bulk, hence fixing its prefactor immediately. This method is different than the one utilized by \cite{D'Hoker:2009mm,D'Hoker:2009bc} for $\mathcal{N}=4$ SYM where the authors match the anomalies of bulk and boundary to fix the normalization of the physical magnetic field. Unfortunately, we cannot follow the strategy of \cite{D'Hoker:2009mm,D'Hoker:2009bc} since in bottom-up AdS/QCD models, the relative normalizations of the bulk and boundary anomalies are not fixed a priori. Usually one achieves this goal by matching the expected (known) QCD anomaly strength with the one derived from the higher-dimensional counterpart. \\

\noindent Before putting these models to work, we want to clarify some further issues related to the two backgrounds given above.

\subsection{Independence of the deconfined phase of $\ell_d$}
\label{indepL}
A curious feature is that anything we might compute in the deconfining black hole phase (\ref{bhmetric}) is actually independent of the value of $\ell_d$. To see this, one has to recall that the physical input parameters of our model are $T$ and $\mathcal{B}$. These determine directly $R_H$ through the Hawking temperature formula (\ref{hawktem}). The horizon function (which is the only place where $\ell_d$ appears) is written as
\begin{equation}
f(r) = 1 - \frac{r^4}{r_h^4} + \frac{2}{3}\frac{\mathcal{B}^2r^4}{1.6^2}\ln\left(\frac{r}{\ell_d}\right),
\end{equation}
where $r_h$ is on its own a function of $\ell_d$, determined by $f(r=R_H)=0$:
\begin{equation}
1 - \frac{R_H^4}{r_h^4} + \frac{2}{3}\frac{\mathcal{B}^2R_H^4}{1.6^2}\ln\left(\frac{R_H}{\ell_d}\right) = 0.
\end{equation}
Solving this equation for $r_h$ and plugging it into the above expression, one finds
\begin{equation}
f(r) = 1 - \frac{r^4}{R_H^4} + \frac{2}{3}\frac{\mathcal{B}^2r^4}{1.6^2}\ln\left(\frac{r}{R_H}\right),
\end{equation}
and all $\ell_d$-dependence has dropped out. \\

\noindent The only important aspect for which $\ell_d$ matters, is whether the above horizon equation can in fact be solved for real $r_h$, which is not always possible. \\
Hence, if one changes $\ell_d$, one changes the range of $\mathcal{B}$ and $T$ for which a black hole geometry is possible. Obviously, we want to maximize this region (as there is no such restriction in QCD), but one has to remember that for sufficiently large $\mathcal{B}$, we cannot trust the geometry anymore and it makes no sense to draw conclusions for higher values of $\mathcal{B}$.

\subsection{Curvature singularities and the validity of the perturbation series}
There is a troublesome feature of the magnetized AdS solution (\ref{thmetric}). The Ricci scalar in both confined and deconfined phases is the same and is equal to
\begin{equation}
R = -\frac{20}{L^2} + \frac{2}{3}\frac{r^4}{L^4}B^2.
\end{equation}
In both cases, this curvature invariant blows up as $r\to\infty$, meaning a singularity is present in the deep interior of AdS, either cloaked in a horizon (for the black hole case), or naked (for the thermal AdS case). \\

\noindent Hence what we thought was just plain magnetized AdS actually contains a naked curvature singularity at $r\to\infty$. The ambiguity with the logarithmic term in $f(r)$ shows that the difference between what we call the black hole and the thermal AdS is actually quite subtle. \\

\noindent Does this mean that this solution is completely useless? In fact, it is not and the artificial (hard or soft) walls that we include will ensure that the naked singularity spacetimes do make sense as thermal AdS as we will demonstrate now. \\

\noindent To that effect, let us better understand the conditions required for the perturbation series in $B$ to make sense. The black hole function is given by
\begin{align}
f(r) &= 1-\frac{r^4}{r_h^4}+\frac{2}{3}\frac{B^2r^4}{L^2}\ln \left(\frac{r}{\ell_d}\right) \nonumber \\
&= 1-\frac{r^4}{R_H^4}+\frac{2}{3}\frac{B^2r^4}{L^2}\ln \left(\frac{r}{R_H}\right).
\end{align}
The perturbation needs to be sufficiently small of course. More precisely, a good criterion is that it is smaller than either of the first two terms separately.\footnote{It is not a contradiction if it were larger than one of them and smaller than the other one, as it would still be valid to call it a perturbation. Also, it is not a good criterion to impose that it is smaller than the sum of the first and second terms, as the sum of the first and second terms vanishes at $r=R_H$ and it is overly restrictive to impose the same thing for the perturbation.}
The logarithm itself is usually $\mathcal{O}(1)$.\footnote{An exception occurs when $r\approx 0$ (the AdS boundary), where the logarithm itself becomes arbitrarily large. For that particular case, the $B^2$-term is though much smaller than the $+1$ term and so there is no problem.} The correction needs to be smaller than either the $+1$ or the black hole $R_H$ term. The second condition gives
\begin{equation}
\mathcal{B}^2 < \frac{1}{R_H^4}.
\end{equation}
Within this same regime, the Hawking temperature is approximated as $T \sim \frac{1}{R_H}$ and hence:
\begin{equation}
\mathcal{B} < T^2,
\end{equation}
which is the criterion D'Hoker and Kraus write down in \cite{D'Hoker:2009bc}. \\
The first condition requires
\begin{equation}
B^2 r^4 < L^2,
\end{equation}
and hence restricts the range of $r$: one cannot trust the perturbative series for too large values of $r$. Luckily, we only care about the solution outside the outer event horizon and we restrict ourselves hence to the range $r<R_H$. This condition is hence precisely the same as the previous one. \\

\noindent Now for the horizonless case (supposedly thermal AdS) the situation is very different. One has instead
\begin{align}
f(r) &= 1+\frac{2}{3}\frac{B^2r^4}{L^2}\ln \left(\frac{r}{\ell_c}\right).
\end{align}
We only have the condition\footnote{This can also be understood without any computation since upon writing the Einstein equations in terms of $\mathcal{B}$, no explicit factor of $L$ is present anymore, and the only dimensionful parameter left in the problem is $r$ itself. Note that the temperature is arbitrary and geometry-independent if there are no horizons.}
\begin{equation}
B^2 r^4 < L^2,
\end{equation}
and hence we should not trust the solution too deep in the interior.\footnote{Note that also the logarithm blows up as $r\to\infty$, but this becomes appreciable only for much larger values of $r$ than the criterion written here.} This time this region is of interest and relevant to our computations. \\
The curvature singularity is hence in a region outside the reach of our perturbative solution and should be resolved upon treating the magnetic field in a non-perturbative fashion. \\
It would seem that we cannot describe the whole space with our constructed metric. This is true, but this is precisely where the walls come in and save the day. \\

\noindent So we find that the naked singularity solutions can be interpreted as magnetized thermal AdS when $r$ is not too large. \\

\noindent In our case, we adjust this model by including either a hard wall or a soft wall in the deep interior of AdS, precisely where the perturbative solution begins to fail. It is particularly transparent to see this in the hard wall case. The range of $r$ is truncated to $r<r_0$ where $r_0 \approx 3$ GeV$^{-1}$. In order to trust the solution all the way to the hard wall, we require
\begin{equation}
\mathcal{B}<  \frac{1}{r_0^2} \approx \frac{1}{9} \text{GeV}^{2}.
\end{equation}
This condition is in fact an order of magnitude less strict than $\mathcal{B} < T^2$ for $T \sim 100 \text{MeV}$. \\

\noindent It should be noted that both the hard wall and the soft wall case sufficiently dampen the curvature singularity contribution (either by excising it or by exponentially damping its contribution) to make the on-shell action finite in the deep interior. This is the reason we will not encounter any pathologies related to this singularity in our answers later on.

\section{Hawking-Page or confinement/deconfinement transition under the influence of a magnetic field}\label{sect4}
As is well-known, the Hawking Page transition is the holographic dual of the confinement/deconfinement phase transition. We will perform a detailed analysis here, first by revisiting the analysis done in \cite{Mamo:2015dea} for the hard wall model, and then by transferring to the soft wall scenario that we are mainly interested in here.

\subsection{Revisiting Mamo's analysis - hard wall model}
As we are interested in the thermodynamics of the system, we need to Euclideanize the on-shell actions \eqref{sbulk} and \eqref{sbndy} from which the free energy $F$ is determined by $S=\beta F$. For the hard wall model, the Euclidean bulk action reads:
\begin{equation}
\label{euclideanbulkaction}
S_{\text{bulk}}=\frac{V_3}{8\pi G_5}\int_0^\beta dt_E\int_{r_\lambda}^{r'}dr\sqrt{g}\left(\frac{4}{L^2}+\frac{2}{3}B^2g^{xx}g^{yy}\right)
\end{equation}
where $r_\lambda$ is the UV cut-off required to regulate the infinite volume available close to the AdS boundary, $r'$ will be $r'=R_H$ in the case of the black hole and $r'=r_0$ in the case of thermal AdS (this is the hard wall cut-off in the IR) and $V_3$ is the volume in the boundary directions. The value of $r_0$ is fixed phenomenologically at $r_0=3.096$ GeV$^{-1}$ by matching with the lowest $\rho$ meson mass \cite{Herzog:2006ra}. The Euclidean boundary action reads:
\begin{equation}
S_{\text{bndy}}=-\frac{V_3}{8\pi G_5}\int_0^\beta dt_E\sqrt{-\gamma}\left(K-\frac{3}{L}-LB^2g^{xx}g^{yy}\left(\ln \frac{r}{L}\right)\right)\Bigg|_{r_\lambda}.
\label{euclideanboundaryaction}
\end{equation}

\noindent We will review the method implemented in \cite{Mamo:2015dea} to compute the on-shell actions in order to analyze the Hawking-Page transition under the influence of a constant magnetic field.

\subsection*{Black hole - deconfined phase}
As the computations are a bit tedious, we present them in Appendix \ref{apphard}. For the hard wall model, we find
\begin{eqnarray}
&S_{\text{bh}}&=S_{\text{bulk}}^{\text{bh}}+S_{\text{bndy}}^{\text{bh}}\\
&=&\frac{V_3L^3}{8\pi G_5}\beta\left[-\frac{1}{R_H^4}+\frac{1}{2r_h^4}+\frac{B^2}{3L^2}+\frac{2B^2}{3L^2}\ln\left(\frac{R_H}{r_\lambda}\right) -\frac{B^2}{3L^2}\ln\left(\frac{r_\lambda}{\ell_d}\right) +\frac{B^2}{L^2}\ln\left(\frac{r_\lambda}{L}\right)\right]+\mathcal{O}(B^4). \nonumber
\end{eqnarray}
Using $f(r=R_H)=0$, we can rewrite this as
\begin{eqnarray}
&S_{\text{bh}}&= \frac{V_3L^3}{8\pi G_5}\beta\left[-\frac{1}{2R_H^4}+\frac{B^2}{3L^2}+\frac{B^2}{L^2}\ln\left(\frac{R_H}{L}\right)\right]+\mathcal{O}(B^4).
\end{eqnarray}

\subsection*{Thermal AdS - confined phase}
The thermal magnetized AdS geometry was written down in equation (\ref{thmetric}) above. \\
Unlike for the black hole geometry, in the thermal AdS space the temperature is not linked to any geometrical quantity and can be chosen at will. \\
The computation of the on-shell action is again deferred to the Appendix \ref{appsoft} and we obtain in this case:
\begin{eqnarray}
&S_{\text{th}}&=S_{\text{bulk}}^{\text{th}}+S_{\text{bndy}}^{\text{th}}\nonumber\\
&=&\frac{V_3L^3}{8\pi G_5}\beta\left[-\frac{1}{r_0^4}+\frac{B^2}{3L^2}+\frac{2B^2}{3L^2}\ln\left(\frac{r_0}{r_\lambda}\right)-\frac{B^2}{3L^2}\ln\left(\frac{r_\lambda}{\ell_c}\right)+\frac{B^2}{L^2}\ln\left(\frac{r_\lambda}{L}\right)\right]+\mathcal{O}(B^4).
\end{eqnarray}
\subsubsection*{Phase transition}
The phase transition occurs when the solution with the lowest free energy switches between the two. Thus we need to find the temperature where $\Delta S=0$:
\begin{eqnarray}
\Delta S&=&S_{\text{bh}}-S_{\text{th}}\nonumber\\
&=&\frac{V_3L^3}{8\pi G_5}\beta\left[\frac{-1}{2R_H^4}+\frac{1}{r_0^4}+\frac{B^2}{3L^2}\ln\left(\frac{R_H^3}{\ell_c r_0^2}\right)\right]+\mathcal{O}(B^4).
\end{eqnarray}
Since $R_H$ is fully determined by $T$ and $B$ by the formula for the Hawking temperature, this equation defines a relation between the Hawking-Page temperature and the applied magnetic field. \\

\begin{figure}[h]
  \centering
   \includegraphics[width=0.75\textwidth]{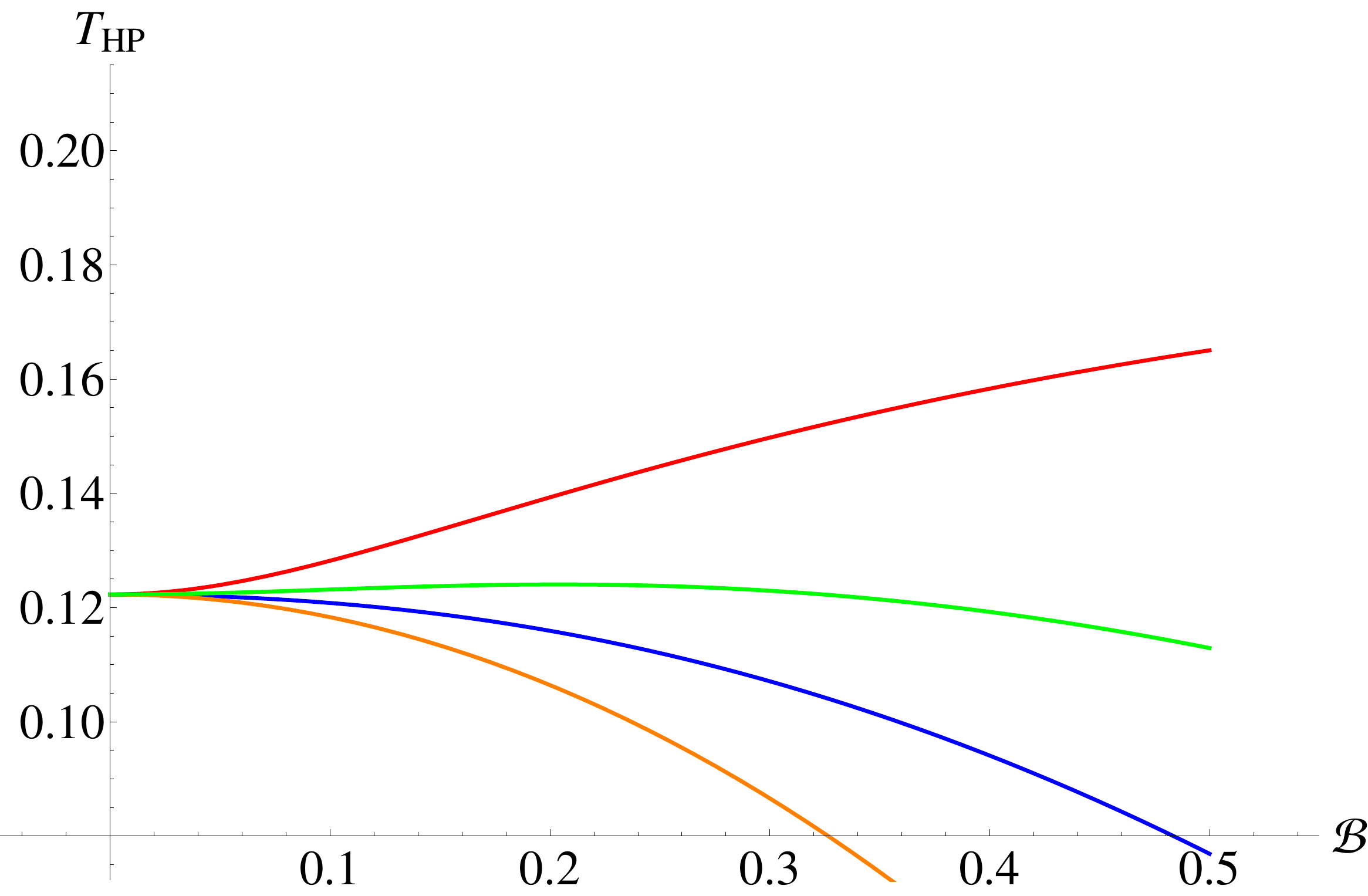}
  \caption{$T_{HP}$ (GeV) in the hard wall model as a function of the applied magnetic field $\mathcal{B}$ (GeV$^2$) with $c=0.151$ GeV$^2$ for various values of $\ell_c$. From top to bottom: $\ell_c=0.1$ GeV$^{-1}$ (red), $\ell_c=0.5$ GeV$^{-1}$ (green), $\ell_c=1$ GeV$^{-1}$ (blue), $\ell_c=2$ GeV$^{-1}$ (orange).}
  \label{hardwallHP}
\end{figure}

\noindent As anticipated earlier, the arbitrary length scale $\ell_c$ of the confined phase leaves a distinct physical imprint on the formulas. Since it is quite difficult to compare the resulting Hawking-Page temperature for various values of $\ell_c$ as a function of $\mathcal{B}$ with lattice results, we will not attempt to do this here. The behavior of the Hawking-Page temperature as a function of $\mathcal{B}$ for various values of $\ell_c$ is shown in Figure \ref{hardwallHP}. \\

\begin{figure}[h]
  \centering
   \includegraphics[width=0.75\textwidth]{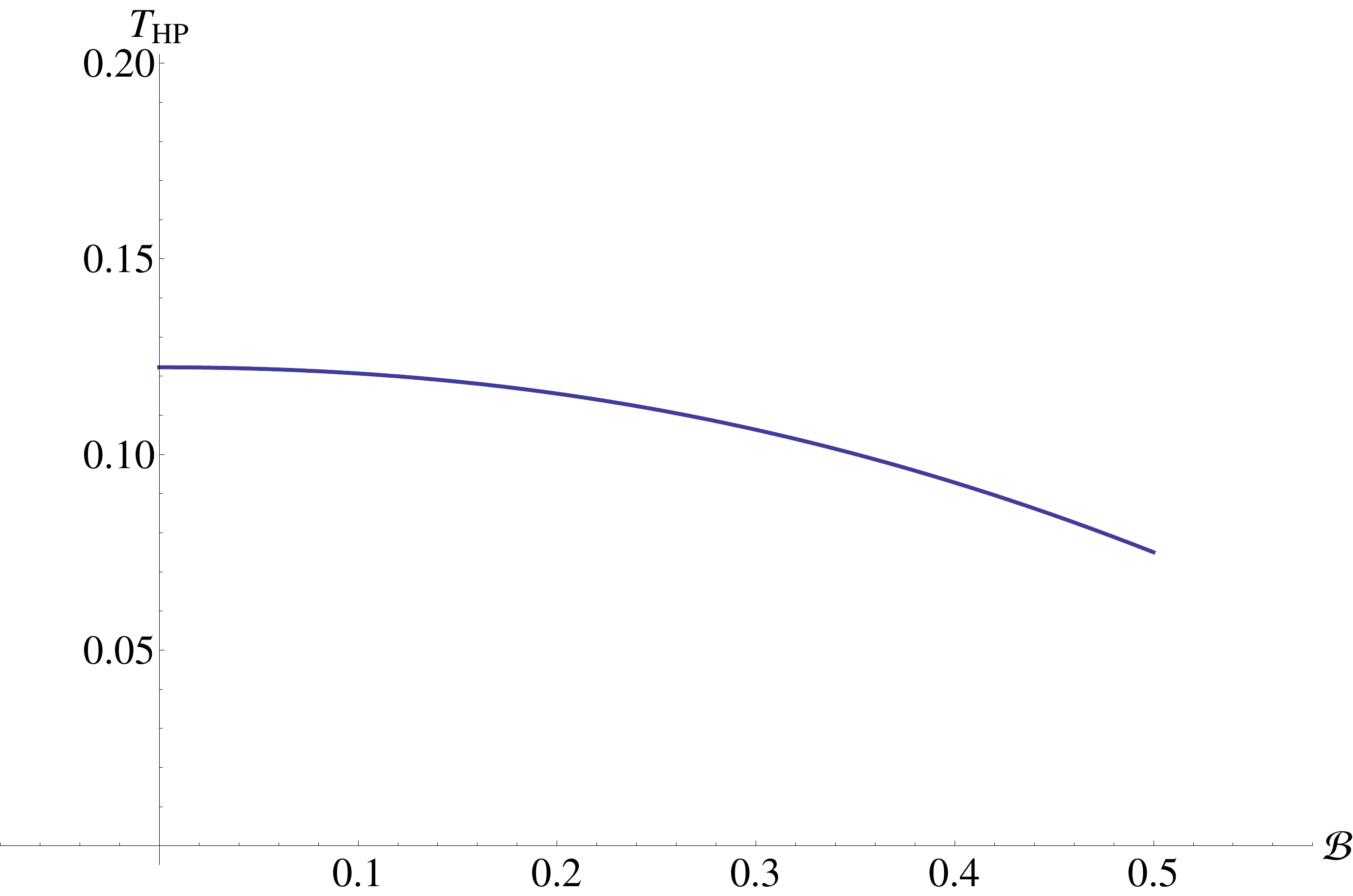}
  \caption{$T_{HP}$ (GeV) in the hard wall model as a function of the applied magnetic field $\mathcal{B}$ (GeV$^2$) with $r_0=3.096$ GeV$^{-1}$ and $\ell_c=1.03$ GeV$^{-1}$. }
  \label{mamoplot}
\end{figure}

\noindent In Section \ref{secchiconf} we will find another way to constrain $\ell_c$ significantly (in the soft wall model) and we will effectively fix it to $\ell_c=1.03$ GeV$^{-1}$. For the purposes of this Section, we will assume this value of $\ell_c$ and make our figures accordingly. We do remark that this value is indeed plausible for the scenario discussed here. \\
In Figure \ref{mamoplot} we can see that the critical temperature $T_{HP}$ decreases with the applied magnetic field $\mathcal{B}$.

\subsection{Hawking-Page transition with a magnetic field in the soft wall model}
\label{HPSOFT}
In this Section we will follow the method implemented in \cite{Mamo:2015dea} to compute the on-shell actions to analyze the Hawking-Page transition for the soft wall model.
\subsection*{Einstein-Maxwell action in the soft wall model}
In order to analyze the Hawking-Page transition under the influence of a magnetic field, we need to first compute the on-shell Euclidean actions for the deconfined phase (black hole geometry) and for the confined phase (thermal AdS) at the same temperature. \\

\noindent As was pointed out in \cite{Herzog:2006ra}, in order to analyze the Hawking-Page transition using the soft wall model we need to assume that the dilaton field does not significantly backreact on the metric, i.e. it does not affect the gravitational dynamics. Thus the equations of motion are \eqref{ricciequation} and \eqref{maxwelleq}.

\subsection*{Black hole - deconfined phase}
The computations themselves are included in Appendix \ref{appsoft}. One finds the on-shell action
\begin{eqnarray}
&S_{\text{bh}}&=S_{\text{bulk}}^{\text{bh}}+S_{\text{bndy}}^{\text{bh}}\\
&=&\frac{V_3L^3}{8\pi G_5}\beta\left[e^{-cR_H^2}\left(\frac{-1}{R_H^4}+\frac{c}{R_H^2}\right)+\left(\frac{B^2}{3L^2}+c^2\right)\text{Ei}(-cR_H^2)-e^{-cr_\lambda^2}\left(\frac{-1}{r_\lambda^4}+\frac{c}{r_\lambda^2}\right)\right.\nonumber\\
&-&\left.\left(\frac{B^2}{3L^2}+c^2\right)\text{Ei}(-cr_\lambda^2)-\frac{1}{r_\lambda^4}+\frac{1}{2r_h^4}+\frac{B^2}{3L^2}-\frac{1}{3}\frac{B^2}{L^2}\ln \left(\frac{r_\lambda}{\ell_d}\right)+\frac{B^2}{L^2}\ln \left(\frac{r_\lambda}{L}\right)\right]+\mathcal{O}(B^4). \nonumber
\end{eqnarray}
\subsection*{Thermal AdS - confined phase}
For the confining phase, the resulting on-shell action is given by
\begin{align}
&S_{\text{th}}=S_{\text{bulk}}^{\text{th}}+S_{\text{bndy}}^{\text{th}}\nonumber\\
&=\frac{V_3L^3}{8\pi G_5}\beta\left[-e^{-cr_\lambda^2}\left(\frac{-1}{r_\lambda^4}+\frac{c}{r_\lambda^2}\right)-\left(\frac{B^2}{3L^2}+c^2\right)\text{Ei}(-cr_\lambda^2)-\frac{1}{r_\lambda^4}+\frac{B^2}{3L^2}-\frac{1}{3}\frac{B^2}{L^2}\ln\left(\frac{r_\lambda}{\ell_c}\right) \right.\nonumber\\&\left.+ \frac{B^2}{L^2}\ln\left(\frac{r_\lambda}{L}\right)\right]+\mathcal{O}(B^4).
\end{align}
\subsubsection*{Phase transition}
The difference in the on-shell action hence becomes (using $f(r=R_H)=0$):
\begin{align}
\Delta S&=S_{\text{bh}}-S_{\text{th}}\nonumber\\
&=\frac{V_3L^3}{8\pi G_5}\beta\left[e^{-cR_H^2}\left(\frac{-1}{R_H^4}+\frac{c}{R_H^2}\right)+\left(\frac{B^2}{3L^2}+c^2\right)\text{Ei}(-cR_H^2)+\frac{1}{2R_H^4}+\frac{1}{3}\frac{B^2}{L^2}\ln\left(\frac{R_H}{\ell_c}\right) \right]\nonumber\\&+\mathcal{O}(B^4).
\end{align}
Again $\ell_c$ appears explicitly in this expression. \\
As a special case, for $B=0$ we retrieve the condition for $\Delta S=0$ as:
\begin{equation}
\frac{1}{2}+e^{-cR_H^2}\left(-1+cR_H^2\right)+\left(c^2R_H^4\right)\text{Ei}(-cR_H^2)=0,
\end{equation}
which is the same expression obtained in \cite{Herzog:2006ra,BallonBayona:2007vp}. \\

\begin{figure}[h]
  \centering
   \includegraphics[width=0.75\textwidth]{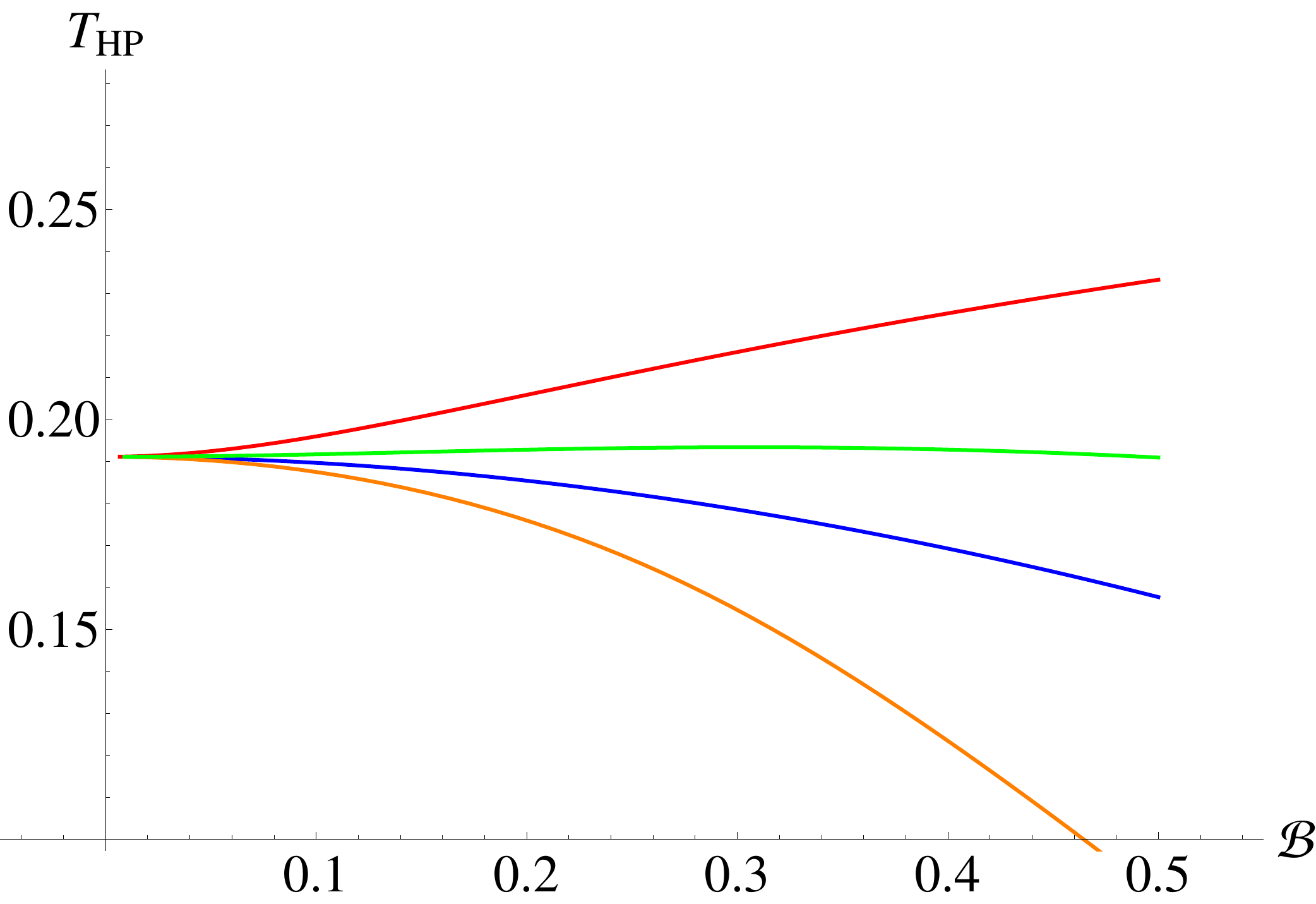}
  \caption{$T_{HP}$ (GeV) in the soft wall model as a function of the applied magnetic field $\mathcal{B}$ (GeV$^2$) with $c=0.151$ GeV$^2$ for various values of $\ell_c$. From top to bottom: $\ell_c=0.1$ GeV$^{-1}$ (red), $\ell_c=0.5$ GeV$^{-1}$ (green), $\ell_c=1$ GeV$^{-1}$ (blue), $\ell_c=2$ GeV$^{-1}$ (orange).}
  \label{softwallHP}
\end{figure}

\noindent The relation of $T_{HP}$ as a function of $\mathcal{B}$ for various values of $\ell_c$ is shown in Figure \ref{softwallHP}.
We remark here already that a qualitative match with the lattice requires that $\ell_c \sim 1$ GeV$^{-1}$, but definitely not much smaller than this. The value we will find later on indeed gives a qualitative nice behavior. One can see that the decreasing critical temperature is not universal in $\ell_c$, as for rather small values of $\ell_c$ we observe an increasing deconfinement temperature with $\mathcal{B}$. \\

\begin{figure}[h]
  \centering
   \includegraphics[width=0.75\textwidth]{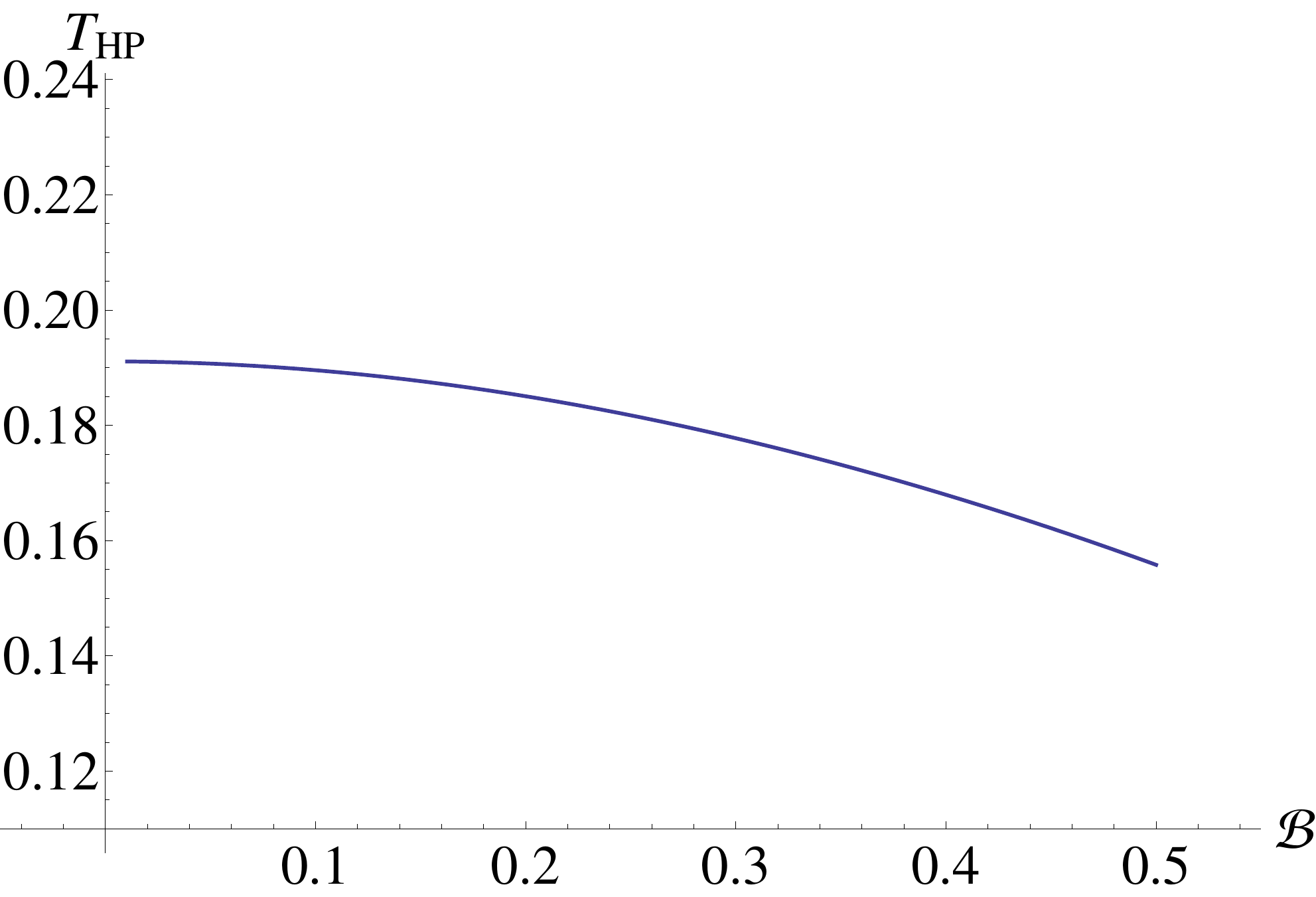}
  \caption{$T_{HP}$ (GeV) in the soft wall model as a function of the applied magnetic field $\mathcal{B}$ (GeV$^2$) with $c=0.151$ GeV$^{2}$ and $\ell_c=1.03$ GeV$^{-1}$.}
  \label{tcb}
\end{figure}

\noindent Specifying again to the value of $\ell_c=1.03$ GeV$^{-1}$, one obtains the profile shown in Figure~\ref{tcb}.

\noindent Again we find a decreasing behavior of the deconfinement temperature with the applied magnetic field $\mathcal{B}$.

\section{Internal energy and thermodynamic stability}\label{sect5}
There is a further thermodynamic stability issue we can discuss using the on-shell action: any stable thermodynamic theory should have a positive heat capacity. We know that black holes in asymptotically flat space violate this stability criterion and they are hence unstable towards either evaporation or growth from the thermal heat bath. In AdS, this does not happen (at least for large AdS black holes) and these are thermodynamically stable. Since we have altered the black hole solution, it seems interesting to reconsider this issue for the current geometry. We anticipate small black holes being unstable (just like in normal AdS). \\
To start off with, we need to find the thermodynamic internal energy of the system. One way of finding the mass contained in this spacetime is to use the thermodynamics of the boundary theory.\footnote{An alternative would be to use $F=E-TS$ where one computes $S$ via the Bekenstein-Hawking entropy of the black hole. We checked however that these expressions do not match when $c\neq0$. This is no surprise, as the soft wall does not solve Einstein's equations. An other alternative would be to use the ADM definition of mass in asymptotically AdS spacetimes \cite{Balasubramanian:1999re}. This however makes crucial use of the background equations of motion as well. Since the free energy as computed holographically in the soft wall model has proven to lead to a very nice criterion on the deconfinement temperature \cite{Herzog:2006ra}, we believe it to be more trustworthy to fully continue in the boundary theory after obtaining $F$ (i.e. to not use any more holographic dictionary entries). The internal energy $E$ is then computed instead using $\partial_{\beta}(\beta F)$.} The on-shell free energy was found in the previous Section and it is given by\footnote{An overall prefactor $L^3/\kappa^2$ with $\kappa^2 = 8\pi G_5$ is left implicit in the following.}
\begin{equation}
F = e^{-cR_H^2}\left(-\frac{1}{R_H^4}+ \frac{c}{R_H^2}\right) + \left(\frac{B^2}{3L^2} + c^2\right) \text{Ei}(-cR_H^2) + \frac{1}{2r_h^4}
\end{equation}
where we have discarded all temperature-independent contributions; these are irrelevant for thermodynamical purposes and include the UV divergent terms that will require holographic renormalization.\footnote{One does have to be a bit careful here, as it seems our result will now depend on $\ell_d$, but this is only as an overall temperature-dependent addition as we will see.} The internal energy $E$ can now be found as
\begin{align}
\label{longformula}
E &= \partial_{\beta}(\beta F) \nonumber \\
&= \exp(-cR_H^2)\left(-\frac{1}{R_H^4}+\frac{c}{R_H^2}\right)+\left(\frac{B^2}{3L^2}+c^2\right)\text{Ei}(-cR_H^2) + \frac{1}{2R_H^4}+\frac{B^2}{3L^2}\ln\left(\frac{R_H}{\ell_d}\right) \nonumber \\
&+\frac{4\pi R_H\left(\frac{1}{27}\frac{\exp(-cR_H^2)(6L^2+B^2R_H^4)(-6L^2+B^2R_H^4)^2}{R_H^5L^4\pi(2L^2+B^2R_H^4)}+\frac{1}{54}\frac{(-6L^2+B^2R_H^4)^3}{R^5L^4\pi(2L^2+B^2R_H^4)}\right)}{4-\frac{2}{3}\frac{B^2R_H^4}{L^2}}.
\end{align}

\noindent We remark that if $c=0$, this complicated formula reduces to
\begin{equation}
E = \frac{1}{2R_H^4}\left[3-\frac{2}{3}\frac{B^2R_H^4}{L^2}\right] +\frac{B^2}{L^2}\ln\left(\frac{R_H}{\ell_d}\right) + (T\text{-indep})= \frac{3}{2R_H^4} + \frac{B^2}{L^2}\ln\left(\frac{R_H}{\ell_d}\right) + (T\text{-indep}),
\end{equation}
where we again have dropped temperature-independent terms. \\

\noindent This energy depends on three dimensionful quantities: $\ell_d$, $c$ and $\mathcal{B}\sim \frac{B}{L}$, from which we can construct two dimensionless numbers. Note that $\ell_d$ only provides a temperature-independent contribution and is hence irrelevant as we have been neglecting such terms throughout. For computational simplicity and without loss of generality, we hence fix $\ell_d=1$ here. The energy hence depends non-trivially on two independent parameters. Numerically analyzing the dependence of equation (\ref{longformula}) on $R_H$ for a selection of the parameters, one learns the following lessons:
\begin{itemize}
\item{If $c=0$ and $\mathcal{B}=0$, the energy decreases monotonically as $R_H$ increases.}
\item{As soon as either $c\neq0$ or $\mathcal{B}\neq0$, the energy only decreases with $R_H$ for sufficiently small $R_H$. It reaches a minimum at some $R_H^*$ after which generically it increases monotonically for all $R_H$ larger than this value. However, for a relatively small subset of the parameter space, it is possible that the energy reaches a maximum and a second minimum, after which it will increase monotonically again. If this happens, it is possible that there exists another stable region within the unstable zone we will discuss below. We will ignore this possibility here.}
\end{itemize}

\begin{figure}[h]
\centering
\includegraphics[width=0.75\textwidth]{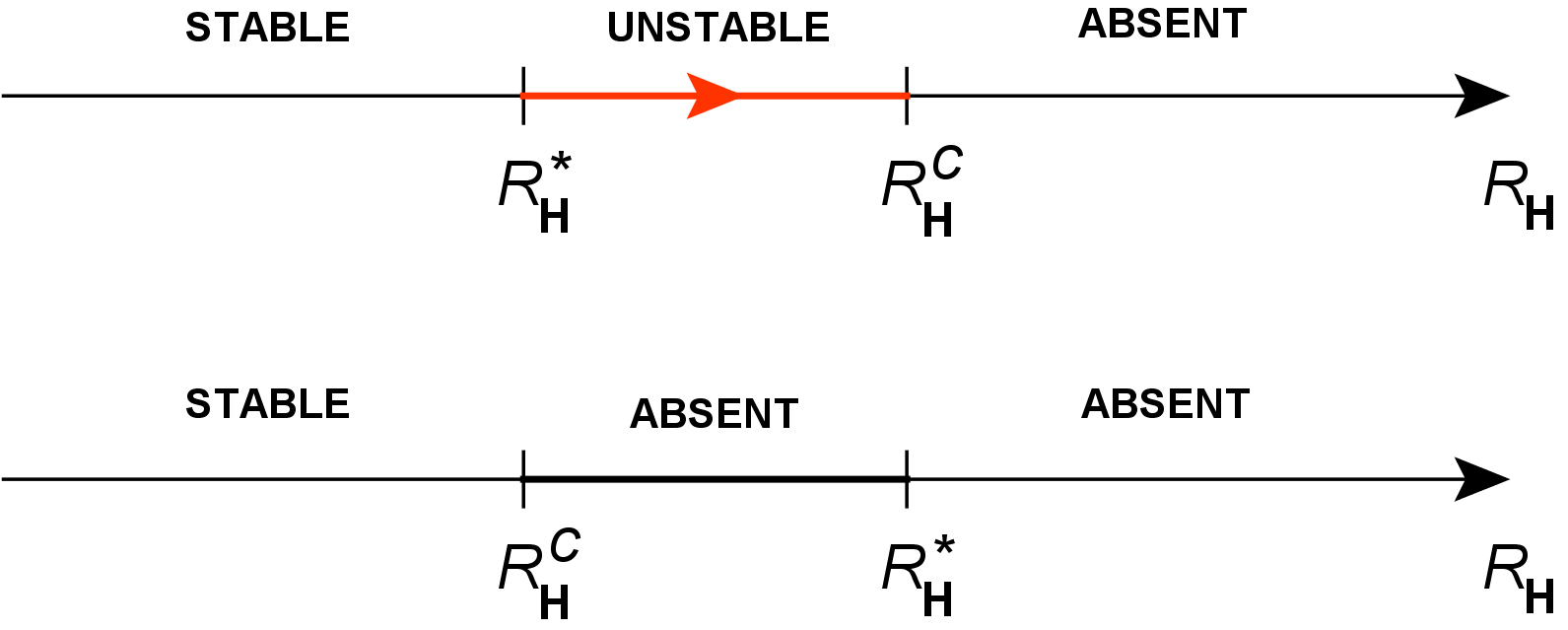}
\caption{The system is thermodynamically unstable in between both special values of $R_H$. For higher values of $R_H$, it is impossible with a given value of $\mathcal{B}$ to construct this black hole geometry with outer horizon $R_H$.}
\label{TDStability}
\end{figure}

\noindent To analyze the thermodynamic stability, we only need to combine this behavior with Figure \ref{THawking} and we can readily reach the following conclusion. If $R < \min(R_H^*, R_H^c)$, the solution is thermodynamically stable, in the sense that $C = \frac{\partial E}{\partial T} > 0$. If $R_H^*$ is smaller than $R_H^c$, the system is thermodynamically unstable in between these values of $R_H$. The instability causes the black hole to shrink (by emitting radiation) until it reaches extremality with $T=0$. In all other cases, the region for larger $R_H$ is not accessible for a given $B$ as shown in Figure \ref{TDStability}.

\begin{figure}[H]
\centering
\includegraphics[width=0.75\textwidth]{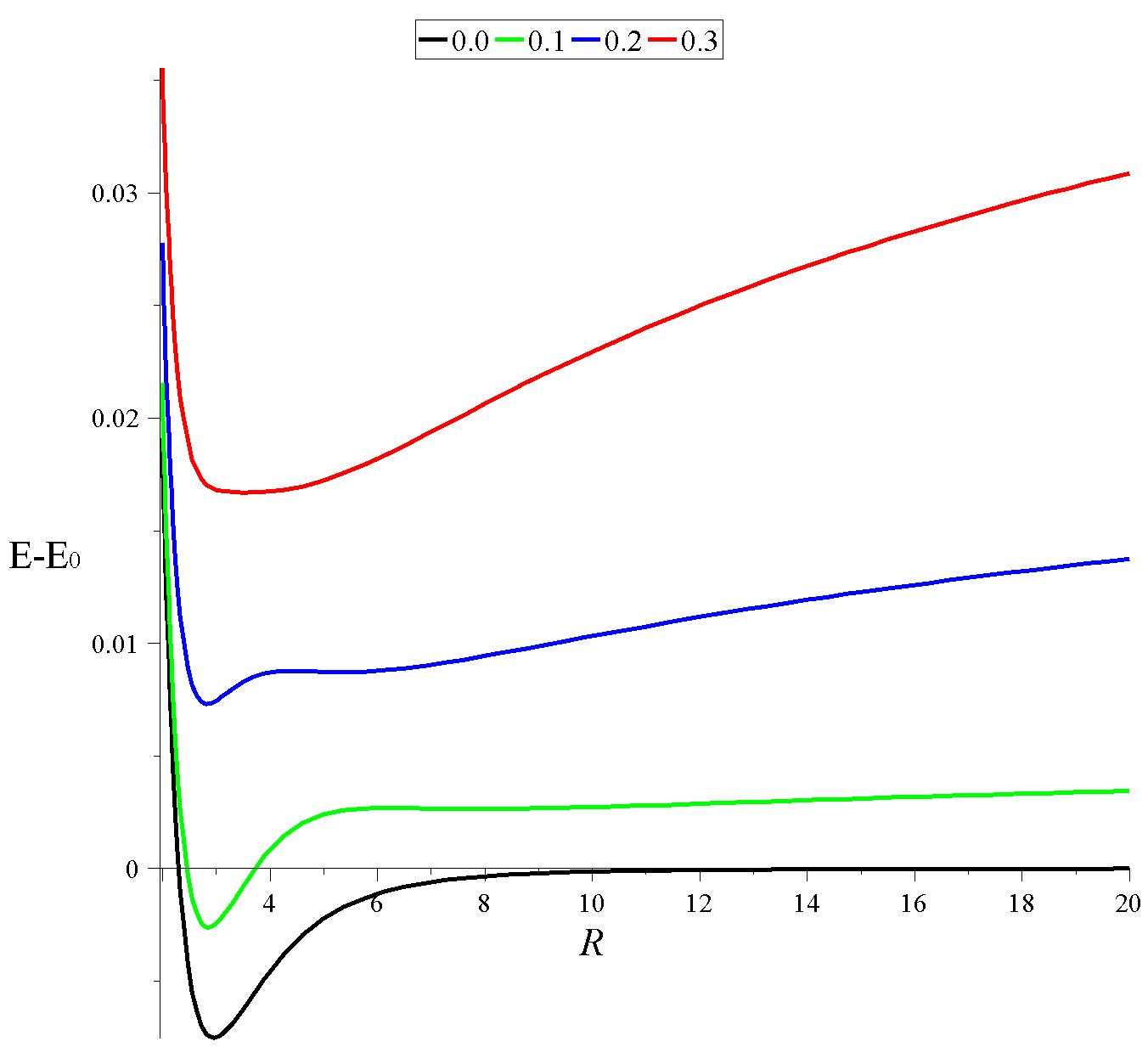}
\caption{Energy $E-E_0$ (GeV) as a function of horizon radius $R_H$ (GeV$^{-1}$) for several values of the applied magnetic field $\mathcal{B}$. Black: $\mathcal{B}=0.0$ GeV$^2$, green: $\mathcal{B}=0.1$ GeV$^2$, Blue: $\mathcal{B}=0.2$ GeV$^2$, Red: $\mathcal{B}=0.3$ GeV$^2$.}
\label{StabilityAn}
\end{figure}

\noindent Next we will apply this general discussion to the case at hand. For our specific case, we take $c=0.151$ GeV$^2$. The value of $\ell_d$ is arbitrary for thermodynamics, as it only provides a temperature-independent shift to the energy. \\ 
With these choices, the behavior of the (temperature-dependent part of the) energy is shown in Figure \ref{StabilityAn}.
It is seen that for larger values of $\mathcal{B}$, this curve has multiple extrema. Since this indeed only happens at larger values of $\mathcal{B}$, we will not discuss this here. \\

\begin{figure}[H]
\centering
\includegraphics[width=0.55\textwidth]{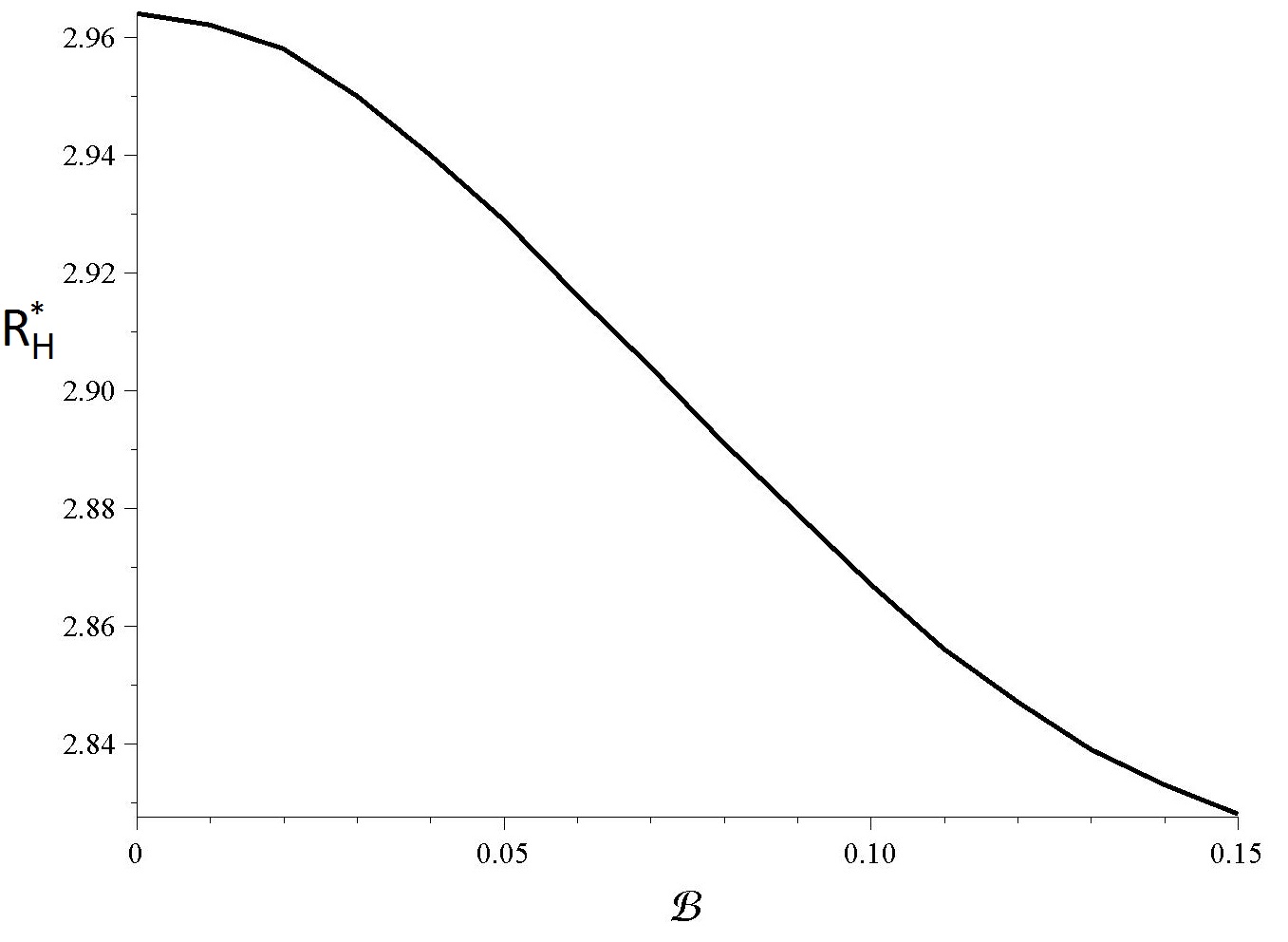}
\caption{Critical horizon radius $R_H^*$ (GeV$^{-1}$) as a function of the applied magnetic field $\mathcal{B}$ (GeV$^{2}$).}
\label{CritRadB}
\end{figure}

\noindent The critical horizon radius, above which an instability occurs is shown for small values of $\mathcal{B}$ in Figure \ref{CritRadB}.

\begin{figure}[H]
\centering
\includegraphics[width=0.55\textwidth]{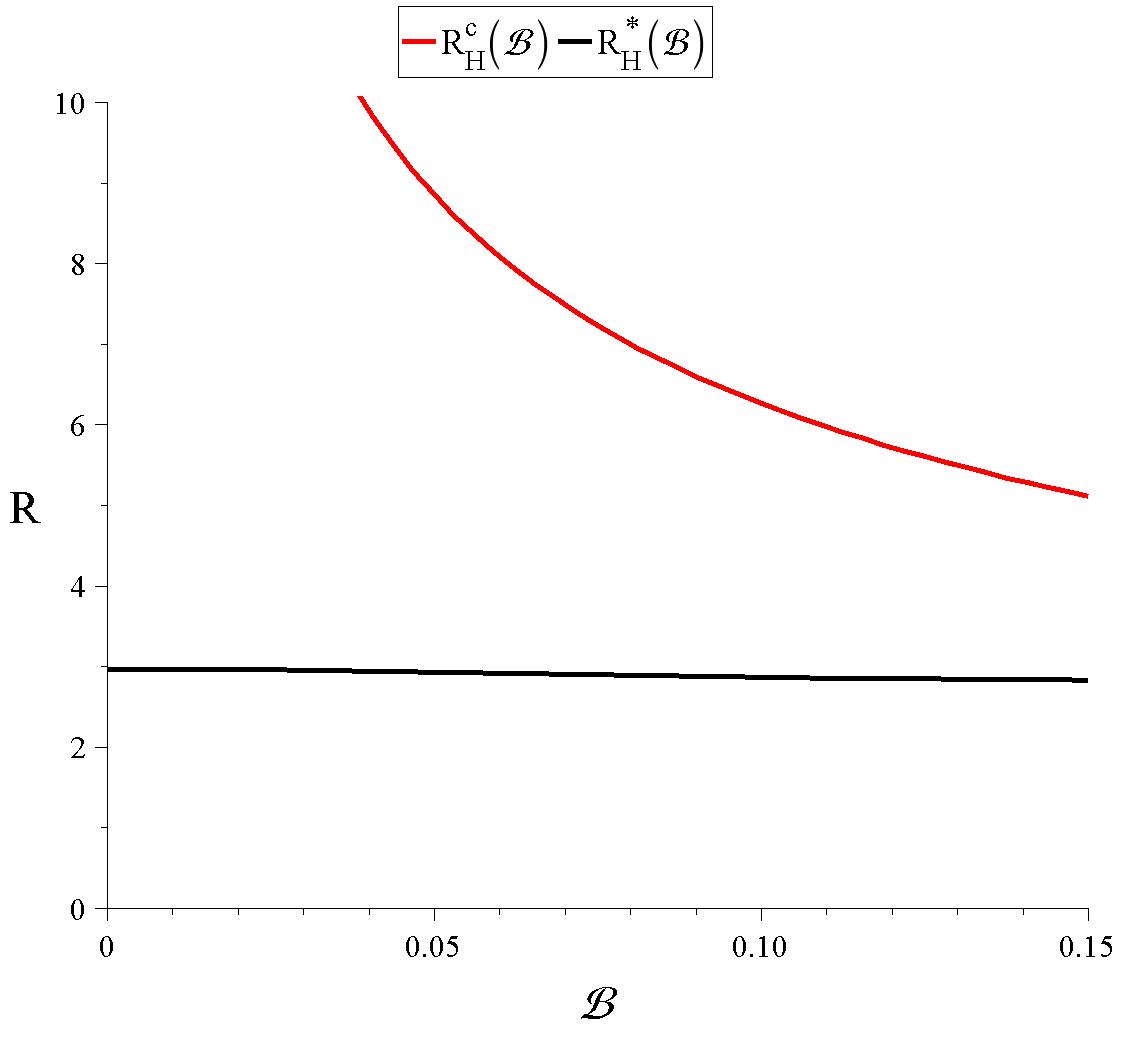}
\caption{Critical horizon radius $R_H^*$ (GeV$^{-1}$) and extremal horizon radius $R_H^c$  (GeV$^{-1}$) as a function of the applied magnetic field $\mathcal{B}$  (GeV$^{2}$).}
\label{comparecurves}
\end{figure}

\begin{figure}[H]
\centering
\includegraphics[width=0.55\textwidth]{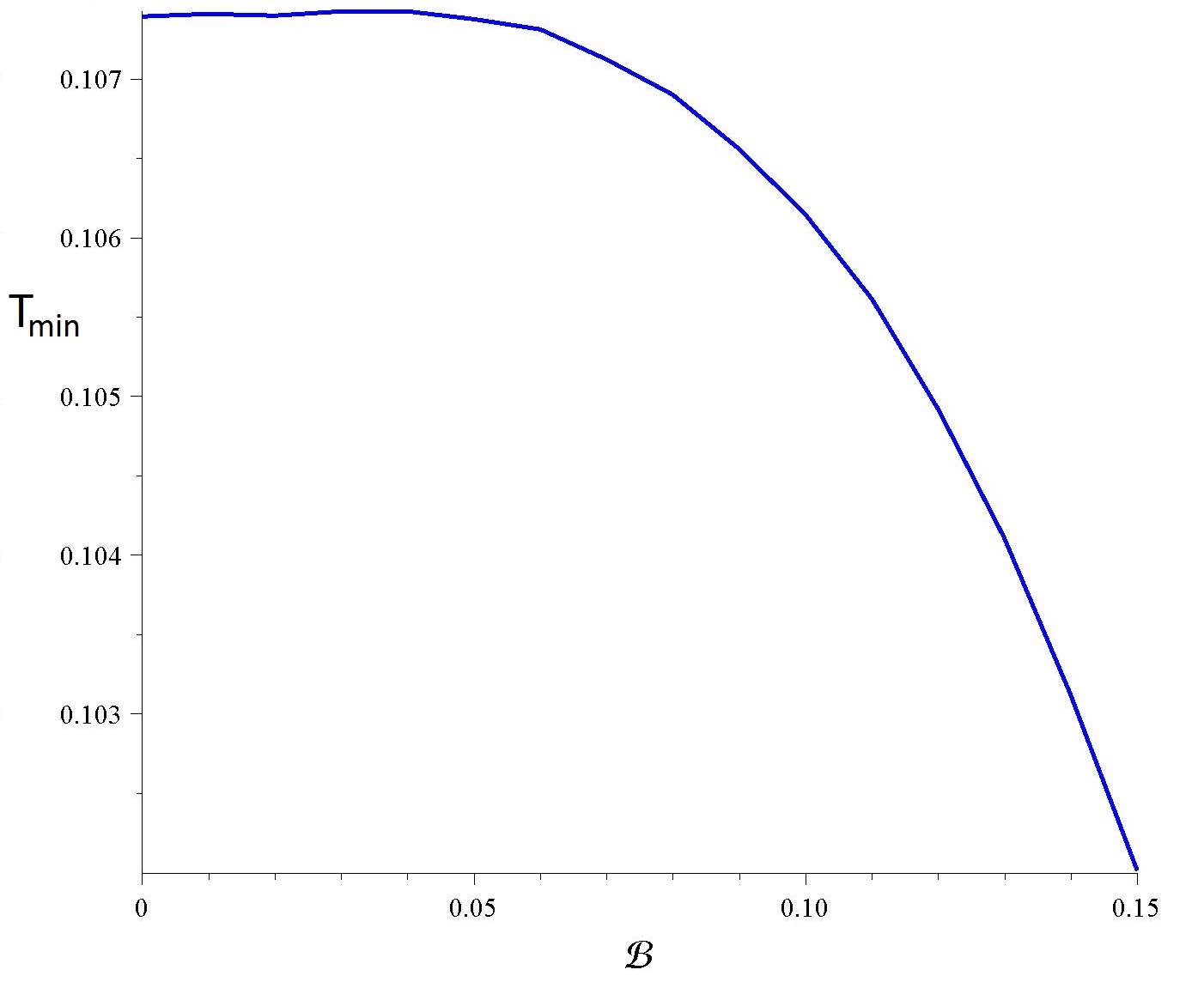}
\caption{Minimal temperature (in GeV) needed to have a stable black hole system as a function of $\mathcal{B}$ (GeV$^{2}$).}
\label{TminB}
\end{figure}

\noindent In Figure \ref{comparecurves}, we combine this with the behavior of $R_H^c$ (determined by equation (\ref{hawktem})) as a function of $\mathcal{B}$.
Clearly, the most stringent condition for these values of $\mathcal{B}$ is that $R_H<R_H^*$; we are in the first situation displayed in Figure \ref{TDStability}. For any such value of $R_H$, the system is stable in the sense discussed above. Finally, we can translate this criterion into one on the temperature $T$. It should be larger than the minimal temperature displayed in Figure \ref{TminB}.
Since for these values of $\mathcal{B}$, this minimal temperature is about 100-108 MeV and this is on its own smaller than the deconfinement (Hawking-Page) temperature, we conclude that the system is indeed thermodynamically stable in the regime we are probing it.

\section{Chiral phase transition under the influence of a magnetic field}\label{sect6}
In this Section, we will proceed with the main goal of this work: to analyze the chiral condensate as a function of the applied external magnetic field and to determine the resulting chiral phase transition temperature. The action relevant for the chiral properties of the dual QCD-like theory was already written down above. Let us retake it here
\begin{equation}
S=\frac{N_c}{16\pi^2}\int d^4x\int^{R_H}_0 dr e^{-\phi}\sqrt{-g}\text{Tr}\left[\left|DX\right|^2-m_5^2\left|X\right|^2-\frac{1}{3}\left(F_L^2+F_R^2\right)\right].
\end{equation}
In this action, $X$ is a complex field in the bifundamental representation of $SU(N_f)_L\otimes SU(N_f)_R$, associated to the chiral symmetry breaking. Its covariant derivative is defined as $D_\mu X=\partial_\mu X-iA_{L,\mu} X+i X A_{R,\mu}$, for two gauge fields whose field strength is $F_{L,R}^{MN}$. The scalar field $\phi(r)=cr^2$ is the dilaton field which is responsible for the phenomenological IR properties of the theory, i.e. confinement \cite{Andreev:2006ct} and the linear Regge behavior of the meson spectrum ($c=0.151$ GeV$^2$ which is fixed by the $\rho$-meson mass \cite{Karch:2006pv}). The mass parameter is fixed at $ m_5^2 L^2 = -3$.\\

\noindent As mentioned in \cite{Colangelo:2011sr}, the dominant behavior of $\braket{\bar{\psi}\psi}$ is expected to be due to the interaction of the field $X$ with the background geometry and dilaton wall. This means that we can restrict ourselves to only the linearized equations of motion for $X$. The scalar field is decomposed as: $X(x^\mu,r)=X_0(r)\mathbf{1}_{N_f}e^{i\pi(x^\mu,r)}$, where $X_0(r)$ is the component independent of the boundary directions and $\pi(x^\mu,r)$ represent chiral fields. As stated before, we work in the approximation of 2 degenerate flavors. In principle, as soon as a magnetic field is turned on, one might expect a different value for the chiral condensate in terms of either the up or down quarks due to their different electromagnetic coupling. This would amount to allowing $X$ to be a diagonal rather than scalar matrix. Though, we shall soon see this would make no difference in our case, so we keep $X$ proportional to the unit matrix $\mathbf{1}_{N_f}$, meaning we can still consider a degenerate quark condensate. \\

\noindent According to the AdS/CFT lore, the boundary expansion of this $X$-field starts with the bare quark mass as the lowest order coefficient. The second term contains the chiral condensate and this is the quantity we are interested in. The main goal is then to solve the equations of motion for $X$ and distill this coefficient of the boundary expansion. Since our computations are done in the Euclidean formalism, we impose our solution to be finite at the black hole horizon.\footnote{We note that we choose $X$ to be time-independent, so our computation borders the Lorentzian and Euclidean methods.} \\

\noindent Before discussing the temperature-dependent chiral condensate in the deconfined regime, we will first discuss it in the low temperature confined region. Note that it is a general property of holographic models that the confined regime is described by thermal AdS. The temperature does \emph{not} emerge from the geometry itself (it is an independent variable) and hence the chiral condensate, as determined by solving bulk equations of motion, is independent of the temperature. This is a general feature of large $N$ holographic models and is something we will not be able to remedy. \\

\noindent Since we will be interested in a spatially homogeneous condensate, we make the ansatz $X(x^\mu,r) = X_0(r)$.
Using the metric \eqref{bhmetric}, the equations of motion are given by:
\begin{equation}
X''_0-\frac{\left(2cr^2+3\right)f(r)-rf'(r)}{rf(r)}X'_0+\frac{3}{r^2f(r)}X_0=0,
\label{eom}
\end{equation}
where $f(r)$ is the horizon function of the black hole. Magnetized AdS can be found by taking $r_h \to \infty$ in $f(r)$. \\
To derive this, we remark that the background magnetic field appears explicitly both in the metric as in the covariant derivatives of $X$. The latter contribution vanishes though for the case at hand, since the magnetic field is modeled into the (diagonal) vector part of the flavor gauge group $A_L = A_R$ and we take $X$ to be proportional to the identity matrix in flavor space.  So
\begin{equation}
D_\mu X=\partial_\mu X-iA_{L,\mu} X+i X A_{R,\mu} = \partial_\mu X,
\end{equation}
hence the only way in which the magnetic field enters, is through its presence in the metric. This demonstrates that we would find no dependence on the magnetic field at all if we were to exclude the backreaction of the $\mathcal{B}$-field on the geometry. Notice here that the foregoing argument stands also were $X$ to be merely diagonal. The equation \eqref{eom} is thus identical for the up and down quark sector, hence there is only need for a single $X_0$. This explains why we maintained from the start $X\propto\mathbf{1}_{N_f}$. \\

\subsection{The chiral condensate for the confining background}
\label{secchiconf}
The $T\to0$ limit of the AdS black hole solution will \emph{not} yield thermal AdS when magnetic fields are turned on; instead it gives the extremal black hole solution. Hence to discuss the confinement behavior of the condensate, we need to numerically solve for the condensate directly in AdS space. \\

\noindent One can show (and we will do so for the deconfining black hole in the next subsection) that the boundary expansion of the $X$-field is of the form:
\begin{eqnarray}
L^{3/2}X_0 = cmr^3\ln \left(\sqrt{c}r\right) +mr+\sigma r^3+\mathcal{O}(r^5),
\end{eqnarray}
where $m$ is the bare quark mass. We demonstrate in Appendix \ref{chirCondens} that the actual (single flavor) condensate is related to the coefficient $\sigma$ in the following sense:
\begin{equation}
\label{phycond}
\braket{\bar{\psi}\psi}_{\mathcal{B},T} -   \braket{\bar{\psi}\psi}_{\mathcal{B}=0,T=0} =\frac{N_c}{2\pi^2} \left(\sigma(\mathcal{B},T) - \sigma(\mathcal{B}=0,T=0)\right).
\end{equation}
The l.h.s.~of the above equation is, by construction, finite. \\

\begin{figure}[h]
    \centering
   \includegraphics[width=0.75\textwidth]{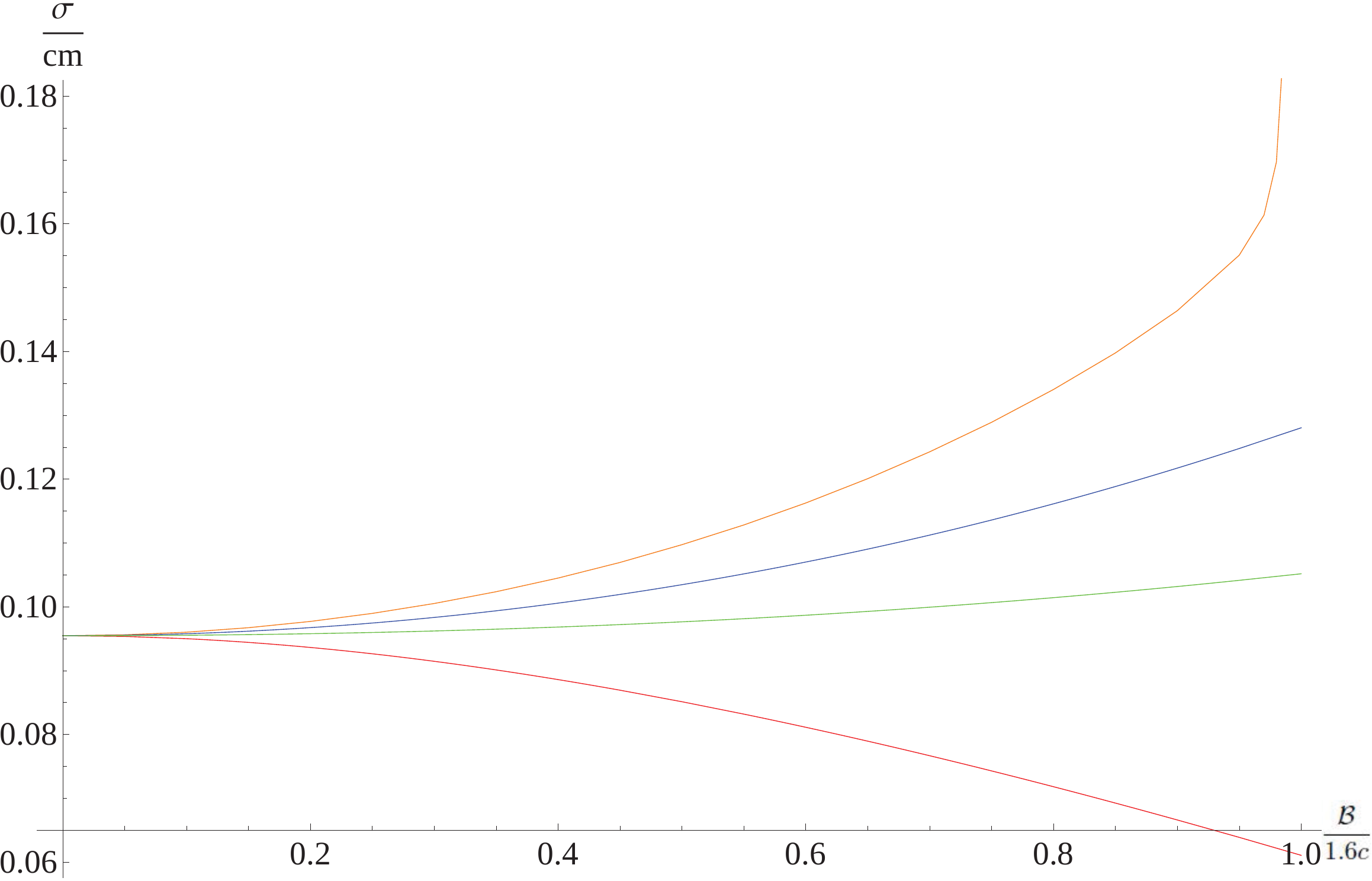}
  \caption{$\frac{\sigma}{cm}$ as a function of applied magnetic field $\mathcal{B}$ for various values of $\ell_c$. From bottom to top: $\ell_c\sqrt{c}=0.1$ (red), $\ell_c\sqrt{c}=0.5$ (green), $\ell_c\sqrt{c}=1$ (blue), $\ell_c\sqrt{c}=2$ (orange).}
  \label{Shooting}
  \end{figure}

\noindent Numerically, we shoot from the boundary until a normalizable solution is found, which fixes $\sigma$. In Figure \ref{Shooting} we show the resulting condensate $\sigma$ (actually $\frac{\sigma}{cm}$) as a function of the external magnetic field for different values of $\ell_c$.\footnote{Since the differential equation contains singular points, we need to perform a Frobenius series expansion near these points. This is presented in the next subsection for the black hole case.} The behavior changes quite drastically. Note that for $\ell_c\sqrt{c}=2$, one finds singular behavior around $\frac{\mathcal{B}}{1.6c}\approx 1$. This is indeed expected, as this thermal magnetized AdS background develops horizons at
\begin{equation}
\frac{\mathcal{B}}{1.6} > \frac{\sqrt{6e}}{\ell_c^2} \approx 1.01 c
\end{equation}
for $\ell_c\sqrt{c}=2$. This again shows that one must choose $\ell_c$ not too large to have a trustworthy model.
It is clear that one needs to have $\ell_c$ not too small, since the condensate would drop with increasing magnetic field then\footnote{And as such, be in contrast with the expected magnetic chiral catalysis at zero temperature.}, but also not too large, since horizons in the geometry might develop.
To determine a suitable value for $\ell_c$, we will compare the results here to those given by the lattice computation of \cite{Bali:2012zg}. In the latter reference, one actually computes the renormalized condensate\footnote{To be more precise, the condensate averaged over up and down flavor. Since we still have a degenerate condensate, the $\Delta\Sigma$ of \cite{Bali:2012zg} does correspond to our $\braket{\bar{\psi}\psi}$ without the need to worry about factors of $2$.} as
\begin{equation}
\Delta \Sigma = \frac{2m}{M_{\pi}^2f_\pi^2}\left(\braket{\bar{\psi}\psi}_{\mathcal{B},T=0} -   \braket{\bar{\psi}\psi}_{\mathcal{B}=0,T=0}\right).
\end{equation}
Using the previous formula (\ref{phycond}), one can write this as
\begin{equation}
\label{formpref}
\Delta \Sigma = \frac{m^2 c N_c}{M_{\pi}^2f_\pi^2\pi^2}\left(\frac{\sigma(\mathcal{B},0)}{cm} - \frac{\sigma(0,0)}{cm}\right).
\end{equation}
Just plugging in the real bare quark mass in this formula, results in a gross underestimation of the chiral condensate. This is a general property of the soft wall model: $\braket{\bar{\psi }\psi} \sim m$ and hence a small bare quark mass leads to a small condensate. We remedy this situation by artificially choosing a much higher value of the bare quark mass $m$ to obtain a reasonable behavior for the chiral condensate at $\mathcal{B}=0$ but $T\neq 0$. This is detailed in Appendix \ref{barequarkmass}, where we obtain $m=2.967$ GeV. Note that this is roughly a factor of 1000 larger than the actual bare quark mass. \\

\noindent With this value of the bare quark mass, and the experimental values of the pion mass and decay constant, one computes the prefactor of (\ref{formpref}) to be $2997.60$, which is gigantic compared to the actual QCD value of this prefactor (0.0085).\footnote{To get these numbers, we used $M_\pi \approx 135 \text{ MeV}$, $f_\pi\approx 86 \text{ MeV}$ and we took the actual QCD bare quark mass to be $m\approx 5 \text{ MeV}$.} Since this means that the curves in Figure \ref{Shooting} are blown up tremendously compared to QCD, the suitable window of $\ell_c$ shrinks substantially. Hence the value of $\ell_c$ that we should take is almost uniquely determined by the condition that the curve as drawn above is almost flat.\footnote{In effect, this is a shooting method to determine $\ell_c$.} Closer scrutiny and comparison with the lattice results for very small applied magnetic field leads to a suitable value of $\ell_c=\frac{0.4}{\sqrt{c}} \approx 1.03$ GeV$^{-1}$ (close to the green curve of figure \ref{Shooting}).\footnote{This value was determined by matching the lattice value of $\Delta\Sigma$ at $\mathcal{B}=0.024$ GeV$^2$ with the behavior above determined numerically by the shooting method.} This value of $\ell_c$ is also in reasonable unison with the Hawking-Page analysis performed above. \\
If one would compute the relative condensate for higher values of $\mathcal{B}$, one would find a discrepancy with the lattice results for \emph{any} $\ell_c$: the curve here roughly follows a parabolic shape, whereas one should obtain linear behavior for larger magnetic fields. But of course, for larger magnetic fields, we should not trust the background in the first place. \\

\noindent We would like to emphasize here that in our set-up we have fixed several holographic parameters for the $\mathcal{B}=0$ case (the quark mass $m$ and the 5D Newton constant $G_5$). We hence have absolutely no predictability in this case. The value of the additional length scale $\ell_c$ in the confining phase was determined using the $T=0$ and $\mathcal{B} \neq0$ regime of the theory. However, once these are all fixed, the most interesting $T\neq0$ and $\mathcal{B}\neq0$ regime is fully determined by our model and it is here (and only here) that we will predict the behavior of the dual QCD-like theory. One might think that all of these additional parameters, that require experimental or lattice results to fix them, is a serious flaw of our approach. In general this is true, but since we constrain our model to fit the data in several explored regions of the ($T$, $\mathcal{B}$) parameter space, it is hence more likely to find the best possible result of these kinds of models for the final remaining parameter region ($T\neq0$ and $\mathcal{B}\neq0$) as well.

\subsection{Chiral condensate in the deconfined phase}
For the black hole case, we are required to solve the differential equation between the boundary ($r=0$) and the black hole horizon ($r=R_H$). Whereas for pure AdS the integration region stretched all the way to $r\to\infty$, here we are solving the differential equation on a finite interval. It actually turns out to be easier if we numerically integrate the differential equation from the horizon of the black hole to the boundary (so we reverse the integration direction compared to the previous subsection). The major benefit from doing this is that we do not have to employ a shooting method. The same problem was studied for the $B=0$ case by \cite{Colangelo:2011sr} where the authors did utilize a shooting method to integrate from boundary to horizon. Both methods obviously agree in the end. \\

\noindent Since the differential equation contains singular points, a Frobenius analysis is required again.

\subsubsection{Frobenius Analysis}

\subsubsection*{The near-Boundary limit}
In the zeroth order expansion around $r\approx0$, the first coefficient of the differential equation \eqref{eom} behaves like $\sim-\frac{3}{r}$ and the second one like $\sim\frac{3}{r^2}$. Utilizing the ansatz $X_0\sim r^\alpha$, we obtain the indicial equation:
\begin{equation}
\alpha^2-4\alpha+3=0
\end{equation}
with solutions: $\alpha_1=3$ and $\alpha_2=1$. From the general Frobenius method, this gives us two solutions:
\begin{eqnarray}
X_{01}&=&r^3\sum_{k=0}^\infty a_kr^k,\\
X_{02}&=&n\ln r X_{01}+\sum_{k=0}^\infty b_kr^{k+1}.
\end{eqnarray}
Proceeding one level further with the Frobenius analysis, one obtains $a_1=0$ and
\begin{eqnarray}
c b_0&=&na_0, \nonumber\\
b_1&=&b_3=0, \nonumber\\
b_2&=&\text{arbitrary}.
\end{eqnarray}
Choosing $b_2=0$ and $a_0=b_0=1$ as the overall normalization, we find $n=c$ such that
\begin{eqnarray}
\label{eq25}
X_{01}&=&r^3+\mathcal{O}(r^5), \\
X_{02}&=& cr^3\ln r +r+\mathcal{O}(r^5).
\end{eqnarray}
The field in the boundary limit is a superposition of these solutions:
\begin{equation}
L^{3/2}X_0=AX_{02}+CX_{01}
\end{equation}
where the condition
\begin{equation}
\frac{L^{3/2}X_0}{r}\Big|_{r\to0}=m
\label{eq28}
\end{equation}
fixes the coefficient of \eqref{eq25} to $A=m$. Therefore we have:
\begin{eqnarray}
L^{3/2}X_0&=&cmr^3\ln r +mr+Cr^3+\mathcal{O}(r^5)\nonumber\\
&=&cmr^3\ln \left(\sqrt{c}r\right) +mr+\sigma r^3+\mathcal{O}(r^5),
\label{boundarysol}
\end{eqnarray}
where we choose to absorb a part of $C$ into the logarithmic term and define $\sigma$ as the remainder. This number $\sigma$ is directly related to the $\braket{\bar{\psi} \psi}$ condensate: the link between this coefficient of the boundary expansion and the actual QCD quark condensate is made clear in Appendix \ref{chirCondens}.\footnote{The number $\sigma$ on its own is ambiguous to define as one can freely absorb portions of it into the logarithmic term. The quark condensate on the other hand luckily does not share this ambiguity and is perfectly well-defined, which we explain in Appendix \ref{chirCondens}.} Clearly, there is no influence of the magnetic field on the near-boundary limit. It is the same analysis as in the case $B=0$ found in \cite{Colangelo:2011sr}.

\subsubsection*{The near-horizon limit}
In the near-horizon ($r\to R_H$) limit, we can expand the horizon function (\ref{horfunction}) generically as:
\begin{equation}
\label{tayexp}
f(r)=A\left(1-\frac{r}{R_H}\right)+\mathcal{O}\left(r-R_H\right)^2,
\end{equation}
for some constant $A$.
Substituting the ansatz $X_0\sim(1-\frac{r}{R_H})^\alpha$ in the differential equation, we obtain the following indicial equation:
\begin{equation}
\alpha(\alpha-1)-\alpha\left(\frac{R_H}{A}\frac{df(r)}{dr}|_{r=R_H}\right)=0
\end{equation}
whose solutions are:
\begin{eqnarray}
\alpha_1=0&\text{and}&\alpha_2=1+\frac{R_H}{A}\frac{df(r)}{dr}\bigg|_{r=R_H} =0
\label{alpha}
\end{eqnarray}
where the second equality in $\alpha_2$ follows directly from the general expansion of $f(r)$ around the horizon (\ref{tayexp}).
At the next order $X_0\sim1+D\left(1-\frac{r}{R_H}\right)$ and the series expansion yields:
\begin{equation}
D=-\frac{3}{4-\frac{2}{3}\frac{B^2R_H^4}{L^2}}
\end{equation}
where we used the condition $f(r=R_H)\equiv0$ to get the result. Therefore the solution in the near-horizon limit reads:
{\begin{equation}
L^{3/2}X_0(r) =1-\frac{3}{4-\frac{2}{3}\frac{B^2R_H^4}{L^2}}\left(1-\frac{r}{R_H}\right)+\mathcal{O}\left(r-R_H\right)^2.
\label{horizonsol}
\end{equation}
In the case $B=0$, we recover the solution found in \cite{Colangelo:2011sr}. Using the solutions \eqref{horizonsol} and \eqref{boundarysol}, we can numerically integrate the differential equation and determined the dependence of the chiral condensate on the applied magnetic field $\mathcal{B}$ and the temperature $T$.

\subsubsection{The results}

\noindent The quantity $\frac{\sigma}{cm}$ as a function of the dimensionless temperature $\frac{T}{\sqrt{c}}$ is shown in Figure \ref{SigmaDeconf} for different value of the magnetic field.
\begin{figure}[h]
    \centering
   \includegraphics[width=0.75\textwidth]{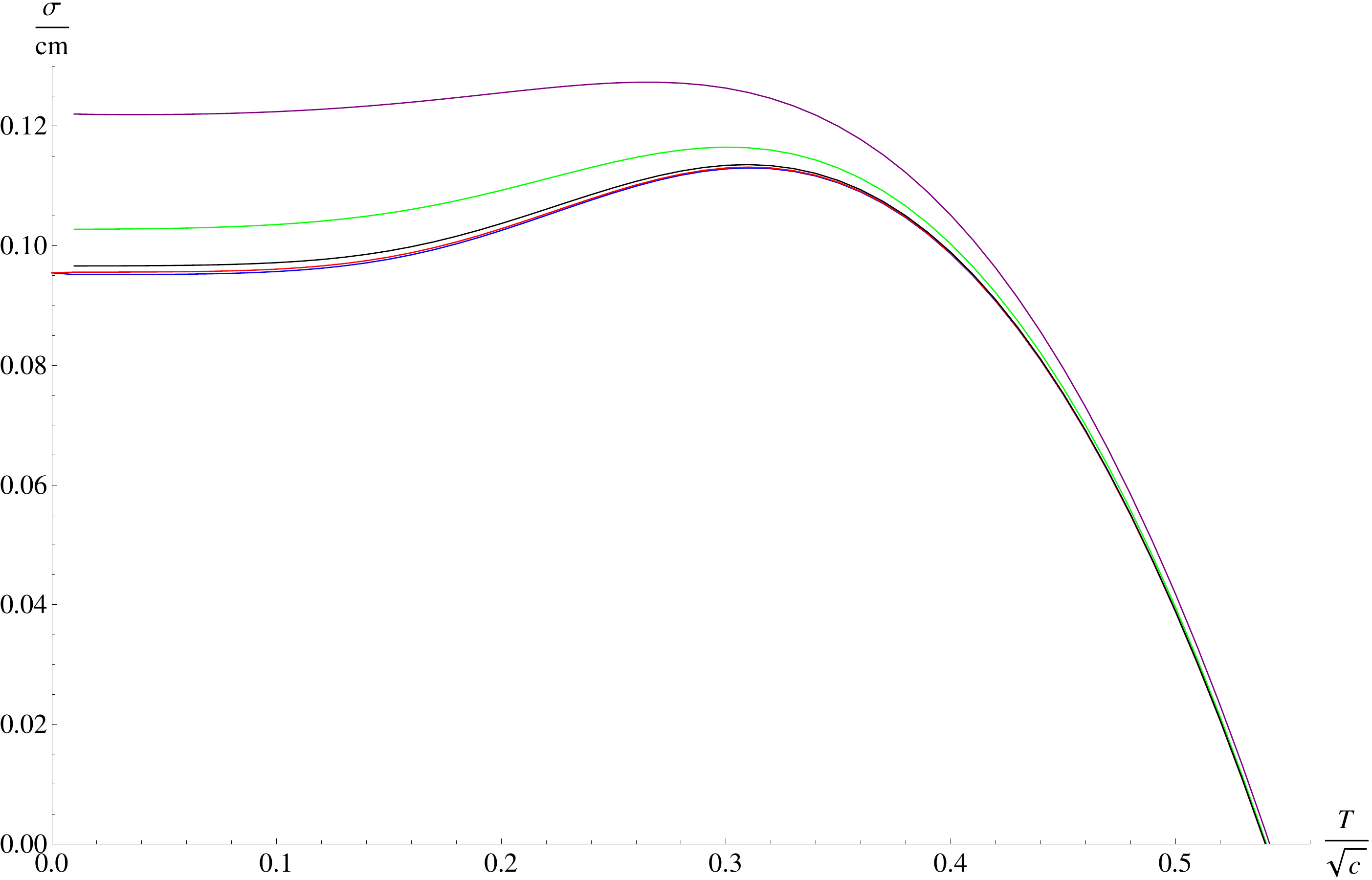}
  \caption{The dimensionless quantity $\frac{\sigma}{cm}$ in terms of $\frac{T}{\sqrt{c}}$ for various values of  $\frac{\mathcal{B}}{c}$. From bottom to top: Blue :  $\frac{\mathcal{B}}{c} = 0$, Red: $\frac{\mathcal{B}}{c} =  0.1$, Black: $\frac{\mathcal{B}}{c} = 0.2$, Green: $\frac{\mathcal{B}}{c}= 0.5$, Purple: $\frac{\mathcal{B}}{c} = 1.0$.}
  \label{SigmaDeconf}
  \end{figure}

\noindent The actual condensate $\braket{\bar{\psi}\psi}$ can then be found as
\begin{equation}
\braket{\bar{\psi}\psi}_{\mathcal{B},T} -   \braket{\bar{\psi}\psi}_{\mathcal{B}=0,T=0} =\frac{N_c mc}{2\pi^2} \left(\frac{\sigma(\mathcal{B},T)}{mc} - \frac{\sigma(\mathcal{B}=0,T=0)}{mc}\right).
\end{equation}
The value of the quark mass $m$ was determined above precisely such that at $\mathcal{B}=0$, the critical temperature is about 210 MeV. Hence all parameters are known in the above equation, and we can readily plot the resulting total chiral condensate (i.e. the sum of up and down condensates) as a function of the applied magnetic field (Figure \ref{CondensateDeconf}).
\begin{figure}[h]
    \centering
   \includegraphics[width=0.75\textwidth]{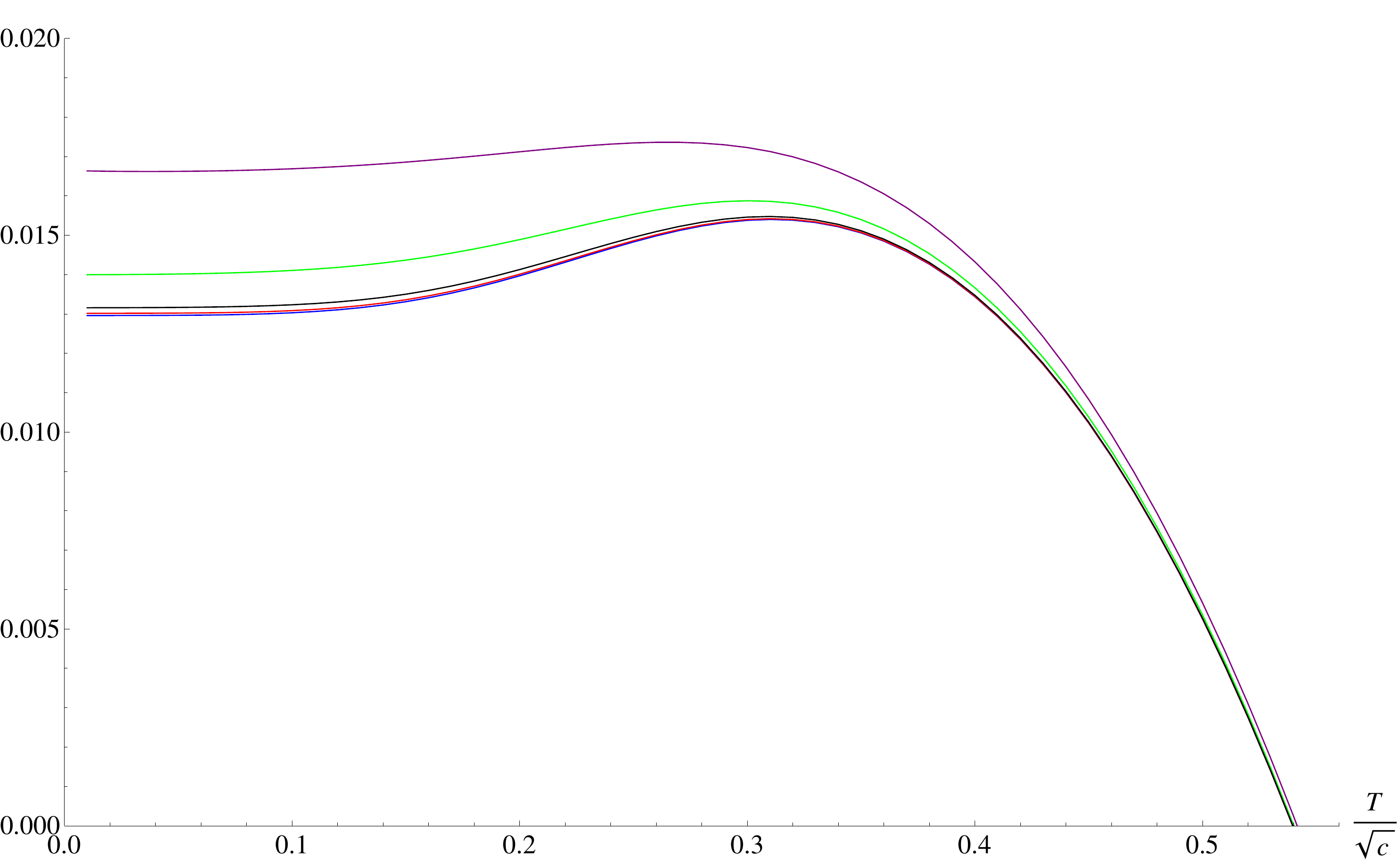}
  \caption{Total 2-flavor condensate $\braket{\bar{Q}Q}$ in terms of $\frac{T}{\sqrt{c}}$ for various values of  $\frac{\mathcal{B}}{c}$. From bottom to top: Blue :  $\frac{\mathcal{B}}{c} = 0$, Red: $\frac{\mathcal{B}}{c} =  0.1$, Black: $\frac{\mathcal{B}}{c} = 0.2$, Green: $\frac{\mathcal{B}}{c}= 0.5$, Purple: $\frac{\mathcal{B}}{c} = 1.0$.}
  \label{CondensateDeconf}
  \end{figure}
	
	\noindent Our main interest in this work lies of course in finding how the critical chiral temperature evolves as the magnetic field is turned on. One can readily distill this relation using the above numerical work, and we find the result of Figure \ref{MagnCata}.
	\begin{figure}[h]
    \centering
   \includegraphics[width=0.75\textwidth]{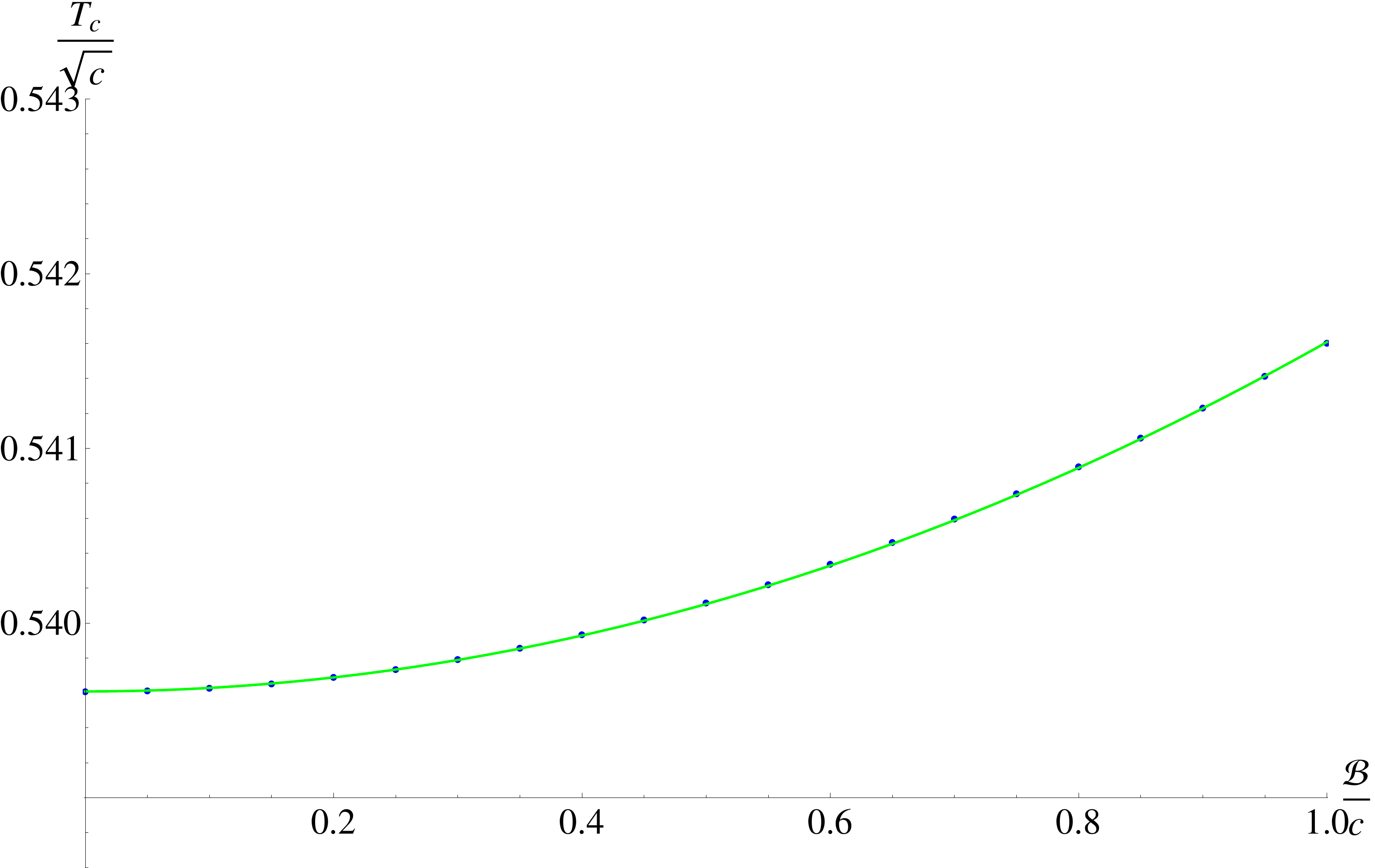}
  \caption{Chiral phase transition temperature $\frac{T_c}{\sqrt{c}}$ as a function of applied magnetic field $\frac{\mathcal{B}}{c}$. The green curve represents a parabolic fit to the data.}
  \label{MagnCata}
  \end{figure}
	Quite surprisingly, the numerical data lie almost perfectly on a parabola of the form $\frac{T_c}{\sqrt{c}} = 0.002 \frac{\mathcal{B}^2}{c^2} + 0.5396$. \\
	
	\noindent For the reader's convenience, we draw the same relation again, but this time with the phenomenological value of $c=0.151$ GeV$^2$ in Figure \ref{MagnCataNoC}.
		\begin{figure}[h]
    \centering
   \includegraphics[width=0.75\textwidth]{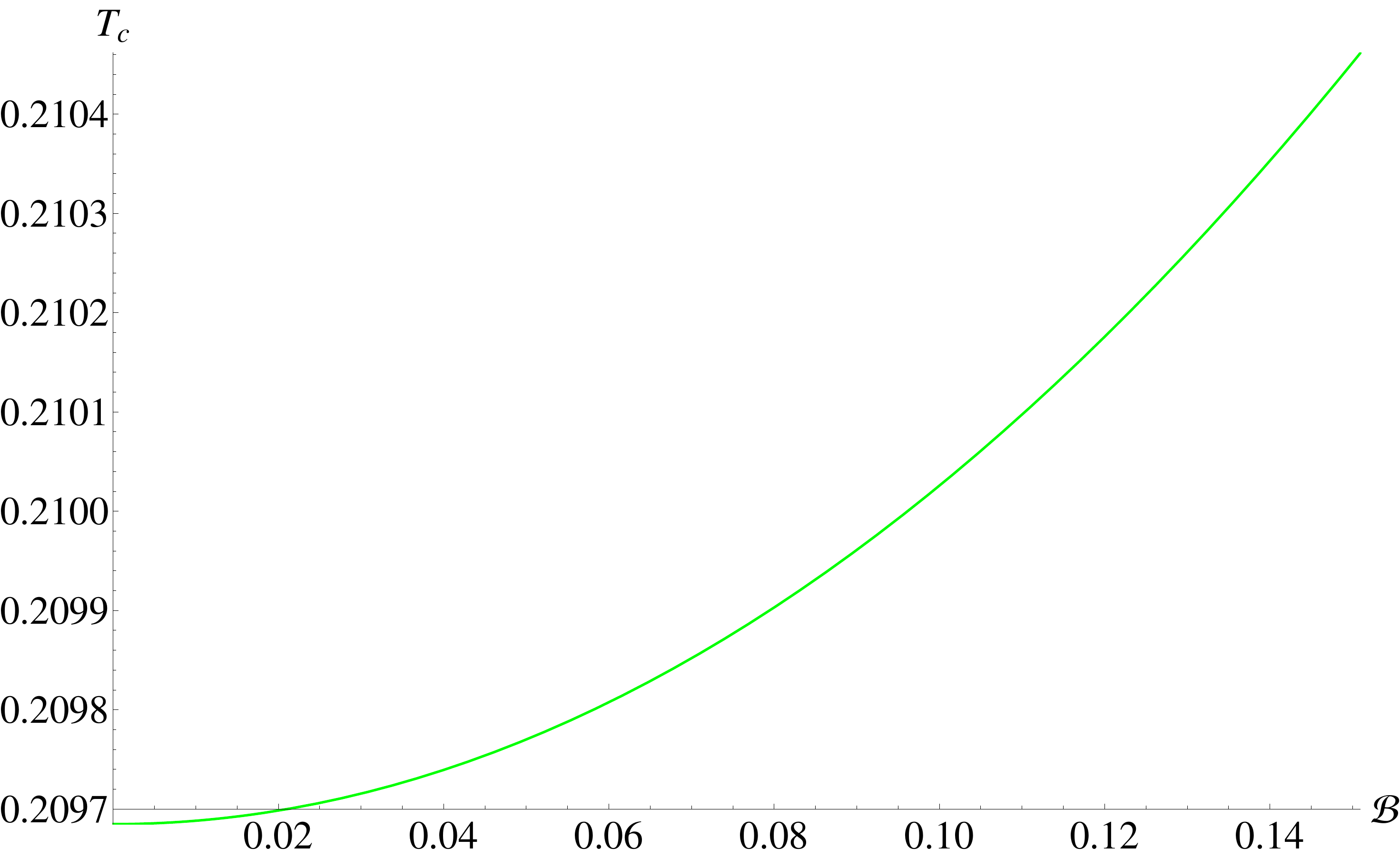}
  \caption{Chiral phase transition temperature $T_c$ (GeV) as a function of applied magnetic field $\mathcal{B}$ (GeV$^2$) for $c=0.151$ GeV$^2$.}
  \label{MagnCataNoC}
  \end{figure}
	
	\noindent Clearly, these Figures show that one finds \emph{magnetic catalysis} for the chiral phase transition. \\
	
\noindent As we are working with the black hole geometry (the deconfined phase), we should in principle only trust these results for $T>T_{HP}$, with $T_{HP}$ the Hawking-Page temperature determined in Section \ref{HPSOFT} as a function of $\mathcal{B}$. In the confined phase geometry (thermal magnetized AdS), the temperature does not figure in the geometry itself. This makes all computations manifestly independent of the temperature, and the chiral condensate would be constant as a function of $T$ all the way up to $T_{HP}$. \\
This property is a generic feature of holographic classical backgrounds (large $N$ approximation) since the only way to properly introduce the temperature into the geometry is by including a black hole horizon. \\
A somewhat uncomfortable consequence here is that the chiral condensate would exhibit a discontinuous jump at $T_{HP}$ where it suddenly starts following the above deconfined curves. Lattice results show no sign of any jump whatsoever. This is a nuisance inherent to holographic QCD models, and something we will have to live with here. This happens for any value of $\mathcal{B}$ and was discussed in \cite{Colangelo:2011sr} as well. Note though that this complication happens at a lower temperature than $T_c$ and hence its effect for our purposes is not really visible.\footnote{One could object here and say that we \emph{chose} $T_c$ to be larger than $T_{HP}$. This is indeed true and this is necessary to have any sensible result at all. If one would not do this, and the chiral temperature $T_c$ would be reached before the deconfinement temperature $T_{HP}$, the condensate would suddenly jump to zero (where we interpret negative values of the condensate to mean that it vanishes). }

Perhaps the critical transition temperatures will develop a different behavior if the magnetic field keeps to grow, but we refrain from speculating about this. As we do know our results are exact at leading order in $\mathcal{B}^2$, they are trustworthy for sufficiently small values of the magnetic field, and already in this region, our holographic predictions are at odds with the lattice predictions for the chiral transition. Another limit that could be probed semi-analytically is the $\mathcal{B}\to\infty$ case, the corresponding metric is also known analytically and presented in the Appendix of \cite{D'Hoker:2009mm}. The deconfinement transition in the hard wall model in this extreme limit was analyzed in \cite{Mamo:2015dea}. We will not generalize that analysis to our current soft wall setting, as the phenomenologically interesting region, potentially realizable during a heavy ion collision, is not that of a very large magnetic field.

\subsection{Revisiting the chiral transition in the hard wall model at zero magnetic field}
\label{hwrevisit}
To clear out some misconceptions about the chiral transition in the hard wall model for $B=0$, let us again go through the analysis here. The relevant equation of motion can be extracted from the one of \eqref{eom} by sending $c\to0$, while keeping in mind that at $r=r_0$, a hard wall is placed.\\
\noindent In the confinement phase, corresponding to $f(r)\to 1$, the relevant solution in the hard wall setting is provided by
\begin{equation}
L^{3/2}X_0(r)=m r + \sigma r^3.
\end{equation}
In this case, the quark mass $m$ and chiral condensate $\sigma$ can both be chosen at will. In the seminal work \cite{Erlich:2005qh}, these and other parameters were fixed by matching a few quantities on top of a preselection of QCD observables.

As soon as a horizon forms, i.e.~when deconfinement sets in, one finds a solution for \eqref{eom} (still with $c\to0$, but now keeping $f(r)$) in terms of hypergeometric functions (see also \cite{Ghoroku:2005kg}),
\begin{equation}
  L^{3/2}X_0(r)= m  r~{}_2F_1(1/4,1/4,1/2;r^4/r_h^4) + \sigma r^3~{}_2F_1(3/4,3/4,3/2;r^4/r_h^4).
\end{equation}
In \cite{Kim:2006ut}, it was then concluded that $m=0$ and $\sigma=0$ as both hypergeometric functions are singular at the horizon $r=r_h$. This means that chiral symmetry is restored maximally (even no quark mass allowed) as soon as the deconfinement phase is considered in the hard wall.
Though, this reasoning is mathematically flawed. Based on general Frobenius analysis arguments, we expect that by taking a suitable linear combination of the foregoing hypergeometric solutions, a regular solution at $r=r_h$ can be obtained. Indeed, one verifies that
\begin{equation}\label{leg}
  L^{3/2}X_0(r)\propto r P(-1/2;r^2/r_h^2)
\end{equation}
in terms of the Legendre function $P(\ell;x)$ renders us with a solution that is nonetheless regular at the horizon. Expanding the solution \eqref{leg} around $r=0$, we find
\begin{equation}\label{leq2}
  L^{3/2}X_0(r)= m r + m\frac{\Gamma^4(3/4)}{\pi^2 r_h^2}r^3 + \mathcal{O}(r^5)
\end{equation}
after a suitable normalization.\\
\noindent Thus, the chiral condensate in the deconfined hard wall model can be read off from expression \eqref{leq2} to be
\begin{equation}\label{leq3}
  \sigma = m\frac{\Gamma^4(3/4)}{\pi^2 r_h^2}\propto mT^2.
\end{equation}
In the last step, we filled in the Hawking temperature related to the horizon at $r=r_h$.\\
\noindent Here, we clearly observe the somewhat pathological behaviour of the chiral dynamics in the deconfined hard wall model: the chiral condensate grows quadratically with the temperature for non-vanishing quark mass. There is thus no obvious chiral restoration in the hard wall model. It thus also makes no sense to identify the deconfinement and chiral transition. Evidently, this is the reason why we chose the soft wall model to begin with to study the possibility of a dynamical chiral transition. Only with $m\equiv0$, meaning in the chiral limit, the chiral transition makes sense in the hard wall model. Though, as soon as an even infinitesimal bare quark mass is coupled on, the unwanted behaviour \eqref{leq3} gives problems for sufficiently large $T$. In the soft wall case, it even makes no sense to work in the chiral limit $m\equiv0$ as otherwise all chiral dynamics would be lost (i.e.~no surviving chiral condensate, even in the confined phase). We end this short digression on chiral dynamics in the hard wall model by noticing that the ensuing conclusion of \cite{Kim:2006ut} can no longer hold: the soft wall chiral dynamics cannot be the same as that of the hard wall case since the hard wall discussion of \cite{Kim:2006ut} needs to be adapted anyhow.

\section{Conclusions}\label{sect7}
On general grounds \cite{Gusynin:1994re}, it is expected that a magnetic field promotes chiral symmetry breaking, said otherwise, it acts as a catalyst. Naively, one would thus also expect that the chiral transition temperature, at which chiral symmetry is restored\footnote{Suitably defined in the presence of massive dynamical quarks.}, increases. Nonetheless, state-of-the-art lattice QCD revealed at sufficiently high temperature an inverse magnetic catalysis in the chiral sector \cite{Bali:2011qj,Bali:2012zg,Ilgenfritz:2012fw}. This has stimulated a lot of research, see e.g.~\cite{Fraga:2012fs,Fraga:2012ev,Ayala:2014gwa,Ayala:2014iba,Ayala:2015bgv,Fukushima:2012kc,Farias:2014eca,Ferreira:2014kpa,Costa:2015bza,Mueller:2015fka}.\\

\noindent As we are to consider QCD around the deconfinement transition, at which instance it is still strongly coupled, we need a suitable tool to access this regime, this in addition of a magnetic field that further complicates matters. One such tool is based on the AdS/CFT correspondence, adapted to the study of strongly coupled QCD questions.\\

\noindent In recent AdS/QCD papers \cite{Mamo:2015dea,Rougemont:2015oea}, the inverse catalysis was reported, though we must remark that these papers solely studied the deconfinement temperature, using different set-ups per paper. No chiral physics was directly included and the faith of a genuine (inverse) magnetic catalysis remained a bit mystified. To our knowledge, there are till today no AdS/QCD papers, be it top-down or bottom-up, on the market that can accommodate for a chiral transition temperature dropping with increasing magnetic field. In this work, we investigated this question into more depth for the first time, this by employing a phenomenological hard and soft wall AdS/QCD model supplemented with a magnetic field in the bulk and with an appropriate asymptotic AdS behavior of the 5D magnetic field-dependent bulk metric \cite{D'Hoker:2009mm,D'Hoker:2009bc}.\\

\noindent Throughout the course of the paper, we obtained several in se interesting results: we studied the black hole horizon structure of the D'Hoker-Kraus solution \cite{D'Hoker:2009mm,D'Hoker:2009bc}; we analyzed the thermodynamic stability of our model in the region of interest; we corroborated on how to introduce a finite chiral condensate; we elaborated on how, at nonzero magnetic field, the AdS length $L$ is no longer completely decoupling from physically relevant quantities.\\

\noindent The main outcome of our work we wish to report is however that, for reasonable values of $\ell_c$, there is indeed ``inverse magnetic catalysis'' for the deconfinement transition as found before in \cite{Mamo:2015dea}, but more importantly, that there is {\textit no} trace of inverse magnetic catalysis observed in the corresponding chiral transition, which is the appropriate quantity to look for it after all. We are thus not able to confirm, within this model at least, the comment of \cite{Rougemont:2015oea}, based on \cite{Fraga:2012fs,Fraga:2012ev}, that inverse magnetic catalysis is more related to a decent description of confinement rather than to a decent description of chiral dynamics, since in the soft wall model, both transitions display the opposite behavior.\\

\noindent This being said, our work is evidently not the final word on this. Two major improvements are in order: we should develop a self-consistent\footnote{That is, at least solving the Einstein gravitational equations of motion. Possibly the tools of \cite{Lindgren:2015lia} can be useful in this context.} dynamical wall model whereby the magnetic field is taken into account \`{a} la D'Hoker-Kraus and next to that, we need to circumvent the undesired chiral condensate properties of the soft wall models (vanishing condensate at vanishing current quark mass). The proceeding goals can, in principle, be achieved by allowing for appropriate potentials for both dilaton $\phi$ that models confinement and the scalar field $X$ that models the chiral condensate. Of course, those will bring gross computational effort with them. We hope to come back to this in future work.

\section*{Acknowledgements}
D.~R.~Granado is grateful for a PDSE scholarship from the ``Coordena\c{c}\~{a}o de Aperfei\c{c}o\-amento de Pessoal de N\'{i}vel Superior'' (CAPES). The work of T.~G.~Mertens was supported by the UGent Special Research Fund, Princeton University, the Fulbright program and a Fellowship of the Belgian American Educational Foundation. We thank R.~Degezelle for a careful reading of the manuscript.

\appendix
\section*{Appendix}
\section{Black hole geometry}\label{sect3}
In this Appendix, we analyze the black hole geometry (\ref{bhmetric}) that we will utilize in the remainder of this work. Since, it has some very peculiar properties from the gravity point of view, we take the time here to perform an elaborate analysis. To appreciate the effects the magnetic field can have on the black hole horizon structure, we will for the moment ignore the fact that $B$ should be sufficiently small, but we shall rather consider the black hole metric \eqref{bhmetric} for arbitrary $B$ for the time being.

\subsection{Locations of the event horizons}
The horizon function is given by
\begin{equation}
\label{horfunction}
f(r) = 1 - \frac{r^4}{r_h^4} + \frac{2}{3}\frac{B^2r^4}{L^2}\ln\frac{r}{\ell_d},
\end{equation}
where $f(R_H)=0$ determines the horizon(s).
This equation can be solved analytically in terms of the Lambert $\mathcal{W}$-functions, where one readily shows that there exist at most two real (physical) solutions given by
\begin{align}
R_{H1}^4 &= \ell_d^4\exp\left(\frac{6L^2}{B^2r_h^4}\right)\exp\left(\mathcal{W}_0\left(-6L^2\frac{e^{-\frac{6L^2}{B^2r_h^4}}}{B^2\ell_d^4}\right)\right) , \\
R_{H2}^4 &= \ell_d^4\exp\left(\frac{6L^2}{B^2r_h^4}\right)\exp\left(\mathcal{W}_{-1}\left(-6L^2\frac{e^{-\frac{6L^2}{B^2r_h^4}}}{B^2\ell_d^4}\right)\right).
\end{align}
For these solutions to exist, the Lambert $\mathcal{W}$-functions have to be real, which is only satisfied if their argument is larger than $-1/e$. This constraint leads to
\begin{equation}
\frac{B^2}{6L^2} < \frac{B^2}{6L^2}\ln\left(\frac{B^2\ell_d^4}{6L^2}\right) + \frac{1}{r_h^4}.
\end{equation}

\noindent At $B=0$, one finds $R_{H1}\to \infty$ and $R_{H2} \to r_h$, using the expansions for $x \ll 1$:
\begin{equation}
\mathcal{W}_0(x) \approx x ~\text{and}~\mathcal{W}_{-1}(x) \approx \ln(-x) - \ln(-\ln(-x)).
\end{equation}

\noindent As $B\to\infty$, one can use the same series expansion and one finds $R_{H1} \to L$ and $R_{H2} \to 0$. Numerically, one can check that this procedure of turning on $B$ causes this transition monotonically. Since the lower $R_H$ is, the \emph{larger} the horizon radius, we find that both outer and inner horizons expand as $B$ is turned on. In the case where $r_h<\ell_d$, the situation is shown in Figures \ref{transition} and \ref{horizonsize}.
\begin{figure}[h]
    \centering
   \includegraphics[width=0.65\textwidth]{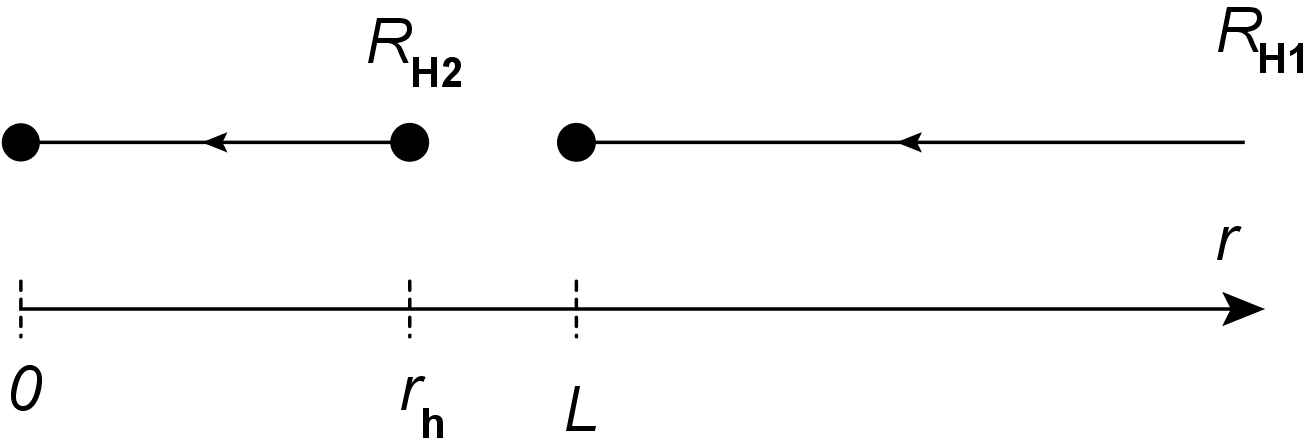}
  \caption{Locations of horizons as $B$ is increased.}
  \label{transition}
  \end{figure}
  \begin{figure}[h]
    \centering
   \includegraphics[width=0.65\textwidth]{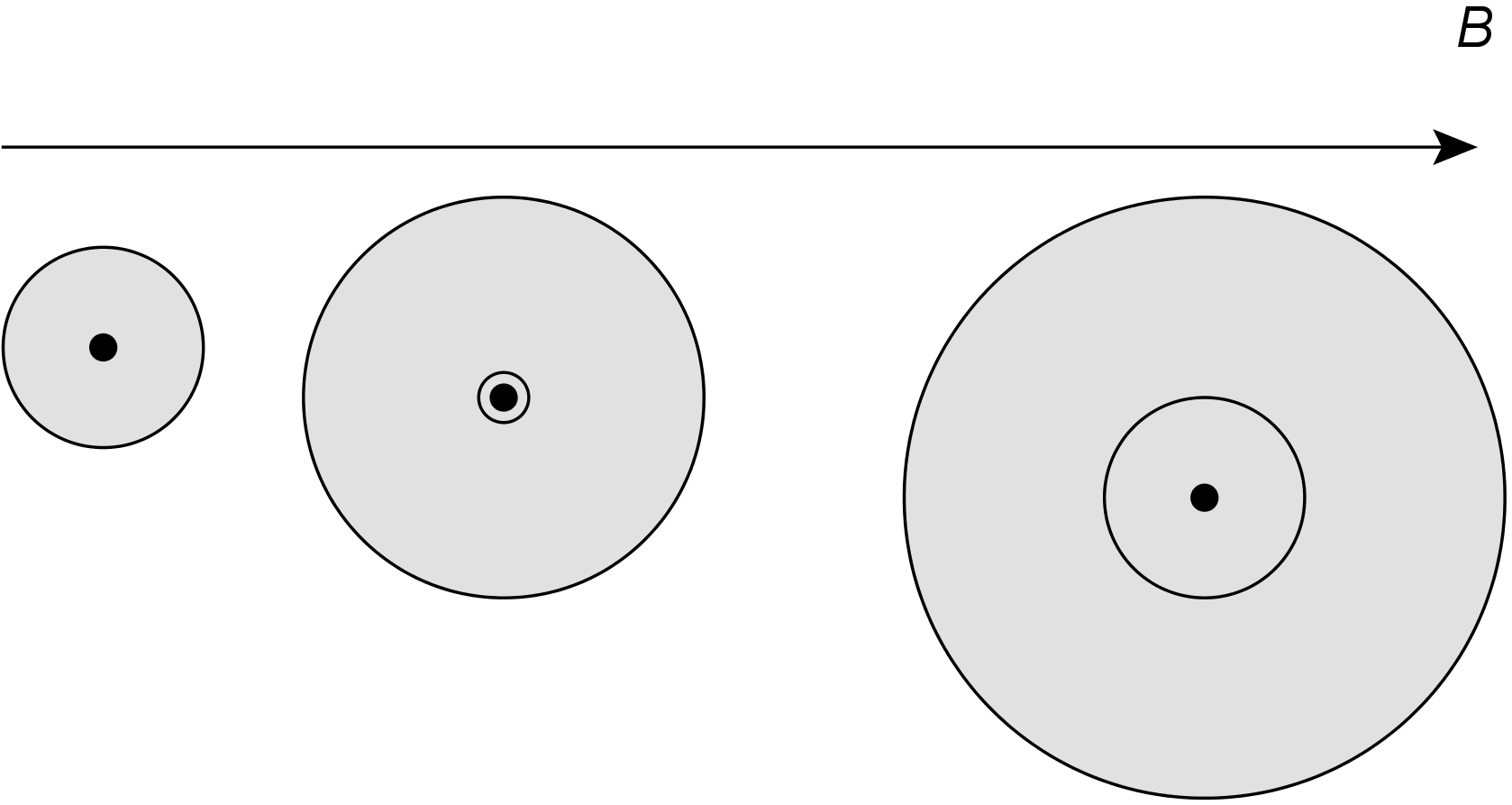}
  \caption{Size of the horizons of the black hole as $B$ is changed. Both horizons grow monotonically.}
  \label{horizonsize}
  \end{figure}

\noindent Of course, in the case where $r_h>\ell_d$, the locations of both horizons have the possibility to join somewhere as shown in Figure \ref{transition2}. Numerically it can be checked that in this case, there \emph{always} exists an intermediate range for $B$ where no horizon is present at all and the singularity is exposed in Figure \ref{horizonsize2}. The full story is quite a bit more complicated in this case, as the horizons no longer move in a monotonic fashion.
\begin{figure}[h]
    \centering
   \includegraphics[width=0.65\textwidth]{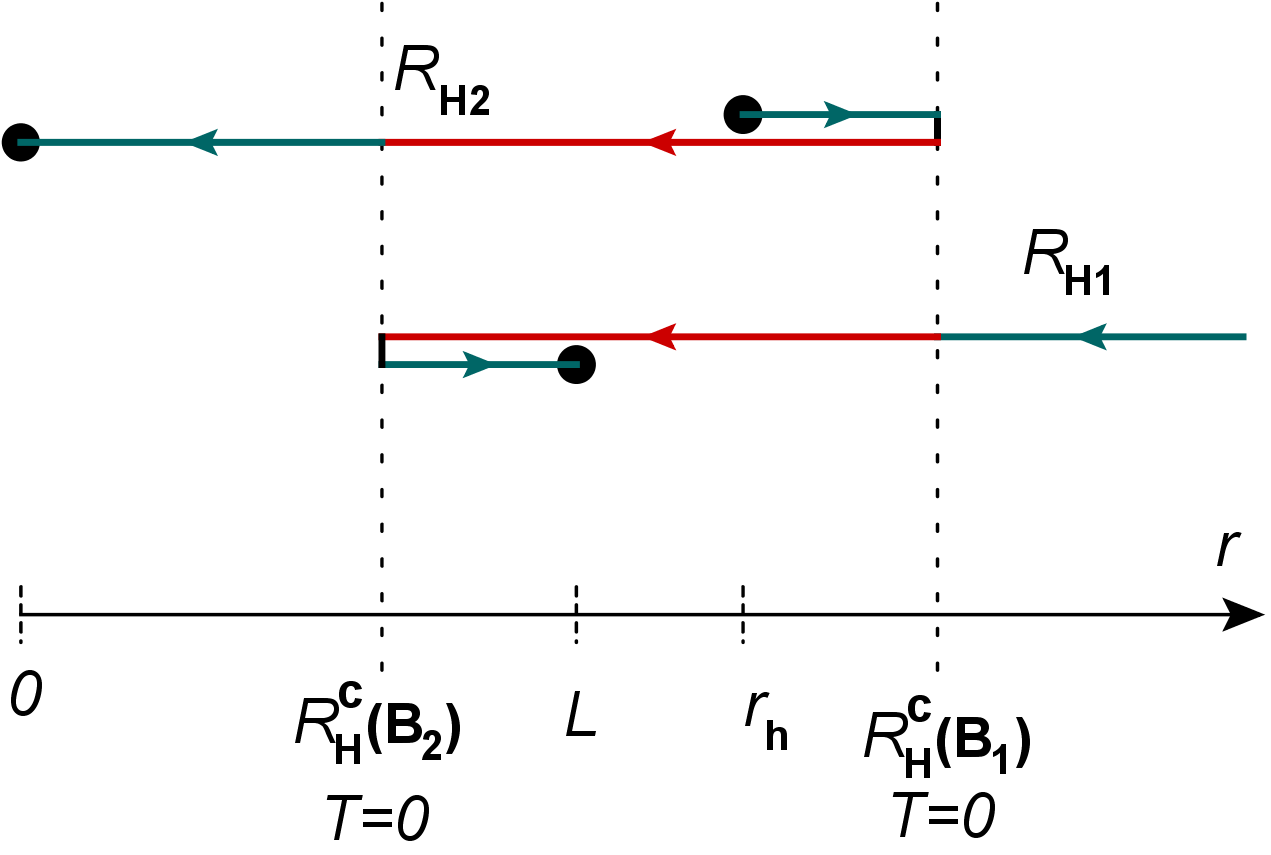}
  \caption{Locations of horizons as $B$ is increased. If $r_h>\ell_d$, the possibility exists that both horizons coincide at some values of $B$. This happens at $B_{1}$ and $B_{2}$. The red zones indicate values of $B$ for which no horizon is present at all.}
  \label{transition2}
  \end{figure}
  \begin{figure}[h]
    \centering
   \includegraphics[width=0.65\textwidth]{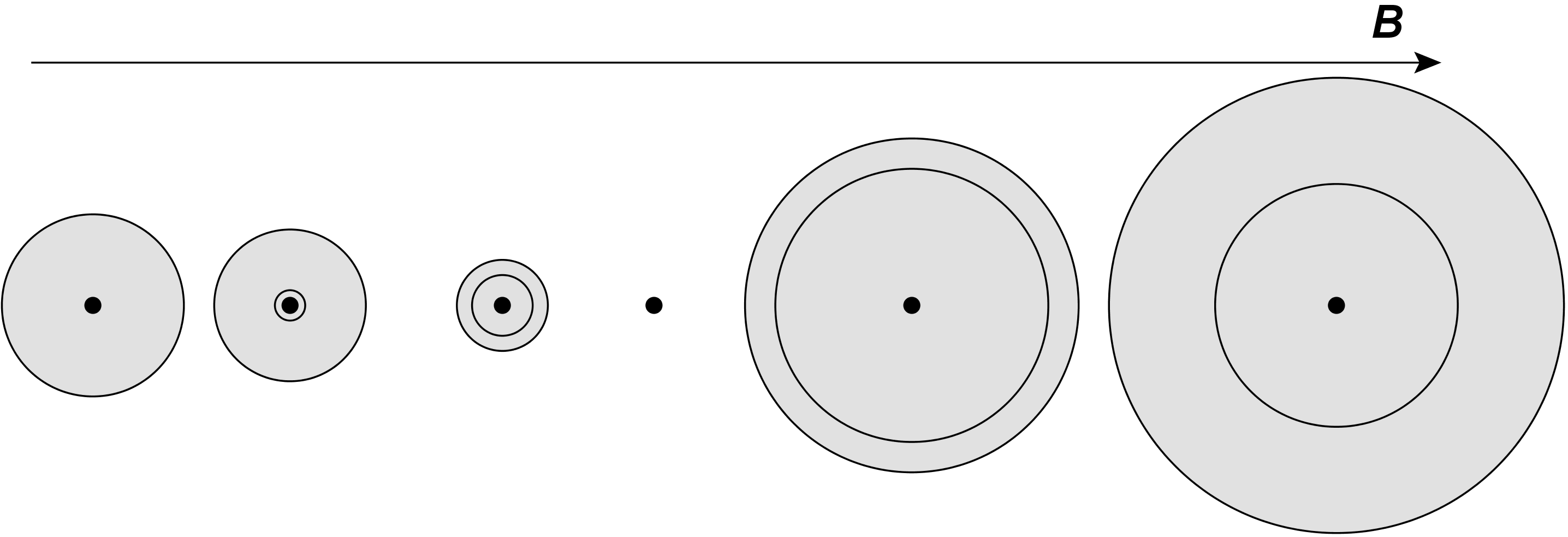}
  \caption{Size of the horizons of the black hole as $B$ is changed. If $r_h>\ell_d$, there always exists a range for $B$ where no horizons are present at all and the singularity becomes naked.}
  \label{horizonsize2}
  \end{figure}
\subsection{Hawking temperature of the black hole}
The Hawking temperature can be readily computed and is given by
\begin{equation}
\label{hawktem}
T_H=\frac{1}{4\pi}\bigg|\frac{4}{R_H}-\frac{2}{3}\frac{B^2R_H^3}{L^2}\bigg|.
\end{equation}
It is shown as a function of $R_H$ in Figure \ref{THawking}.
\begin{figure}[h]
\centering
\includegraphics[width=0.65\textwidth]{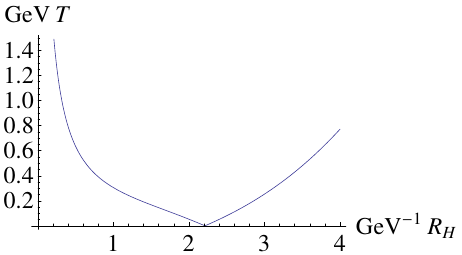}
\caption{$T_H$ as a function of horizon radius $R_H$ for $\mathcal{B} = 0.8$ GeV$^{2}$.}
\label{THawking}
\end{figure}
The Hawking temperature vanishes at a critical value of $R_H$ that we will henceforth call $R_H^c$. \\

\noindent As a sidenote, we remark that in comparing the free energy as computed in the bulk with that of the boundary, in the large temperature limit one should reproduce the free energy of a weakly interacting gluon gas. We will use this in Appendix \ref{normalization} to find the 5D Newton constant in terms of the AdS length $L$. \\
Turning on a magnetic field, the large $T$-limit can be found by taking $R_H\to0$. After some straightforward computations with the on-shell action
\begin{equation}
F =\frac{L^3}{8\pi  G_5}\left[ e^{-cR_H^2}\left(-\frac{1}{R_H^4}+ \frac{c}{R_H^2}\right) + \left(\frac{B^2}{3L^2} + c^2\right) \text{Ei}(-cR_H^2) + \frac{1}{2r_h^4}\right],
\end{equation}
one finds that all $B$-dependent terms are subdominant and the large $T$ asymptotics follows the same Stefan-Boltzmann $F\sim T^4$ result.\footnote{We discarded temperature-independent terms when writing this expression.} This is expected, since at high temperatures, the average kinetic energy of the particles is high enough such that the influence of the $B$-field becomes negligible.

\subsection{Inequality on the horizon radius and extremal black holes}
In our case, we start with the temperature $T$ and the physical magnetic field $\mathcal{B}$ as imposed by the boundary QCD-like theory. The above temperature relation (\ref{hawktem}) then allows us to distill two possible values of $R_H$. The horizon condition $f(R_H)=0$ on its turn then gives us a unique value of the parameter $r_h$. Having determined all of the parameters, we must finally check that our $R_H$ is indeed the outer horizon of the black hole by determining both solutions of $f(r)=0$ with the now known value of $r_h$. \\
The order of determining the black hole parameters given above is very important for discerning the dependent from the independent variables in our story. \\
\noindent It turns out that if one should choose the lowest value of $R_H$ in the first step, this always leads to an outer horizon. Conversely, choosing the highest value of $R_H$ always leads to an inner horizon. So we can only use the first descending part of the $T(R_H)$ curve.\footnote{This solves an initial worry one might have in that large $T$ could also imply large $R_H$. In that case, one would have found instead for the free energy at high temperatures:
\begin{equation}
F \sim \frac{L}{G_5}B^2\ln\left(\frac{6\pi TL^2}{B^2\ell_d^3}\right),
\end{equation}
which is unphysical, as it disagrees with the Stefan-Boltzmann prediction. Fortunately, this regime is absent altogether.}
This immediately imposes an upper bound on $R_H$ for a given $\mathcal{B}$ as
\begin{equation}
R_H(\mathcal{B}) \leq R_H^c(\mathcal{B}),
\end{equation}
which implies small black holes are incompatible with turning on a $\mathcal{B}$-field.
We will see below that it is possible to saturate this bound. \\

\noindent Some conclusions.
\begin{itemize}
\item{For both cases, Figures \ref{transition} and \ref{transition2} demonstrate that for any given value of $R$, there exists at most one value of $\mathcal{B}$ for which this $R$ is a horizon.}
\item{
Both horizons coincide when
\begin{equation}
\frac{B^2}{6L^2} = \frac{B^2}{6L^2}\ln\left(\frac{B^2\ell_d^4}{6L^2}\right) + \frac{1}{r_h^4}.
\end{equation}
This equation has two solutions when $r_h>\ell_d$, which we call $B_1$ and $B_2$ with associated physical magnetic fields $\mathcal{B}_1$ and $\mathcal{B}_2$. For these values of $\mathcal{B}$, the horizon locations are respectively $R_H^c(\mathcal{B}_1)$ and $R_H^c(\mathcal{B}_2)$, saturating the inequality
\begin{equation}
R_H(\mathcal{B}) \leq R_H^c(\mathcal{B}).
\end{equation}
Hence at these values of $\mathcal{B}$, both horizons coincide, and the Hawking temperature becomes zero.
}
\item{The converse statement is also true. If $T=0$, then both horizons should coincide and the black hole becomes extremal. One can easily demonstrate this by substituting $B^2=6L^2/R_H^4$ into the horizon condition $f(R_H)=0$. Rewriting this in terms of $B$, one finds
\begin{equation}
\frac{B^2}{6L^2} = \frac{B^2}{6L^2}\ln\left(\frac{B^2\ell_d^4}{6L^2}\right) + \frac{1}{r_h^4},
\end{equation}
precisely the condition for a doubly degenerate horizon.}

\item{From the previous remark, it is clear that taking $T\to 0$ does \emph{not} yield thermal AdS, but instead the extremal versions of these black holes. This is of course a general property of charged black holes.}
\end{itemize}

\subsection{Thermal AdS}
Thermal AdS can be obtained by letting $r_h \to \infty$. Hence,
\begin{equation}
f(r) = 1 + \frac{2}{3}\frac{B^2r^4}{L^2}\ln\frac{r}{\ell_c}.
\end{equation}
Curiously, this background can also develop horizons if $B$ is too large. For $B=0$ there are obviously no horizons present. However for $B$ large enough, i.e.
\begin{equation}
B > \frac{\sqrt{6e}L}{\ell_c^2},
\end{equation}
a degenerate horizon forms that immediately splits into an inner and outer horizon. As $B$ increases further, the inner horizon moves inwards and the outer horizon moves outwards in a monotonic fashion. \\

\noindent If we want to interpret this as a confining background, we are hence restricted to studying this space for sufficiently small values of $B$, which is indeed the range of validity of the solution in the first place.

\section{On the normalization of the magnetic field $B$}
\label{normalization}
In the $\mathcal{N}=4$ SYM theory studied by D'Hoker and Kraus \cite{D'Hoker:2009bc}, the relation between the physical magnetic field, $q\mathcal{B}$, and the magnetic field in their action, $B$, can be found by matching R-current anomalies in bulk and boundary. In the bulk this is given by the contribution of a Chern-Simons term whose prefactor is fully fixed from supersymmetry. On the boundary, one relies on the triangle anomaly computed between R-current operators. Upon reinstating units of the AdS length, they obtain
\begin{equation}
\label{n4Bfys}
q\mathcal{B} = \frac{\sqrt{3}}{L} B \approx  \frac{1.73}{L} B.
\end{equation}

\noindent For QCD, or at least the AdS/QCD wall models under consideration, we cannot follow the same logic, as the normalization of the Chern-Simons term in the bulk is not fixed by supersymmetry, and in fact is determined by demanding equality between the anomalies in bulk and boundary. This leaves no further information to be distilled from this and one hence cannot fix the normalization of the magnetic field using this method. \\

\noindent Instead, we will rely on the normalization of the gauge term in the bulk. The action of the gauge fields in the soft wall model, normalized by comparing with the QCD flavor-flavor correlators, is given by \cite{Jugeau:2013zza,Colangelo:2008us}
\begin{equation}
S= - \frac{N_c}{48\pi^2 L}\int d^5x e^{-\phi}\sqrt{-g}\text{Tr}\left[F_L^2+F_R^2\right].
\end{equation}
This gauge field is holographically dual to the conserved $SU(N_f)$ flavor currents of the boundary QCD-like theory. A background magnetic field is modeled by turning on a vacuum expectation value for the vector-part of the gauge fields by setting:
\begin{equation}
V = A_L = A_R
\end{equation}
and choosing $F_{12} = Q q\mathcal{B}$ where $Q$ is the $2\times2$ diagonal matrix in flavor space with entries $\left(+\frac{2}{3},-\frac{1}{3}\right)$ representing the electric charges of the $u$ and $d$ quark. Here $q$ denotes the elementary electric charge. \\
Plugging this ansatz into the action, we obtain
\begin{equation}
\label{ac1}
S= - \frac{N_c}{12\pi^2 L}\int d^5x e^{-\phi}\sqrt{-g}\text{Tr}\left[Q^2\right] \left(q\mathcal{B}\right)^2  g^{xx}g^{yy}.
\end{equation}

\noindent In \cite{D'Hoker:2009bc}, D'Hoker and Kraus choose a different normalization of the Maxwell part of their action, consistent with $D=5$ gauged supergravity. Their Maxwell action is normalized as\footnote{We have inserted the contribution from the dilaton here, even though it is turned off in the solution obtained in \cite{D'Hoker:2009bc}.}
\begin{equation}
S =- \frac{1}{16\pi G_5} \int d^5x e^{-\phi}\sqrt{-g}F^2.
\end{equation}
The magnetic field introduced by D'Hoker and Kraus is simply the magnitude of the non-zero component of $F$ and leads to
\begin{equation}
\label{ac2}
S =- \frac{1}{8\pi G_5} \int d^5x e^{-\phi}\sqrt{-g}B^2 g^{xx}g^{yy}.
\end{equation}

\noindent Comparing the actions (\ref{ac1}) and (\ref{ac2}), one can find the rescaling of $B$ necessary to obtain the physical magnetic field $\mathcal{B}$:
\begin{equation}
\label{mageq}
B = \sqrt{\frac{8\pi G_5 N_c \text{Tr}\left[Q^2\right]}{12\pi^2 L}} q\mathcal{B}.
\end{equation}

\noindent To proceed, we need the relation between the 5D Newton constant and the AdS length $L$ determined previously. The ratio $L^3/G_5$ can be found by demanding that the high temperature limit of the free energy approaches the Stefan-Boltzmann result and hence matches between the bulk and the boundary gluon gas, see e.g.~\cite{Veschgini:2010ws,Gursoy:2008bu}. Comparing these expressions, one readily finds
\begin{equation}
G_5 = \frac{45\pi L^3}{16\left(N_c^2-1\right)}.
\end{equation}
Finally inserting this expression in equation (\ref{mageq}) and setting $N_c=3$ and $\text{Tr}\left[Q^2\right]=\frac{5}{9}$, we obtain
\begin{equation}
B = \frac{5}{8}L q\mathcal{B} = 0.625 L q\mathcal{B},
\end{equation}
or
\begin{equation}
\label{qcdbfys}
q\mathcal{B}  = \frac{1.6}{L} B.
\end{equation}
One can readily compare the $\mathcal{N}=4$ result (\ref{n4Bfys}) (obtained through anomaly matching) and the QCD result (\ref{qcdbfys}) (obtained by matching the normalization of the action using flavor-flavor correlators), which are remarkably close. \\
Clearly, the obtained 4D physical magnetic field $q\mathcal{B}$ has the correct dimension of GeV$^2$. In all remaining sections of this work, we will omit writing the elementary charge $q$.

\section{Computational details on the Hawking-Page transition for the hard wall model}
\label{apphard}
We collect the details to determine the on-shell actions for the hard wall model.
\subsection{Black hole - deconfined phase}

\subsubsection*{Bulk action}
Using the Euclidean version of the black hole metric \eqref{bhmetric} we can compute the black hole bulk action:
\begin{eqnarray}
S^{bh}_{bulk}&=&\frac{V_3}{8\pi G_5}\int^\beta_0dt_E\int^{R_H}_{r_\lambda}dr\sqrt{g}\left(\frac{4}{L^2}+\frac{2}{3}B^2g^{xx}g^{yy}\right)\nonumber\\
&=&\frac{V_3L^3}{8\pi G_5}\beta\int^{R_H}_{r_\lambda}dr\left(\frac{4}{r^5}+\frac{2B^2}{3L^2r}\right)+\mathcal{O}(B^4),
\end{eqnarray}
where $R_H$ is the horizon location, $r_\lambda$ is the UV-cutoff, $\beta=\frac{1}{T}$, $V_3=\int d^3x$  and $\sqrt{g}=\sqrt{\det g_{\mu\nu}}=\frac{L^5}{r^5}+\mathcal{O}(B^4)$. Computing the integral we get:
\begin{equation}
S_{\text{bulk}}^{\text{bh}}=\frac{V_3L^3}{8\pi G_5}\beta\left[-\frac{1}{R_H^4}+\frac{1}{r_\lambda^4}+\frac{2B^2}{3L^2}\ln\left(\frac{R_H}{r_\lambda}\right)\right].
\end{equation}
For $B=0$ we have:
\begin{equation}
S_{\text{bulk}}^{\text{bh}}=\frac{V_3L^3}{8\pi G_5}\beta\left[-\frac{1}{R_H^4}+\frac{1}{r_\lambda^4}\right],
\end{equation}
which is the same result obtained by \cite{BallonBayona:2007vp}.
\subsubsection*{Boundary action}
From \eqref{euclideanboundaryaction} we have:
\begin{eqnarray}
S^{\text{bh}}_{\text{bndy}}&=&\frac{-V_3}{8\pi G_5}\int^\beta_0dt_E\sqrt{\gamma}\left(-\frac{\sqrt{g^{rr}}\partial_r\sqrt{\gamma}}{\sqrt{\gamma}}-\frac{3}{L}-LB^2g^{xx}g^{yy}\ln \left(\frac{r}{L}\right)\right)\Bigg|_{r_\lambda}\nonumber\\
&=&\frac{-V_3L^3}{8\pi G_5}\beta\left[\frac{1}{r^4}-\frac{1}{2r_h^4}-\frac{B^2}{3L^2}+\frac{1}{3}\frac{B^2}{L^2}\ln\left(\frac{r}{\ell_d}\right)-\frac{B^2}{L^2}\ln\left(\frac{r}{L}\right)\right]\Bigg|_{r_\lambda}+\mathcal{O}(B^4).
\end{eqnarray}
For $B=0$ we have:
\begin{equation}
S^{\text{bh}}_{\text{bndy}}=\frac{-V_3L^3}{8\pi G_5}\beta\left[\frac{1}{r^4}-\frac{1}{2r_h^4}\right]\Bigg|_{r_\lambda},
\end{equation}
which is the same result found in \cite{BallonBayona:2007vp}.
\subsection{Thermal AdS - confined phase}
\subsubsection*{Bulk action}
\begin{eqnarray}
S^{\text{th}}_{\text{bulk}}&=&\frac{V_3}{8\pi G_5}\int^{\beta}_0dt_E\int^{r_0}_{r_\lambda}dr\sqrt{g}\left(\frac{4}{L^2}+\frac{2B^2}{3}g^{xx}g^{yy}\right)\nonumber\\
&=&\frac{V_3L^3}{8\pi G_5}\beta\int^{r_0}_{r_\lambda}dr\left(\frac{4}{r^5}+\frac{2B^2}{3L^2r}\right)+\mathcal{O}(B^4)\nonumber\\
&=&\frac{V_3L^3}{8\pi G_5}\beta\left[\frac{1}{r_\lambda^4}-\frac{1}{r_0^4}+\frac{2B^2}{3L^2}\ln\left(\frac{r_0}{r_\lambda}\right)\right]+\mathcal{O}(B^4),
\end{eqnarray}
where $\beta$ is the periodicity of the compactified time direction.
\subsubsection*{Boundary action}
\begin{eqnarray}
 S^{\text{th}}_{\text{bndy}}&=&-\frac{V_3}{8\pi G_5}\int^{\beta}_0dt_E\sqrt{\gamma}\left(-\frac{\sqrt{g^{rr}}\partial_r\sqrt{\gamma}}{\sqrt{\gamma}}-\frac{3}{L}-LB^2g^{xx}g^{yy}\ln \left(\frac{r}{L}\right)\right)\Bigg|_{r_\lambda}\nonumber\\
&=&\frac{V_3L^3}{8\pi G_5}\beta\left[-\frac{1}{r_\lambda^4}+\frac{B^2}{3L^2}-\frac{B^2}{3L^2}\ln\left(\frac{r_\lambda}{\ell_c}\right) + \frac{B^2}{L^2}\ln\left(\frac{r_\lambda}{L}\right)\right]+\mathcal{O}(B^4).
\end{eqnarray}

\section{Computational details on the Hawking-Page transition in the soft wall model}
\label{appsoft}
Here we present some computational details to determine the on-shell actions for the soft wall model.

\subsection{Black hole - deconfined phase}
\subsubsection*{Bulk action}
Using the Euclidean version of the black-hole metric \eqref{bhmetric} we can compute the black hole bulk action:
\begin{eqnarray}
S^{\text{bh}}_{\text{bulk}}&=&\frac{V_3}{8\pi G_5}\int^\beta_0dt_E\int^{R_H}_{r_\lambda}dr\sqrt{g}\left(\frac{4}{L^2}+\frac{2}{3}B^2g^{xx}g^{yy}\right)e^{-cr^2}\nonumber\\
&=&\frac{V_3L^3}{8\pi G_5}\beta\int^{R_H}_{r_\lambda}dre^{-cr^2}\left(\frac{4}{r^5}+\frac{2B^2}{3L^2r}\right)+\mathcal{O}(B^4),
\end{eqnarray}
where $R_H$ is the horizon location, $r_\lambda$ is the UV-cutoff, $\beta=\frac{1}{T}$  and $\sqrt{g}=\sqrt{\det g_{\mu\nu}}=\frac{L^5}{r^5}+\mathcal{O}(B^4)$. Solving the integral we get:
\begin{equation}
S_{\text{bulk}}^{\text{bh}}=\frac{V_3L^3}{8\pi G_5}\beta\left[e^{-cr^2}\left(\frac{-1}{r^4}+\frac{c}{r^2}\right)+\left(\frac{B^2}{3L^2}+c^2\right)\text{Ei}(-cr^2)\right]\Bigg|_{r_\lambda}^{R_H}+\mathcal{O}(B^4),
\end{equation}
where $\text{Ei}(x)\equiv-\int_{-x}^{\infty}\frac{e^{-t}}{t}dt$. For $B=0$ we have
\begin{equation}
S_{\text{bulk}}^{\text{bh}}=\frac{V_3L^3}{8\pi G_5}\beta\left[e^{-cr^2}\left(\frac{-1}{r^4}+\frac{c}{r^2}\right)+c^2Ei(-cr^2)\right]\Bigg|_{r_\lambda}^{RH},
\end{equation}
which is the same result obtained by \cite{BallonBayona:2007vp}.
\subsubsection*{Boundary action}
From \eqref{euclideanboundaryaction} we have:
\begin{eqnarray}
S^{\text{bh}}_{\text{bndy}}&=&-\frac{V_3}{8\pi G_5}\int^\beta_0dt_E\sqrt{\gamma}\left(-\frac{\sqrt{g^{rr}}\partial_r\sqrt{\gamma}}{\sqrt{\gamma}}-\frac{3}{L}-LB^2g^{xx}g^{yy}\ln \left(\frac{r}{L}\right)\right)\Bigg|_{r_\lambda}\nonumber\\
&=&\frac{-V_3L^3}{8\pi G_5}\beta\left[\frac{1}{r^4}-\frac{1}{2r_h^4}-\frac{B^2}{3L^2}+\frac{1}{3}\frac{B^2}{L^2}\ln\left(\frac{r}{\ell_d}\right)-\frac{B^2}{L^2}\ln\left(\frac{r}{L}\right)\right]\Bigg|_{r_\lambda}+\mathcal{O}(B^4).
\end{eqnarray}
For $B=0$ we have:
\begin{equation}
S^{\text{bh}}_{\text{bndy}}=-\frac{V_3L^3}{8\pi G_5}\beta\left[\frac{1}{r^4}-\frac{1}{2r_h^4}\right]\Bigg|_{r_\lambda},
\end{equation}
which is the same result found in \cite{BallonBayona:2007vp}.
\subsection{Thermal AdS - confined phase}
\subsubsection*{Bulk action}
\begin{eqnarray}
S^{\text{th}}_{\text{bulk}}&=&\frac{V_3}{8\pi G_5}\int^{\beta}_0dt_E\int^{\infty}_{r_\lambda}dr\sqrt{g}\left(\frac{4}{L^2}+\frac{2}{3}B^2g^{xx}g^{yy}\right)e^{-cr^2}\nonumber\\
&=&\frac{V_3L^3}{8\pi G_5}\beta\int^{\infty}_{r_\lambda}dre^{-cr^2}\left(\frac{4}{r^5}+
\frac{2B^2}{3L^2r}\right)+\mathcal{O}(B^4)\nonumber\\
&=&\frac{V_3L^3}{8\pi G_5}\beta\left[e^{-cr^2}\left(\frac{-1}{r^4}+\frac{c}{r^2}\right)+
\left(\frac{B^2}{3L^2}+c^2\right)\text{Ei}(-cr^2)\right]\Bigg|_{r_\lambda}^{\infty}+\mathcal{O}(B^4),
\end{eqnarray}
where $\beta$ is the periodicity of the compactified time direction.
\subsubsection*{Boundary action}
\begin{eqnarray}
S^{\text{th}}_{\text{bndy}}&=&-\frac{V_3}{8\pi G_5}\int^{\beta}_0dt_E\sqrt{\gamma}\left(-\frac{\sqrt{g^{rr}}\partial_r\sqrt{\gamma}}{\sqrt{\gamma}}-\frac{3}{L}-LB^2g^{xx}g^{yy}\ln\left(\frac{r}{L}\right)\right)\Bigg|_{r_\lambda}\nonumber\\
&=&\frac{V_3L^3}{8\pi G_5}\beta\left[-\frac{1}{r_\lambda^4}+\frac{B^2}{3L^2}+\frac{B^2}{3L^2}\ln\left(\frac{r_\lambda}{\ell_c}\right) - \frac{B^2}{L^2}\ln\left(\frac{r_\lambda}{L}\right)\right]+\mathcal{O}(B^4).
\end{eqnarray}

\section{Chiral condensate in holography}
\label{chirCondens}
The $\braket{\bar{\psi}\psi}$ condensate can be determined by differentiating $W = \log(Z)$ with respect to $m$, the bare quark mass, as\footnote{For clarity, we focus on a single quark flavor at this time. The final result has to be taken twice to account for both degenerate up and down quarks. The field theory Lagrangian is given by $\mathcal{L} = \bar{\psi}\left(\gamma^{\mu}\partial_{\mu} - m \right)\psi$.}
\begin{equation}
\frac{1}{Z}\frac{d Z}{d m} = \frac{\int \left[\mathcal{D} \psi \mathcal{D}\bar{\psi}\right]\left(\int d^4\mathbf{x}\bar{\psi}\psi\right) e^{-\int d^4\mathbf{x}\mathcal{L}}}{\int \left[\mathcal{D} \psi \mathcal{D}\bar{\psi}\right] e^{-\int d^4\mathbf{x}\mathcal{L}}}.
\end{equation}
Since in holography the path integrals are identified in bulk and boundary, one actually obtains
\begin{equation}
V_4 \braket{{\bar\psi}\psi} = -\frac{d}{dm}\left(\frac{N_c}{16\pi^2}\int d^5\mathbf{x}\sqrt{-g}e^{-\Phi}\left(g^{\mu\nu}\partial_{\mu}X\partial_{\nu}X + m^2X^2\right)\right).
\end{equation}
Restricting to a homogeneous condensate requires in the bulk $X(r,x^{\mu}) = X(r)$. A partial integration in the kinetic term, and using the equations of motion of $X(r)$, one retrieves
\begin{equation}
\braket{\bar{\psi}\psi} = -\frac{N_c}{16\pi^2}\frac{d}{dm}\left(\left.\sqrt{-g}e^{-\Phi}g^{r r}\partial_{r}X X\right|^{r=R_H}_{r=0}\right),
\end{equation}
when considering the black hole (deconfining) case. Since $g^{rr} = \frac{r^2}{L^2}f(r)$, and $f(R_H)=0$ by definition, the horizon contribution vanishes.\footnote{In the confining phase, one would have instead
\begin{equation}
\braket{\bar{\psi}\psi} = -\frac{N_c}{16\pi^2}\frac{d}{dm}\left(\left.\sqrt{-g}e^{-\Phi}g^{r r}\partial_{r}X X\right|^{r=+\infty}_{r=0}\right),
\end{equation}
where the contribution from the upper value $r=+\infty$ vanishes also here due to $e^{-\Phi}g^{rr}\sqrt{-g}\sim \frac{e^{-cr^2}}{r^3}$ and $X$ is assumed finite as $r\to\infty$. The argument also holds for the deconfining phase of the hard wall model. But note that making this argument in the confining phase of the hard wall model seems more subtle. Luckily, we will not need this in this work.} Moreover, $m$ is determined by the boundary expansion of $X$ as
\begin{equation}
L^{3/2}X(r) = m r + c m r^3 \log(\sqrt{c}r) + \sigma r^3 + \hdots.
\end{equation}
A closer look at the differential equation shows that the full solution $X \sim m$ as an overall prefactor.\footnote{This is also true for the deconfining phase in the hard wall model as we make explicit in subsection \ref{hwrevisit}. In fact, it is true as long as the differential equation is linear, as it simply represents an overall scaling of the solution. One would need to add terms of cubic or higher order in $X$ in the action to break this (unwanted) property. But then of course, the analysis presented here would have to be redone as evaluating the derivative w.r.t.~$m$ might not be so simple anymore. As far as we know, cf.~\cite{Gherghetta:2009ac,Fang:2015ytf,Chelabi:2015gpc}, the precise connection between the chiral condensate $\braket{\bar\psi\psi}$ and $\sigma$ is not considered in case higher order terms in $X$ are added and the chiral symmetry is primarily probed via $\sigma$ itself.} Hence the derivative w.r.t.~$m$ is readily performed and one finds
\begin{equation}
\braket{\bar{\psi}\psi} = \frac{N_c}{16\pi^2}\frac{2}{m}\left.\sqrt{-g}e^{-\Phi}g^{r r}\partial_{r}X X\right|_{r=0}.
\end{equation}
Inserting the explicit expansion, one obtains
\begin{eqnarray}
&&\frac{16\pi^2}{N_c} \frac{m}{2}\braket{\bar{\psi}\psi} =\nonumber\\ &&\left.\frac{1}{r^3}\left(m + 3mcr^2\log(\sqrt{c}r) + mcr^2 + 3\sigma r^2 + \hdots\right)\left( m r + m c r^3 \log(\sqrt{c}r) + \sigma r^3 + \hdots\right)\right|_{r=0} \nonumber\\
&=& \left.\frac{1}{r^3}\left(m^2 r + m^2 c r^3 \log(\sqrt{c}r) + m\sigma r^3 + 3m^2c r^3\log(\sqrt{c}r) + m^2c r^3 + 3m\sigma r^3 + \hdots\right)\right|_{r=0} \nonumber\\
&=& \left(m^2 \frac{1}{\epsilon^2} + 4m^2c\log(\sqrt{c}\epsilon) + m^2c + 4m\sigma \right).
\end{eqnarray}
Clearly, one needs holographic renormalization to proceed. We will however consider only differences between the $T\neq 0$ and the $T=0$ condensate, for which these divergent terms cancel out. Indeed, in thermal field theory one encounters no additional UV divergences besides those already present at $T=0$. The same is true when including non-zero $\mathcal{B}$: all divergences remain the same. One finds
\begin{equation}
\braket{\bar{\psi}\psi}_{\mathcal{B},T} -   \braket{\bar{\psi}\psi}_{\mathcal{B}=0,T=0} =\frac{N_c}{2\pi^2} \left(\sigma(\mathcal{B},T) - \sigma(\mathcal{B}=0,T=0)\right).
\end{equation}
The expression in the l.h.s.~is completely similar to the subtracted definition of the $B$-dependent chiral condensate on the lattice, see e.~g.~\cite{Bali:2011qj,Bali:2012zg}.

\section{Numerical value of $m$}
\label{barequarkmass}
In the soft wall model, one of the major disadvantages is that $\braket{\bar\psi\psi} \sim m$, and hence as the bare quark mass vanishes, so does the condensate, in direct opposition to QCD. The Gell-Mann-Oakes-Renner relation \cite{GellMann:1968rz},\footnote{Remember that $\braket{\bar Q Q} = 2 \braket{\bar\psi\psi}$.}
\begin{equation}\label{gmor}
  m_\pi^2\approx -(m_u+m_d)\frac{\braket{\bar Q Q}}{f_\pi^2},
\end{equation}
while keeping the pion mass $m_\pi$ and decay constant $f_\pi$ fixed at their experimental values, dictates, with $m_u=m_d=m$, that $\braket{\bar Q Q} \sim 1/m$ which conceivably leads to a large quark condensate. Roughly speaking, we might then also expect that, since $m$ in real life is quite small, the value of the condensate will get grossly underestimated in the soft wall model. This is indeed the case here. \\

\noindent To get a handle on this issue, we will artificially impose a very high bare quark mass to get a realistic value of the quark condensate. We will determine this artificial bare quark mass, by comparing with known lattice results at $B=0$. After this, we will use this same value of $m$ to look at the $B\neq0$ case. \\

\noindent The real quark condensate at finite temperature $T$ and $B=0$ is given by (equation (\ref{phycond}))
\begin{equation}
\braket{\bar{\psi}\psi}_T = \braket{\bar{\psi}\psi}_{T=0} + \frac{N_c mc}{2\pi^2}\left(\left.\frac{\sigma}{mc}\right|_T - \left.\frac{\sigma}{mc}\right|_{T=0}\right).
\end{equation}
The authors of \cite{Colangelo:2011sr} showed that the dimensionless combination $\frac{\sigma}{mc}$ vanishes at a temperature $T \approx 210$ MeV. As this is a physically reasonable value, we will \emph{impose} this value as the critical temperature for the real condensate as well. Note that this is an external and somewhat arbitrary choice that is used as further input in our model to constrain the parameters. Plugging in the numerical values of the quantities appearing here,\footnote{We use the value of the total condensate $\left|\left\langle \bar{Q}Q\right\rangle\right|\approx 0.013 \text{GeV}^3$ as given on page 151 of \cite{Shuryak:1988ck}. We also use the fact that $\left.\frac{\sigma}{mc}\right|_{T=0} = \frac{1+\gamma}{2}-\ln(2) \approx 0.095$.} we get
\begin{equation}
0.013~ \text{GeV}^3=  \frac{3\cdot 0.151 \cdot 2m}{2\pi^2}0.095~ \text{GeV}^2
\end{equation}
which leads to $m=2.967$ GeV.  The factor of 2 in the above expression originates from comparing with the full condensate (i.e. sum of up and down condensates). \\


\begin{thebibliography}{99}

\bibitem{Mamo:2015dea}
  K.~A.~Mamo,
  ``Inverse magnetic catalysis in holographic models of QCD,''
  JHEP {\bf 1505} (2015) 121
  [arXiv:1501.03262 [hep-th]].

\bibitem{maldacena}
 J.~M.~Maldacena,
  ``The Large N limit of superconformal field theories and supergravity,''
  Int.\ J.\ Theor.\ Phys.\  {\bf 38} (1999) 1113
  [Adv.\ Theor.\ Math.\ Phys.\  {\bf 2} (1998)  231]
  [hep-th/9711200].

 \bibitem{Aharony:1999ti}
  O.~Aharony, S.~S.~Gubser, J.~M.~Maldacena, H.~Ooguri and Y.~Oz,
  ``Large N field theories, string theory and gravity,''
  Phys.\ Rept.\  {\bf 323} (2000) 183
  [hep-th/9905111].

 \bibitem{Karch:2002sh}
 A.~Karch and E.~Katz, ``Adding flavor to AdS / CFT,''
  JHEP {\bf 0206} (2002) 043
  [hep-th/0205236].

\bibitem{polchinski}
J.~Polchinski and M.~J.~Strassler,
  ``The String dual of a confining four-dimensional gauge theory,''
  hep-th/0003136.


\bibitem{Sakai:2004cn}
  T.~Sakai and S.~Sugimoto,``Low energy hadron physics in holographic QCD,''
  Prog.\ Theor.\ Phys.\  {\bf 113} (2005) 843
  [hep-th/0412141].

\bibitem{Sakai:2005yt}
  T.~Sakai and S.~Sugimoto,``More on a holographic dual of QCD,''
  Prog.\ Theor.\ Phys.\  {\bf 114} (2005) 1083
  [hep-th/0507073].

\bibitem{Kruczenski:2003be}
 M.~Kruczenski, D.~Mateos, R.~C.~Myers and D.~J.~Winters,
  ``Meson spectroscopy in AdS / CFT with flavor,''
  JHEP {\bf 0307} (2003) 049
  [hep-th/0304032].

\bibitem{Kruczenski:2003uq}
 M.~Kruczenski, D.~Mateos, R.~C.~Myers and D.~J.~Winters,
  ``Towards a holographic dual of large N(c) QCD,''
  JHEP {\bf 0405} (2004) 041
  [hep-th/0311270].

\bibitem{Erdmenger:2007cm}
 J.~Erdmenger, N.~Evans, I.~Kirsch and E.~Threlfall,
  ``Mesons in Gauge/Gravity Duals - A Review,''
  Eur.\ Phys.\ J.\ A {\bf 35} (2008) 81
  [arXiv:0711.4467 [hep-th]].

\bibitem{Erlich:2005qh}
J.~Erlich, E.~Katz, D.~T.~Son and M.~A.~Stephanov,``QCD and a holographic model of hadrons,''
  Phys.\ Rev.\ Lett.\  {\bf 95} (2005) 261602
  [hep-ph/0501128].

\bibitem{Karch:2006pv}
  A.~Karch, E.~Katz, D.~T.~Son and M.~A.~Stephanov,
  ``Linear confinement and AdS/QCD,''
  Phys.\ Rev.\ D {\bf 74} (2006) 015005
  [hep-ph/0602229].

\bibitem{Karch:2010eg}
A.~Karch, E.~Katz, D.~T.~Son and M.~A.~Stephanov,
  ``On the sign of the dilaton in the soft wall models,''
  JHEP {\bf 1104} (2011) 066
  [arXiv:1012.4813 [hep-ph]].

\bibitem{dePaula:2008fp}
  W.~de Paula, T.~Frederico, H.~Forkel and M.~Beyer,
  ``Dynamical AdS/QCD with area-law confinement and linear Regge trajectories,''
  Phys.\ Rev.\ D {\bf 79} (2009) 075019
  [arXiv:0806.3830 [hep-ph]].

\bibitem{Gursoy:2008bu}
  U.~Gursoy, E.~Kiritsis, L.~Mazzanti and F.~Nitti,
  ``Deconfinement and Gluon Plasma Dynamics in Improved Holographic QCD,''
  Phys.\ Rev.\ Lett.\  {\bf 101} (2008) 181601
  [arXiv:0804.0899 [hep-th]].

\bibitem{Gursoy:2007cb}
 U.~Gursoy and E.~Kiritsis,
  ``Exploring improved holographic theories for QCD: Part I,''
  JHEP {\bf 0802} (2008) 032
  [arXiv:0707.1324 [hep-th]].

\bibitem{Gursoy:2007er}
U.~Gursoy, E.~Kiritsis and F.~Nitti,
  ``Exploring improved holographic theories for QCD: Part II,''
  JHEP {\bf 0802} (2008) 019
  [arXiv:0707.1349 [hep-th]].

\bibitem{Gursoy:2009jd}
 U.~Gursoy, E.~Kiritsis, L.~Mazzanti and F.~Nitti,
  ``Improved Holographic Yang-Mills at Finite Temperature: Comparison with Data,''
  Nucl.\ Phys.\ B {\bf 820} (2009) 148
  [arXiv:0903.2859 [hep-th]].

\bibitem{Jarvinen:2011qe}
 M.~Jarvinen and E.~Kiritsis,
  ``Holographic Models for QCD in the Veneziano Limit,''
  JHEP {\bf 1203} (2012) 002
  [arXiv:1112.1261 [hep-ph]].

\bibitem{Alho:2012mh}
T.~Alho, M.~Jarvinen, K.~Kajantie, E.~Kiritsis and K.~Tuominen,
  ``On finite-temperature holographic QCD in the Veneziano limit,''
  JHEP {\bf 1301} (2013) 093
  [arXiv:1210.4516 [hep-ph]].
	
	\bibitem{Johnson:2008vna}
  C.~V.~Johnson and A.~Kundu, ``External Fields and Chiral Symmetry Breaking in the Sakai-Sugimoto Model,''
  JHEP {\bf 0812} (2008) 053
  [arXiv:0803.0038 [hep-th]].

	
\bibitem{Callebaut:2013ria}
  N.~Callebaut and D.~Dudal, ``On the transition temperature(s) of magnetized two-flavour holographic QCD,''
  Phys.\ Rev.\ D {\bf 87} (2013) 106002
  [arXiv:1303.5674 [hep-th]].

\bibitem{Ballon-Bayona:2013cta}
  A.~Ballon-Bayona,``Holographic deconfinement transition in the presence of a magnetic field,''
  JHEP {\bf 1311} (2013) 168
  [arXiv:1307.6498 [hep-th]].
	
	
	\bibitem{Callebaut:2011ab}
  N.~Callebaut, D.~Dudal and H.~Verschelde, ``Holographic rho mesons in an external magnetic field,''
  JHEP {\bf 1303} (2013) 033
  [arXiv:1105.2217 [hep-th]].

\bibitem{Callebaut:2013wba}
  N.~Callebaut and D.~Dudal, ``A magnetic instability of the non-Abelian Sakai-Sugimoto model,''
  JHEP {\bf 1401} (2014) 055
  [arXiv:1309.5042 [hep-th]].
	
	\bibitem{Dudal:2014jfa}
  D.~Dudal and T.~G.~Mertens,
  ``Melting of charmonium in a magnetic field from an effective AdS/QCD model,''
  Phys.\ Rev.\ D {\bf 91} (2015) 086002
  [arXiv:1410.3297 [hep-th]].

\bibitem{Dudal:2015kza}
  D.~Dudal and T.~G.~Mertens,
  ``Radiation Gauge in AdS/QCD: inadmissibility and implications on spectral functions in the deconfined phase,''
  Phys.\ Lett.\ {\bf B} 751 (2015) 352
  [arXiv:1510.05490 [hep-th]].

\bibitem{Sadofyev:2015hxa}
  A.~V.~Sadofyev and Y.~Yin,
  ``The charmonium dissociation in an ``anomalous wind'',''  JHEP {\bf 1601} (2016) 052
    [arXiv:1510.06760 [hep-th]].

\bibitem{Ali-Akbari:2013txa}
  M.~Ali-Akbari and H.~Ebrahim,
  ``Chiral symmetry breaking: To probe anisotropy and magnetic field in quark-gluon plasma,''
  Phys.\ Rev.\ D {\bf 89} (2014) 6,  065029
  [arXiv:1309.4715 [hep-th]].

\bibitem{Ali-Akbari:2015bha}
  M.~Ali-Akbari, F.~Charmchi, A.~Davody, H.~Ebrahim and L.~Shahkarami,
  ``Time-dependent meson melting in an external magnetic field,''
  Phys.\ Rev.\ D {\bf 91} (2015) 106008
  [arXiv:1503.04439 [hep-th]].


\bibitem{Rougemont:2014efa}
 R.~Rougemont, R.~Critelli and J.~Noronha,
  ``Anisotropic heavy quark potential in strongly-coupled $\mathcal{N}=4$ SYM in a magnetic field,''
  Phys.\ Rev.\ D {\bf 91} (2015) 6,  066001
  [arXiv:1409.0556 [hep-th]].

\bibitem{Rougemont:2015oea}
 R.~Rougemont, R.~Critelli and J.~Noronha,
  ``Holographic calculation of the QCD crossover temperature in a magnetic field,''   Phys.\ Rev.\ D {\bf 93} (2016) 4,  045013
  [arXiv:1505.07894 [hep-th]].

\bibitem{D'Hoker:2009mm}
  E.~D'Hoker and P.~Kraus,
  ``Magnetic Brane Solutions in AdS,''
  JHEP {\bf 0910} (2009) 088
  [arXiv:0908.3875 [hep-th]].
	
\bibitem{D'Hoker:2009bc}
  E.~D'Hoker and P.~Kraus,
  ``Charged Magnetic Brane Solutions in AdS (5) and the fate of the third law of thermodynamics,''
  JHEP {\bf 1003} (2010) 095
  [arXiv:0911.4518 [hep-th]].

\bibitem{Herzog:2006ra}
  C.~P.~Herzog,
  ``A Holographic Prediction of the Deconfinement Temperature,''
  Phys.\ Rev.\ Lett.\  {\bf 98} (2007) 091601
  [hep-th/0608151].

\bibitem{Gherghetta:2009ac}
  T.~Gherghetta, J.~I.~Kapusta and T.~M.~Kelley,
  ``Chiral symmetry breaking in the soft-wall AdS/QCD model,''
  Phys.\ Rev.\ D {\bf 79} (2009) 076003
  [arXiv:0902.1998 [hep-ph]].

\bibitem{Bartz:2014oba}
  S.~P.~Bartz and J.~I.~Kapusta,
  ``Dynamical three-field AdS/QCD model,''
  Phys.\ Rev.\ D {\bf 90} (2014) 7,  074034
  [arXiv:1406.3859 [hep-ph]].

\bibitem{Fang:2015ytf}
Z.~Fang, S.~He and D.~Li,
  ``Chiral and Deconfining Phase Transitions from Holographic QCD Study,''
  arXiv:1512.04062 [hep-ph].

\bibitem{Chelabi:2015gpc}
K.~Chelabi, Z.~Fang, M.~Huang, D.~Li and Y.~L.~Wu,
  ``Chiral Phase Transition in the Soft-Wall Model of AdS/QCD,''
  arXiv:1512.06493 [hep-ph].

 \bibitem{Gusynin:1994re}
  V.~P.~Gusynin, V.~A.~Miransky and I.~A.~Shovkovy,
  ``Catalysis of dynamical flavor symmetry breaking by a magnetic field in (2+1)-dimensions,''
  Phys.\ Rev.\ Lett.\  {\bf 73} (1994) 3499
  [Phys.\ Rev.\ Lett.\  {\bf 76} (1996) 1005]
  [hep-ph/9405262].

\bibitem{Miransky:2002rp}
  V.~A.~Miransky and I.~A.~Shovkovy,
  ``Magnetic catalysis and anisotropic confinement in QCD,''
  Phys.\ Rev.\ D {\bf 66} (2002) 045006
  [hep-ph/0205348].

\bibitem{Mizher:2010zb}
  A.~J.~Mizher, M.~N.~Chernodub and E.~S.~Fraga,
  ``Phase diagram of hot QCD in an external magnetic field: possible splitting of deconfinement and chiral transitions,''
  Phys.\ Rev.\ D {\bf 82} (2010) 105016
  [arXiv:1004.2712 [hep-ph]].

\bibitem{Fraga:2008um}
  E.~S.~Fraga and A.~J.~Mizher,
  ``Can a strong magnetic background modify the nature of the chiral transition in QCD?,''
  Nucl.\ Phys.\ A {\bf 820} (2009) 103C
  [arXiv:0810.3693 [hep-ph]].

\bibitem{Gatto:2010pt}
  R.~Gatto and M.~Ruggieri,
  ``Deconfinement and Chiral Symmetry Restoration in a Strong Magnetic Background,''
  Phys.\ Rev.\ D {\bf 83} (2011) 034016
  [arXiv:1012.1291 [hep-ph]].

\bibitem{Gatto:2010qs}
  R.~Gatto and M.~Ruggieri,
  ``Dressed Polyakov loop and phase diagram of hot quark matter under magnetic field,''
  Phys.\ Rev.\ D {\bf 82} (2010) 054027
  [arXiv:1007.0790 [hep-ph]].

\bibitem{Osipov:2007je}
  A.~A.~Osipov, B.~Hiller, A.~H.~Blin and J.~da Providencia,
  ``Dynamical chiral symmetry breaking by a magnetic field and multi-quark interactions,''
  Phys.\ Lett.\ B {\bf 650} (2007) 262
  [hep-ph/0701090].

\bibitem{Kashiwa:2011js}
  K.~Kashiwa,
  ``Entanglement between chiral and deconfinement transitions under strong uniform magnetic background field,''
  Phys.\ Rev.\ D {\bf 83} (2011) 117901
  [arXiv:1104.5167 [hep-ph]].

\bibitem{Klimenko:1992ch}
  K.~G.~Klimenko,
  ``Three-dimensional Gross-Neveu model at nonzero temperature and in an external magnetic field,''
  Theor.\ Math.\ Phys.\  {\bf 90} (1992) 1
  [Teor.\ Mat.\ Fiz.\  {\bf 90} (1992) 3].

\bibitem{Fraga:2013ova}
E.~S.~Fraga, B.~W.~Mintz and J.~Schaffner-Bielich,
  ``A search for inverse magnetic catalysis in thermal quark-meson models,''
  Phys.\ Lett.\ B {\bf 731} (2014) 154
  [arXiv:1311.3964 [hep-ph]].
.
\bibitem{Alexandre:2000yf}
  J.~Alexandre, K.~Farakos and G.~Koutsoumbas,
  ``Magnetic catalysis in QED(3) at finite temperature: Beyond the constant mass approximation,''
  Phys.\ Rev.\ D {\bf 63} (2001) 065015
  [hep-th/0010211].

\bibitem{Filev:2007gb}
  V.~G.~Filev, C.~V.~Johnson, R.~C.~Rashkov and K.~S.~Viswanathan,
  ``Flavoured large N gauge theory in an external magnetic field,''
  JHEP {\bf 0710} (2007) 019
  [hep-th/0701001].

\bibitem{Albash:2007bk}
  T.~Albash, V.~G.~Filev, C.~V.~Johnson and A.~Kundu,
  ``Finite temperature large N gauge theory with quarks in an external magnetic field,''
  JHEP {\bf 0807} (2008) 080
  [arXiv:0709.1547 [hep-th]].

\bibitem{Bergman:2008sg}
  O.~Bergman, G.~Lifschytz and M.~Lippert,
  ``Response of Holographic QCD to Electric and Magnetic Fields,''
  JHEP {\bf 0805} (2008) 007
  [arXiv:0802.3720 [hep-th]].

\bibitem{Evans:2010xs}
  N.~Evans, T.~Kalaydzhyan, K.~y.~Kim and I.~Kirsch,
  ``Non-equilibrium physics at a holographic chiral phase transition,''
  JHEP {\bf 1101} (2011) 050
  [arXiv:1011.2519 [hep-th]].

\bibitem{Alam:2012fw}
  M.~S.~Alam, V.~S.~Kaplunovsky and A.~Kundu,
  ``Chiral Symmetry Breaking and External Fields in the Kuperstein-Sonnenschein Model,''
  JHEP {\bf 1204} (2012) 111
  [arXiv:1202.3488 [hep-th]].

\bibitem{Preis:2010cq}
  F.~Preis, A.~Rebhan and A.~Schmitt,
  ``Inverse magnetic catalysis in dense holographic matter,''
  JHEP {\bf 1103} (2011) 033
  [arXiv:1012.4785 [hep-th]].

\bibitem{Filev:2010pm}
  V.~G.~Filev and R.~C.~Raskov,
  ``Magnetic Catalysis of Chiral Symmetry Breaking. A Holographic Prospective,''
  Adv.\ High Energy Phys.\  {\bf 2010} (2010) 473206
  [arXiv:1010.0444 [hep-th]].

\bibitem{Callebaut:2011zz}
  N.~Callebaut, D.~Dudal and H.~Verschelde,
  ``Holographic study of magnetically induced QCD effects,''
  Acta Phys.\ Polon.\ Supp.\  {\bf 4} (2011) 671.

\bibitem{Bolognesi:2011un}
  S.~Bolognesi and D.~Tong,
  ``Magnetic Catalysis in AdS4,''
  Class.\ Quant.\ Grav.\  {\bf 29}  (2012) 194003
  [arXiv:1110.5902 [hep-th]].

\bibitem{Bali:2011qj}
  G.~S.~Bali, F.~Bruckmann, G.~Endr\"{o}di, Z.~Fodor, S.~D.~Katz, S.~Krieg, A.~Sch\"afer and K.~K.~Szabo,
  ``The QCD phase diagram for external magnetic fields,''
  JHEP {\bf 1202} (2012) 044
  [arXiv:1111.4956 [hep-lat]].

\bibitem{Bali:2012zg}
  G.~S.~Bali, F.~Bruckmann, G.~Endr\"{o}di, Z.~Fodor, S.~D.~Katz and A.~Sch\"afer,
  ``QCD quark condensate in external magnetic fields,''
  Phys.\ Rev.\ D {\bf 86} (2012) 071502
  [arXiv:1206.4205 [hep-lat]].

\bibitem{Bali:2014kia}
G.~S.~Bali, F.~Bruckmann, G.~Endr\"{o}di, S.~D.~Katz and A.~Sch\"afer,
  ``The QCD equation of state in background magnetic fields,''
  JHEP {\bf 1408} (2014) 177
  [arXiv:1406.0269 [hep-lat]].

\bibitem{Bonati:2014ksa}
  C.~Bonati, M.~D'Elia, M.~Mariti, M.~Mesiti, F.~Negro and F.~Sanfilippo,
  ``Anisotropy of the quark-antiquark potential in a magnetic field,''
  Phys.\ Rev.\ D {\bf 89} (2014) 11,  114502
  [arXiv:1403.6094 [hep-lat]].

\bibitem{Bonati:2013lca}
C.~Bonati, M.~D'Elia, M.~Mariti, F.~Negro and F.~Sanfilippo,
  ``Magnetic Susceptibility of Strongly Interacting Matter across the Deconfinement Transition,''
  Phys.\ Rev.\ Lett.\  {\bf 111} (2013) 182001
  [arXiv:1307.8063 [hep-lat]].

\bibitem{D'Elia:2011zu}
   M.~D'Elia and F.~Negro,
  ``Chiral Properties of Strong Interactions in a Magnetic Background,''
  Phys.\ Rev.\ D {\bf 83} (2011) 114028
  [arXiv:1103.2080 [hep-lat]].

\bibitem{D'Elia:2010nq}
M.~D'Elia, S.~Mukherjee and F.~Sanfilippo,
  ``QCD Phase Transition in a Strong Magnetic Background,''
  Phys.\ Rev.\ D {\bf 82} (2010) 051501
  [arXiv:1005.5365 [hep-lat]].

\bibitem{Ilgenfritz:2013ara}
  E.-M.~Ilgenfritz, M.~M\"uller-Preussker, B.~Petersson and A.~Schreiber,
  ``Magnetic catalysis (and inverse catalysis) at finite temperature in two-color lattice QCD,''
  Phys.\ Rev.\ D {\bf 89} (2014) 5,  054512
  [arXiv:1310.7876 [hep-lat]].

\bibitem{Ilgenfritz:2012fw}
   E.-M.~Ilgenfritz, M.~Kalinowski, M.~M\"uller-Preussker, B.~Petersson and A.~Schreiber,
  ``Two-color QCD with staggered fermions at finite temperature under the influence of a magnetic field,''
  Phys.\ Rev.\ D {\bf 85} (2012) 114504
  [arXiv:1203.3360 [hep-lat]].

\bibitem{Frasca:2011zn}
M.~Frasca and M.~Ruggieri,
  ``Magnetic Susceptibility of the Quark Condensate and Polarization from Chiral Models,''
  Phys.\ Rev.\ D {\bf 83} (2011) 094024
  [arXiv:1103.1194 [hep-ph]].

\bibitem{Fukushima:2012xw}
 K.~Fukushima and J.~M.~Pawlowski,
  ``Magnetic catalysis in hot and dense quark matter and quantum fluctuations,''
  Phys.\ Rev.\ D {\bf 86} (2012) 076013
  [arXiv:1203.4330 [hep-ph]].

\bibitem{Ferreira:2013oda}
M.~Ferreira, P.~Costa and C.~Provid\^{e}ncia,
  ``Deconfinement, chiral symmetry restoration and thermodynamics of (2+1)-flavor hot QCD matter in an external magnetic field,''
  Phys.\ Rev.\ D {\bf 89} (2014) 3,  036006
  [arXiv:1312.6733 [hep-ph]].

\bibitem{McInnes:2015kec}
B.~McInnes,
  ``Inverse Magnetic/Shear Catalysis,''
  arXiv:1511.05293 [hep-th].

\bibitem{Kharzeev:2012ph}
  D.~E.~Kharzeev, K.~Landsteiner, A.~Schmitt and H.~U.~Yee,
  ``'Strongly interacting matter in magnetic fields': an overview,''
  Lect.\ Notes Phys.\  {\bf 871} (2013) 1
  [arXiv:1211.6245 [hep-ph]].

\bibitem{Miransky:2015ava}
 V.~A.~Miransky and I.~A.~Shovkovy,
  ``Quantum field theory in a magnetic field: From quantum chromodynamics to graphene and Dirac semimetals,''
  Phys.\ Rept.\  {\bf 576} (2015) 1
  [arXiv:1503.00732 [hep-ph]].

\bibitem{Kharzeev:2007jp}
D.~E.~Kharzeev, L.~D.~McLerran and H.~J.~Warringa,
  ``The Effects of topological charge change in heavy ion collisions: 'Event by event P and CP violation',''
  Nucl.\ Phys.\ A {\bf 803} (2008) 227
  [arXiv:0711.0950 [hep-ph]].

\bibitem{Skokov:2009qp}
V.~Skokov, A.~Y.~Illarionov and V.~Toneev,
  ``Estimate of the magnetic field strength in heavy-ion collisions,''
  Int.\ J.\ Mod.\ Phys.\ A {\bf 24} (2009) 5925
  [arXiv:0907.1396 [nucl-th]].

\bibitem{Bzdak:2011yy}
A.~Bzdak and V.~Skokov,
  ``Event-by-event fluctuations of magnetic and electric fields in heavy ion collisions,''
  Phys.\ Lett.\ B {\bf 710} (2012) 171
  [arXiv:1111.1949 [hep-ph]].

\bibitem{Deng:2012pc}
 W.~T.~Deng and X.~G.~Huang,
  ``Event-by-event generation of electromagnetic fields in heavy-ion collisions,''
  Phys.\ Rev.\ C {\bf 85} (2012) 044907
  [arXiv:1201.5108 [nucl-th]].

\bibitem{Tuchin:2013apa}
 K.~Tuchin,
  ``Time and space dependence of the electromagnetic field in relativistic heavy-ion collisions,''
  Phys.\ Rev.\ C {\bf 88} (2013) 2,  024911
  [arXiv:1305.5806 [hep-ph]].

\bibitem{Tuchin:2013ie}
 K.~Tuchin,
  ``Particle production in strong electromagnetic fields in relativistic heavy-ion collisions,''
  Adv.\ High Energy Phys.\  {\bf 2013} (2013) 490495
  [arXiv:1301.0099].

\bibitem{McLerran:2013hla}
 L.~McLerran and V.~Skokov,
  ``Comments About the Electromagnetic Field in Heavy-Ion Collisions,''
  Nucl.\ Phys.\ A {\bf 929} (2014) 184
  [arXiv:1305.0774 [hep-ph]].

\bibitem{Fraga:2012fs}
 E.~S.~Fraga and L.~F.~Palhares,
  ``Deconfinement in the presence of a strong magnetic background: an exercise within the MIT bag model,''
  Phys.\ Rev.\ D {\bf 86} (2012) 016008
  [arXiv:1201.5881 [hep-ph]].

\bibitem{Fraga:2012ev}
 E.~S.~Fraga, J.~Noronha and L.~F.~Palhares,
  ``Large $N_c$ Deconfinement Transition in the Presence of a Magnetic Field,''
  Phys.\ Rev.\ D {\bf 87} (2013) 11,  114014
  [arXiv:1207.7094 [hep-ph]].

\bibitem{Fukushima:2012kc}
K.~Fukushima and Y.~Hidaka,
  ``Magnetic Catalysis Versus Magnetic Inhibition,''
  Phys.\ Rev.\ Lett.\  {\bf 110} (2013) 3,  031601
  [arXiv:1209.1319 [hep-ph]].

\bibitem{Ayala:2014iba}
 A.~Ayala, M.~Loewe, A.~J.~Mizher and R.~Zamora,
  ``Inverse magnetic catalysis for the chiral transition induced by thermo-magnetic effects on the coupling constant,''
  Phys.\ Rev.\ D {\bf 90} (2014) 3,  036001
  [arXiv:1406.3885 [hep-ph]].

\bibitem{Ayala:2014gwa}
 A.~Ayala, M.~Loewe and R.~Zamora,
  ``Inverse magnetic catalysis in the linear sigma model with quarks,''
  Phys.\ Rev.\ D {\bf 91} (2015) 1,  016002
  [arXiv:1406.7408 [hep-ph]].

\bibitem{Ayala:2015bgv}
 A.~Ayala, C.~A.~Dominguez, L.~A.~Hernandez, M.~Loewe and R.~Zamora,
  ``Inverse magnetic catalysis from the properties of the QCD coupling in a magnetic field,''
  arXiv:1510.09134 [hep-ph].
	
	\bibitem{Farias:2014eca}
  R.~L.~S.~Farias, K.~P.~Gomes, G.~I.~Krein and M.~B.~Pinto,
  ``Importance of asymptotic freedom for the pseudocritical temperature in magnetized quark matter,''
  Phys.\ Rev.\ C {\bf 90} (2014) 2,  025203
  [arXiv:1404.3931 [hep-ph]].

\bibitem{Ferreira:2014kpa}
 M.~Ferreira, P.~Costa, O.~Lourenço, T.~Frederico and C.~Provid\^{e}ncia,
  ``Inverse magnetic catalysis in the (2+1)-flavor Nambu-Jona-Lasinio and Polyakov-Nambu-Jona-Lasinio models,''
  Phys.\ Rev.\ D {\bf 89} (2014) 11,  116011
  [arXiv:1404.5577 [hep-ph]].
	
	\bibitem{Costa:2015bza}
 P.~Costa, M.~Ferreira, D.~P.~Menezes, J.~Moreira and C.~Provid\^{e}ncia,
  ``Influence of the inverse magnetic catalysis and the vector interaction in the location of the critical end point,''
  Phys.\ Rev.\ D {\bf 92} (2015) 3,  036012
  [arXiv:1508.07870 [hep-ph]].

\bibitem{Mueller:2015fka}
 N.~Mueller and J.~M.~Pawlowski,
  ``Magnetic catalysis and inverse magnetic catalysis in QCD,''
  Phys.\ Rev.\ D {\bf 91} (2015) 116010
  [arXiv:1502.08011 [hep-ph]].

\bibitem{Colangelo:2011sr}
  P.~Colangelo, F.~Giannuzzi, S.~Nicotri and V.~Tangorra,
  ``Temperature and quark density effects on the chiral condensate: An AdS/QCD study,''
  Eur.\ Phys.\ J.\ C {\bf 72} (2012) 2096
  [arXiv:1112.4402 [hep-ph]].

\bibitem{Jugeau:2013zza}
  F.~Jugeau, S.~Narison and H.~Ratsimbarison,
  ``SVZ+1/q$^2$-expansion versus some QCD holographic models,''
  Phys.\ Lett.\ B {\bf 722} (2013) 111
  [arXiv:1302.6909 [hep-ph]].
	
	\bibitem{Colangelo:2008us}
  P.~Colangelo, F.~De Fazio, F.~Giannuzzi, F.~Jugeau and S.~Nicotri,
  ``Light scalar mesons in the soft-wall model of AdS/QCD,''
  Phys.\ Rev.\ D {\bf 78} (2008) 055009
  [arXiv:0807.1054 [hep-ph]].


\bibitem{Krikun:2008tf}
 A.~Krikun,
  ``On two-point correlation functions in AdS/QCD,''
  Phys.\ Rev.\ D {\bf 77} (2008) 126014
  [arXiv:0801.4215 [hep-th]].
	
	\bibitem{BallonBayona:2007vp}
  C.~A.~Ballon Bayona, H.~Boschi-Filho, N.~R.~F.~Braga and L.~A.~Pando Zayas,
  ``On a Holographic Model for Confinement/Deconfinement,''
  Phys.\ Rev.\ D {\bf 77} (2008) 046002
  [arXiv:0705.1529 [hep-th]].

\bibitem{Balasubramanian:1999re}
  V.~Balasubramanian and P.~Kraus,
  ``A Stress tensor for Anti-de Sitter gravity,''
  Commun.\ Math.\ Phys.\  {\bf 208} (1999) 413
  [hep-th/9902121].
		
	\bibitem{Andreev:2006ct}
  O.~Andreev and V.~I.~Zakharov,
  ``Heavy-quark potentials and AdS/QCD,''
  Phys.\ Rev.\ D {\bf 74} (2006) 025023
  [hep-ph/0604204].
	
	\bibitem{Ghoroku:2005kg}
K.~Ghoroku and M.~Yahiro,
  ``Holographic model for mesons at finite temperature,''
  Phys.\ Rev.\ D {\bf 73} (2006) 125010
    [hep-ph/0512289].

\bibitem{Kim:2006ut}
K.~Jo, Y.~Kim, H.~K.~Lee and S.~J.~Sin,
  ``Quark number Susceptibility and Phase Transition in hQCD Models,''
  JHEP {\bf 0811} (2008) 040
  [arXiv:0810.0063 [hep-ph]].
	
	\bibitem{Lindgren:2015lia}
J.~Lindgren, I.~Papadimitriou, A.~Taliotis and J.~Vanhoof,
  ``Holographic Hall conductivities from dyonic backgrounds,''
  JHEP {\bf 1507} (2015) 094
  [arXiv:1505.04131 [hep-th]].
	
		\bibitem{Veschgini:2010ws}
  K.~Veschgini, E.~Megias and H.~J.~Pirner,
  ``Trouble Finding the Optimal AdS/QCD,''
  Phys.\ Lett.\ B {\bf 696} (2011) 495
  [arXiv:1009.4639 [hep-th]].

\bibitem{GellMann:1968rz}
  M.~Gell-Mann, R.~J.~Oakes and B.~Renner,
  ``Behavior of current divergences under SU(3) x SU(3),''
  Phys.\ Rev.\  {\bf 175} (1968) 2195.
	
	\bibitem{Shuryak:1988ck}
  E.~V.~Shuryak,
  ``The QCD vacuum, hadrons and the superdense matter,''
  World Sci.\ Lect.\ Notes Phys.\  {\bf 71} (2004) 1
   [World Sci.\ Lect.\ Notes Phys.\  {\bf 8} (1988) 1].

\end{thebibliography}
\end{document}